\pdfoutput=1
\documentclass[10pt,amsmath,amssymb,nofootinbib,twoside,twocolumn,superscriptaddress,
floats,floatfix,longbibliography,aps,prd]{revtex4-2}
\usepackage[british]{babel}
\usepackage{txfonts,mathrsfs,tensor}
\usepackage{natbib,textcase}
\usepackage[caption=false]{subfig}
\usepackage{graphicx,color,xcolor,soul}
\usepackage{enumitem}
\setlist[description]{labelindent=0em,leftmargin=0em,font=\normalfont\itshape}

\usepackage{etoolbox}
\apptocmd{\sloppy}{\hbadness 10000\relax}{}{}

\usepackage[pdftex,pdfusetitle]{hyperref}
\hypersetup{colorlinks=true,linkcolor=red,citecolor=blue,filecolor=black,urlcolor=black,
pdfauthor={Edgardo Franzin, Stefano Liberati and Mauro Oi}}
\usepackage[capitalize]{cleveref}
\crefname{subequation}{Eqs.}{Eqs.}
\Crefname{subequation}{Equations}{Equations}
\labelcrefformat{subequation}{#2(#1)#3}
\usepackage[exponent-product=\cdot]{siunitx}

\makeatletter\g@addto@macro\bfseries{\boldmath}\makeatother%

\def\be#1\ee{\begin{align}#1\end{align}}

\allowdisplaybreaks

\newcommand{\ie}{i.e.}
\newcommand{\eg}{e.g.}
\renewcommand{\d}{\text{d}}
\newcommand{\e}{\text{e}}

\newcommand{\0}{\nonumber}
\newcommand{\iu}{\mathrm{i}\mkern1mu}

\renewcommand{\geq}{\geqslant}

\renewcommand{\leq}{\leqslant}

\DeclareMathOperator\const{const}

\begin{document}

\title{Superradiance in Kerr-like black holes}


\newcommand{\CBPF}{\affiliation{Department of Astrophysics, Cosmology and Fundamental Interactions (COSMO), Centro Brasileiro de Pesquisas F\'isicas (CBPF), rua Dr.\ Xavier Sigaud 150, Urca, Rio de Janeiro -- RJ, 22290-180 Brazil}}
\newcommand{\SISSA}{\affiliation{SISSA, International School for Advanced Studies, via Bonomea 265, 34136 Trieste, Italy}}
\newcommand{\InfnTS}{\affiliation{INFN, Sezione di Trieste, via Valerio 2, 34127 Trieste, Italy}}
\newcommand{\IFPU}{\affiliation{IFPU, Institute for Fundamental Physics of the Universe, via Beirut 2, 34014 Trieste, Italy}}
\newcommand{\UCagliari}{\affiliation{Dipartimento di Fisica, Universit\`a di Cagliari, Cittadella Universitaria, 09042 Monserrato, Italy}}
\newcommand{\InfnCA}{\affiliation{INFN, Sezione di Cagliari, Cittadella Universitaria, 09042 Monserrato, Italy}}

\author{Edgardo Franzin}
\CBPF\SISSA\IFPU\InfnTS

\author{Stefano Liberati}
\SISSA\IFPU\InfnTS

\author{Mauro Oi}
\UCagliari\InfnCA

\begin{abstract}
Recent strong-field regime tests of gravity are so far in agreement with general relativity.
In particular, astrophysical black holes appear all to be consistent with the Kerr spacetime, but the statistical error on current observations allows for small yet detectable deviations from this description.
Here we study superradiance of scalar and electromagnetic test fields around the Kerr-like Konoplya--Zhidenko black hole and we observe that for large values of the deformation parameter superradiance is highly suppressed with respect to the Kerr case. Surprisingly, there exists a range of small values of the deformation parameter for which the maximum amplification factor is larger than the Kerr one.
We also provide a first result about the superradiant instability of these non-Kerr spacetimes against massive scalar fields.
\end{abstract}

\maketitle%

\section{Introduction}

General relativity has been extensively and successfully tested~\cite{Will:2014kxa,Yagi:2016jml} from the weak to the strong regime --- the most recent results being the detection of gravitational waves produced by the merger of two black holes~\cite{Abbott:2016blz} and the observation of the shadow of the supermassive black hole M87*~\cite{Akiyama:2019cqa}.
Nowadays black holes are widely accepted as astrophysical objects~\cite{Celotti:1999tg,Bambi:2015kza} compatible with the Kerr metric~\cite{Kerr:1963ud}, yet, we still do not have the ultimate evidence for such black holes to exactly match this general-relativistic solution, as their defining property --- the event horizon --- is intrinsically not \emph{directly} observable~\cite{Visser:2014zqa,Abramowicz:2016lja,Cardoso:2016rao,Cardoso:2019rvt}.

There exists a number of alternative theories of gravity as well as exotic compact objects proposed to compete or substitute black holes.
These black-hole mimickers typically share the same features at large distances, while they present qualitative differences close to the event horizon.
Current and future gravitational-wave observations are and will be able to test general relativity, the no-hair theorem, the near-horizon geometry, distinguish the Kerr spacetime from putative alternatives, and even probe quantum gravity effects~\cite{TheLIGOScientific:2016src,Johannsen:2015mdd,Yunes:2016jcc,Cardoso:2016ryw,Johannsen:2016uoh,Cardoso:2019rvt,Carballo-Rubio:2018jzw}.
These effects in a consistent setup are commonly invoked to regularize spacetime singularities, which are inevitable in classical general relativity~\cite{Penrose:1964wq}.

While nowadays observations agree with numerical simulations based on Einstein gravity, the current uncertainties on the measurements of the black-hole parameters leave room for alternatives.
A possible framework is to describe this freedom by introducing suitable parametrized deviations from the Kerr geometry. The observed interval values for the black-hole mass $M$ and angular momentum $J=aM$ can be therefore translated in an allowed range for the deviation parameters.
Of course, we do not expect these deviations to be large or they would be observable in the weak-field regime as well.
But, for instance, one can consider non-negligible deviations from Kerr and obtain the same quasinormal frequencies.
If the geometry of the spacetime is different from Kerr only in a small region near the would-be horizon, asymptotically the geometry would be barely distinguishable from Kerr, leaving a weak signature in the form of gravitational-wave echoes at late times~\cite{Ferrari:2000sr,Cardoso:2016rao,Abedi:2020ujo,Cardoso:2016oxy,Cardoso:2017cqb,Berti:2018vdi}.

From this point of view, instead of testing a specific theory against general relativity case by case and/or a specific black-hole alternative, it could be more convenient to work in a model-independent framework describing the most generic black holes in any metric theory of gravity.
The idea of this framework is similar to the parametrized post-Newtonian (PPN) formalism~\cite{Will:2014kxa} but in this case it is valid in the whole space outside the event horizon.

In Refs.~\cite{Johannsen:2011dh,Johannsen:2015pca,Johannsen:2015hib}, deviations from general relativity and the general-relativistic black-hole geometry are written in terms of an expansion in $M/r$ being $r$ some radial coordinate. Some coefficients are easily constrained with the PPN parameters, while a very large number of equally important coefficients remains undetermined in the near-horizon region, with the additional drawback of a lack of a hierarchy among them.
Even if this formulation works well for small deviations from general relativity, it fails for \eg\ Einstein--dilaton--Gauss--Bonnet with large coupling constants~\cite{Cardoso:2014rha}.

A more robust general parametrization to describe, respectively, spherically symmetric and axisymmetric asymptotically flat black holes has been introduced by Konoplya, Rezzolla and Zhidenko in Refs.~\cite{Rezzolla:2014mua,Konoplya:2016jvv}, and tested to constrain deviations from the Kerr hypothesis with the iron-line method~\cite{Ni:2016uik,Cardenas-Avendano:2016zml,Nampalliwar:2019iti} and to produce black-hole shadows simulations~\cite{Younsi:2016azx,Mizuno:2018lxz}.
In this framework, deviations from general relativity and the Kerr metric are given again as an expansion whose coefficient values can be fixed from observations in the strong-gravity regime (close to the horizon) and in the post-Newtonian region (far from the black hole).
This parametrization also allows for non-spherical deformations of the horizon, provides a faster convergence of the series, and typically requires a small number of parameters to approximate known solutions to the desired precision. Besides, there exists a hierarchy among the parameters.

A different perspective is to modify each mass and spin term in the Kerr metric and test whether the magnitude of the spacetime curvature matches with that predicted by general relativity~\cite{Ghasemi-Nodehi:2016wao}.
More recently, the work of Ref.~\cite{Johannsen:2015pca} has been extended to the most general stationary, axisymmetric and asymptotically flat spacetime with separable geodesic equations~\cite{Carson:2020dez}.

However, even if these parametrizations may depend on a large number of parameters to be fixed with data, it is natural to think that astrophysical observables --- \eg\ quasinormal frequencies, orbits of particles, accretion, parameters of the shadow, electromagnetic radiation --- depend only on a few of them~\cite{Konoplya:2020hyk}.

A common feature of rotating spacetimes is the multifaceted phenomenon of superradiance~\cite{Zeldovich:1971,Bekenstein:1998nt,Brito:2015oca}: in a gravitational system and under certain conditions, the scattering of radiation off absorbing rotating objects produces waves with amplitude larger than the incident one. For a monochromatic wave of frequency $\omega$ scattering off a body with angular velocity $\Omega$, the superradiant condition is satisfied as long as $\omega<m\Omega$, being $m$ the azimuthal number with respect to the rotation axis.

When rotating black holes are surrounded with matter, superradiance gives rise to exponentially growing modes, \ie\ black-hole bombs~\cite{Press:1972zz,Cardoso:2004nk}.
The scattering of massive fields produces a similar effect: the mass term can effectively confine the field giving rise to floating orbits and superradiant instabilities which extract rotational energy away from the black hole~\cite{Damour:1976kh,Detweiler:1980uk,Cardoso:2011xi}.
The observation or the absence of effects related to these instabilities can be used to impose bounds on the mass of ultralight bosons, see e.g.\ Refs.~\cite{Witek:2012tr,Brito:2013wya,Brito:2017wnc,Brito:2017zvb,Cardoso:2018tly}.

Similarly to the Kerr black hole, Kerr-like spacetimes dissipate energy as well as any classical dissipative system, and the aim of this paper is to investigate differences and analogies for these objects with respect to the superradiant scattering around Kerr black holes.
We stress that these spacetimes are not solutions to the field equations of any specific gravitational theory, meaning that we can only study test fields propagating in these backgrounds while the gravitational-wave dynamics is excluded.
However, in extended theories of gravity exact rotating solutions are difficult to derive and in some cases they are known only perturbatively in the spin parameter, or numerically.
To our knowledge, there are no studies of superradiant amplification in these extended theories, neither for those which admit general-relativistic solutions~\cite{Yunes:2013dva,Berti:2015itd} but predict different dynamics.

In the most general parametrization, there is no reason to believe that the separability property of the Kerr metric is guaranteed, not even for the Klein--Gordon equation.
In particular, the class of Kerr-like spacetimes which allows for the separation of variables in the Klein--Gordon and Hamilton--Jacobi equations has been derived in Ref.~\cite{Konoplya:2018arm}, which is a subclass of the Johannsen metrics~\cite{Johannsen:2015pca}.
In this paper we show that, under given conditions, a subclass of the metrics presented in Ref.~\cite{Konoplya:2018arm} also allows for the separation of variables in the Maxwell equation.

The results presented in this paper are mostly relative to the Konoplya--Zhidenko black hole~\cite{Konoplya:2016pmh}, which introduces a single extra parameter. Despite its simplicity, this model preserves a lot of features of the Kerr spacetime: the asymptotic properties, the post-Newtonian expansion coefficients, the relation between quadrupole moment and mass, the spherical horizon, and the mirror symmetry.
Yet, it allows for significant differences in the near-horizon region~\cite{Wang:2016paq,Wang:2017hjl,Konoplya:2019xmn}.

The scope of this paper is twofold: first we analyze the structure of the Konoplya--Zhidenko spacetime, and second we study superradiant scattering of test fields.
In particular, the paper is organized as follows.
In \cref{s:deformed} we review the family of spacetimes which admits separability of the perturbative equations for massless spin-0 and \mbox{spin-1} fields, with a particular focus on the Konoplya--Zhidenko rotating black hole.
In \cref{s:results} we present our results regarding the superradiant emission in the Konoplya--Zhidenko spacetime for massless and massive bosonic test fields.
Finally, we conclude with a discussion and prospects in \cref{s:conclusion}.
In \cref{app:maxwell} we derive the angular and radial equations for a general non-Kerr black-hole parametrization and we study their boundary conditions.
In \cref{app:mattercontent} we provide useful formulas for the Konoplya--Zhidenko spacetime, namely the Einstein tensor, the geodesic equations and the four-velocity of a zero-angular-momentum observer.
In \cref{app:instability} we study the instability of the Konoplya--Zhidenko black hole against massive scalar fields in the low-frequency, small-mass and small-deformation limit.
Throughout this work we use $G=c=1$ units.

\section{Parametrized Kerr-like spacetimes and the Konoplya--Zhidenko black hole\label{s:deformed}}

The metric of a generic axially symmetric, stationary and asymptotically flat spacetime can be written as
\be
\d s^2 = - \frac{N^2 - W^2\sin^2\theta}{K^2}\,\d t^2
-2Wr \sin^2\theta\,\d t\,\d\varphi\0\\
+K^2 r^2 \sin^2\theta\,\d\varphi^2 
+ \frac{\Sigma}{r^2} \left(\frac{B^2}{N^2}\,\d r^2 + r^2\,\d\theta^2\right),\label{eq:gen_axi_st}
\ee
where $N$, $W$, $K$, $\Sigma$ and $B$ are in general functions of \mbox{$r$ and $\theta$}. In this paper we focus on parametrized Kerr-like spacetimes which possess Kerr-like symmetries and admit separable Klein--Gordon equations for test fields~\cite{Konoplya:2018arm}.
As in Ref.~\cite{Konoplya:2018arm}, we are not interested in the general conditions for the separability of variables, which are related to the symmetry of the background and the choice of appropriate coordinates.
Being our pragmatic objective to test strong-gravity effects in an asymptotically flat and axisymmetric spacetime describing a Kerr-like black hole, we can simplify the above spacetime metric leaving only three arbitrary functions of the radial coordinate, so that
\begin{subequations}\label{eq:sep_functions}
\be
B(r,\theta)&=R_B(r),\quad
\Sigma(r,\theta)=r^2 R_\Sigma(r) + a^2\cos^2\theta,\\
W(r,\theta)&=\frac{a R_M(r)}{\Sigma(r,\theta)},\quad
N^2(r,\theta)=R_\Sigma(r)-\frac{R_M(r)}{r}+\frac{a^2}{r^2},\\
K^2(r,\theta)&=\frac{1}{\Sigma(r,\theta)}\left[r^2 R_\Sigma^2(r) + a^2 R_\Sigma(r) + a^2\cos^2\theta\,N^2(r,\theta)\right]\0\\
&\phantom{=}+\frac{a\,W(r,\theta)}{r}.
\ee
\end{subequations}
For further convenience we define $\Delta\equiv r^2 N^2 = r^2 R_\Sigma - R_M r + a^2$ and we observe that for this class of spacetimes the event horizon is defined by the largest positive root of $\Delta=0$.

Asymptotic flatness and current PPN parameters imply $R_M \to 2M + O\,(1/r^2)$ as $r\to\infty$. With a suitable change of the radial coordinate it is possible to set $R_B$ or $R_\Sigma$ to 1, so only two of the three radial functions are independent.
The Kerr metric is recovered for $R_\Sigma=R_B=1$ and $R_M=2M$.
\Cref{eq:sep_functions} describe a Petrov D spacetime, and as a consequence, the Hamilton--Jacobi equation is separable with a generalized Carter constant~\cite{Konoplya:2018arm} --- see also \cref{app:mattercontent}.
In \cref{app:maxwell}, we show that the subclass of this spacetime such that $R_B=1$ and $R_\Sigma=(1+\xi/r)^2$ also admits separable Maxwell equations for test fields.

A minimal deformation for the Kerr spacetime was introduced by Konoplya and Zhidenko in Ref.~\cite{Konoplya:2016pmh} and can be obtained from \cref{eq:sep_functions} by setting $R_\Sigma=R_B=1$ and $R_M = 2M + \eta/r^2$.
For the rest of the paper we consider this background geometry.

\subsection{Event horizons and causal structure\label{KZblackhole:horizons}}

For the Konoplya--Zhidenko metric the event horizon radius is given by the largest positive real root of $\Delta = r^2 - 2Mr + a^2 - \eta/r = 0$, which in general admits three (possibly complex-valued) solutions
\be
r_k &= \frac{2M}{3} + \frac{2}{3} \sqrt{4 M^2-3 a^2} \cos\left(\beta-\frac{2k\pi}{3}\right),\label{horizons}\\
\beta &= \frac{1}{3} \cos^{-1}\frac{16 M^3 - 18 M a^2 + 27\eta}{2 \left(4 M^2-3 a^2\right)^{3/2}}\,,\quad k=0,1,2\,.\0
\ee
We immediately notice that the Kerr limit $\eta\to0$ is not continuous, as in looking for the roots of $\Delta=0$ we pass from solving a cubic to a quadratic equation.
Nevertheless, for $a<M$ and in the small $\eta/M^3$ limit, we have
\be\label{r0smalleta}
r_0 = r_+ + \frac{\eta}{r_+ (r_+-r_-)} - \frac{\eta^2 (2r_+ - r_-)}{r_+^3 (r_+-r_-)^3} + O\left(\eta^3\right),
\ee
where $r_\pm = M \pm \sqrt{M^2-a^2}$ are the radii of the event and Cauchy horizon for the Kerr spacetime.
For $|\eta|/M^3\lesssim0.07$ the difference between $r_0$ calculated as a linear correction to $r_+$ and the exact value as in \cref{horizons} is less than 1\% for values of $a\lesssim0.9M$.
\Cref{r0smalleta} does not apply in the extremal limit, which must be treated separately, as in this case the leading order correction is $O\left(\eta^{1/2}\right)$ and $r_0$ is given by
\be
r_0 = M+\sqrt{\frac{\eta}{M}}-\frac{\eta}{2M^2}+O\left(\eta^{3/2}\right).
\ee
Under these assumptions, the compactness of the spacetime for $a<M$ is
\be\label{compactness}
\mathcal{C} = \mathcal{C}_\text{Kerr}\left(1 - \frac{\eta}{r_+^2(r_+-r_-)}\right) + O\left(\eta^2\right),
\ee
being $\mathcal{C}_\text{Kerr}=M/r_+$ the compactness of the Kerr black hole, while in the extremal case ($\mathcal{C}_\text{Kerr}=1$)
\be
\mathcal{C}=1-\sqrt{\frac{\eta}{M^3}} + \frac{3\eta}{2M^3}+O\left(\eta^{3/2}\right).
\ee
\Cref{compactness} indicates that positive (negative) values of $\eta$ corresponds to less (more) compact configurations.

For $a<M$, instead of working with $\eta$, deviations from the Kerr spacetime can be parametrized in terms of the quantity $\delta r$, such that the position of the event horizon can be written as $r_0 = r_+ + \delta r$ --- cfr.\ \cref{r0smalleta}, although $\delta r$ can account for large values of $\eta/M^3$ and it is not limited to a perturbative expansion.
This writing is obviously coordinate-dependent but since we are using asymptotic Boyer--Lindquist coordinates, a significant deviation from Kerr should be similarly acknowledged by different observers.

Differently from the Kerr case, in the Konoplya--Zhidenko spacetime there exists no maximum value for $a$ beyond which the spacetime always describes a naked singularity.
As $a$ always enters quadratically in \cref{horizons}, without loss of generality, in the following we consider positive $a$. 

Although this spacetime belongs to a class of metrics which are constructed to describe the spacetime \emph{outside} the event horizon, it is instructive to explore the implications \emph{inside} the horizon. This should be taken with great care and interpreted prudently, but it might give insights about what a small difference at infinity entails about the structure of the spacetime inside the horizon. This being said, in what follows we do not limit our analysis to the largest positive real root of $\Delta=0$ but we give a more comprehensive discussion.

The Ricci scalar of the Konoplya--Zhidenko metric is non-vanishing, $R=2\eta/\left[r^3 \left(r^2 + a^2 \cos^2\theta\right)\right]$, from which we infer that $r=0$ is a physical singularity.

To classify the solutions of $\Delta=0$ it is helpful to introduce
\be\label{eq:hors}
\eta_\pm = \frac{2}{27} \left[9M a^2 -8 M^3 \pm\left(4 M^2-3 a^2\right)^{3/2}\right],
\ee
and to define three separate parameter regions as (I)~$\eta<\eta_-$; (II)~$\eta_-\leq\eta\leq\eta_+$; (III)~$\eta\geq\eta_+$.
Then we sort configurations according to the value of the spin parameter: below the Kerr bound, $a<M$; highly spinning $M\leq a< a_\ast \equiv 2M/\sqrt{3}$; and ultra spinning $a\geq a_\ast$.

\begin{figure*}[t]
\subfloat[Region (II) with $\eta_-<\eta<0$.]{\includegraphics[width=0.3\textwidth]{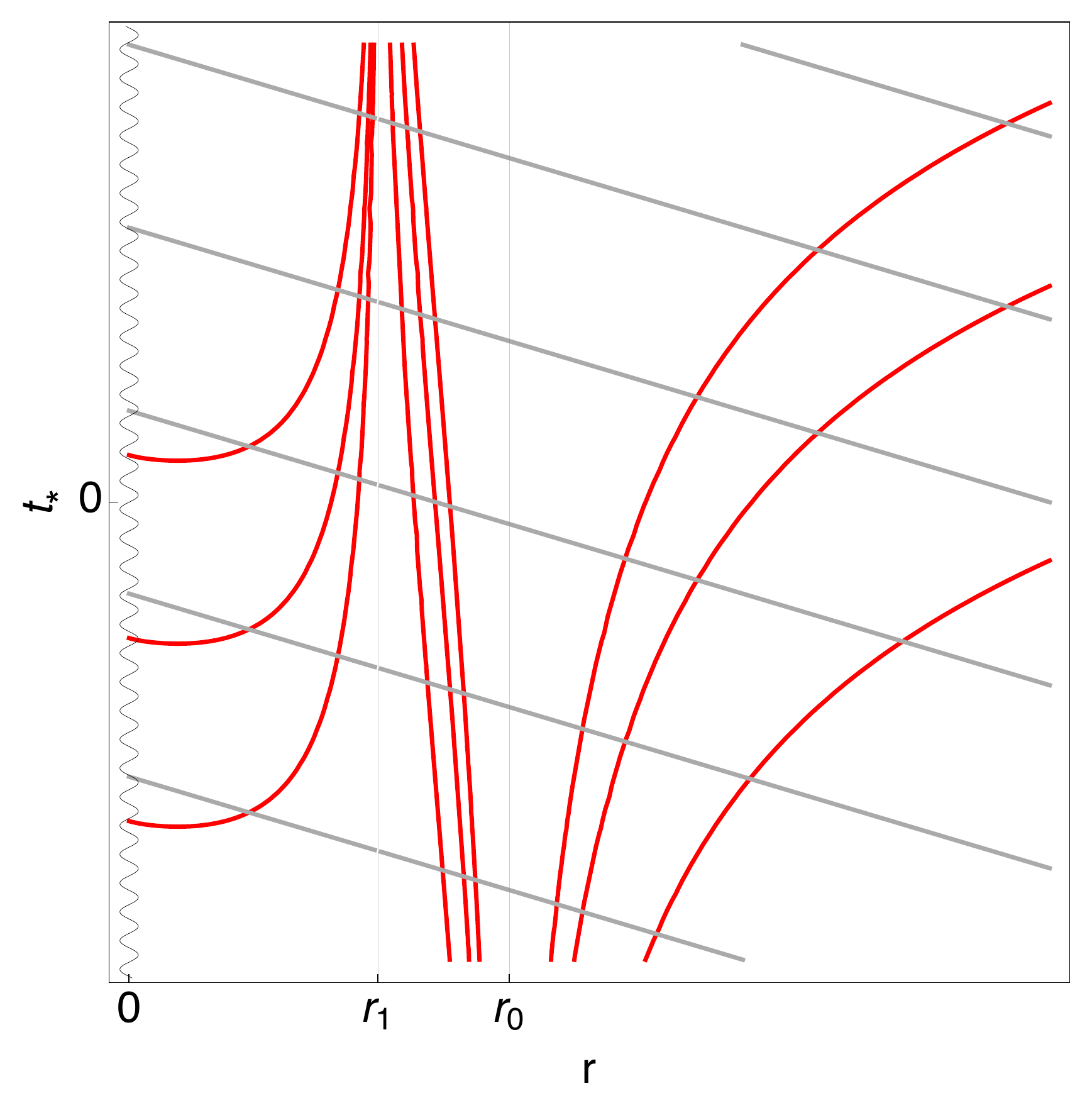}}
\quad
\subfloat[Region (II) with $0<\eta<\eta_+$.]{\includegraphics[width=0.3\textwidth]{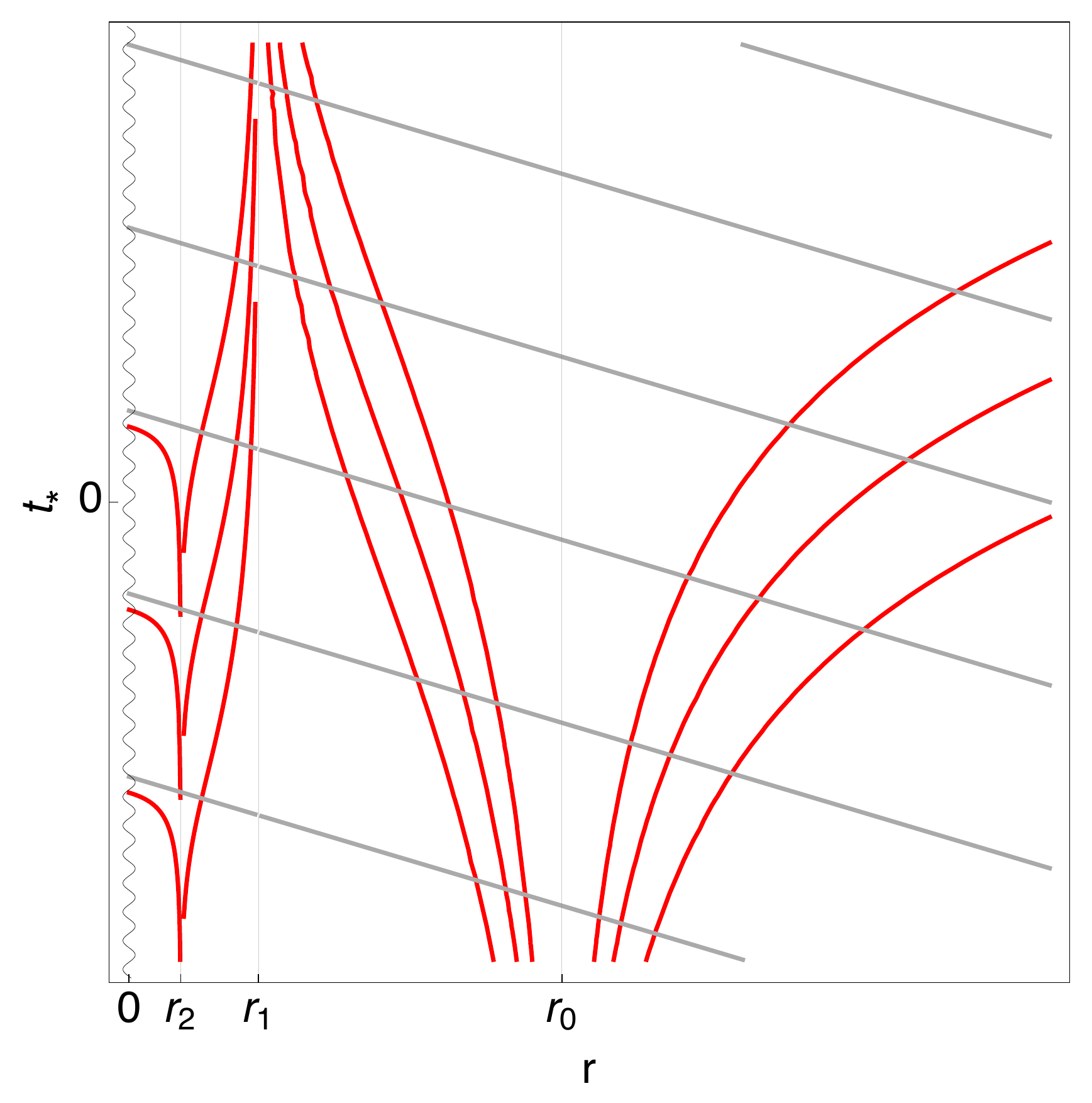}}
\quad
\subfloat[Region (III).]{\includegraphics[width=0.3\textwidth]{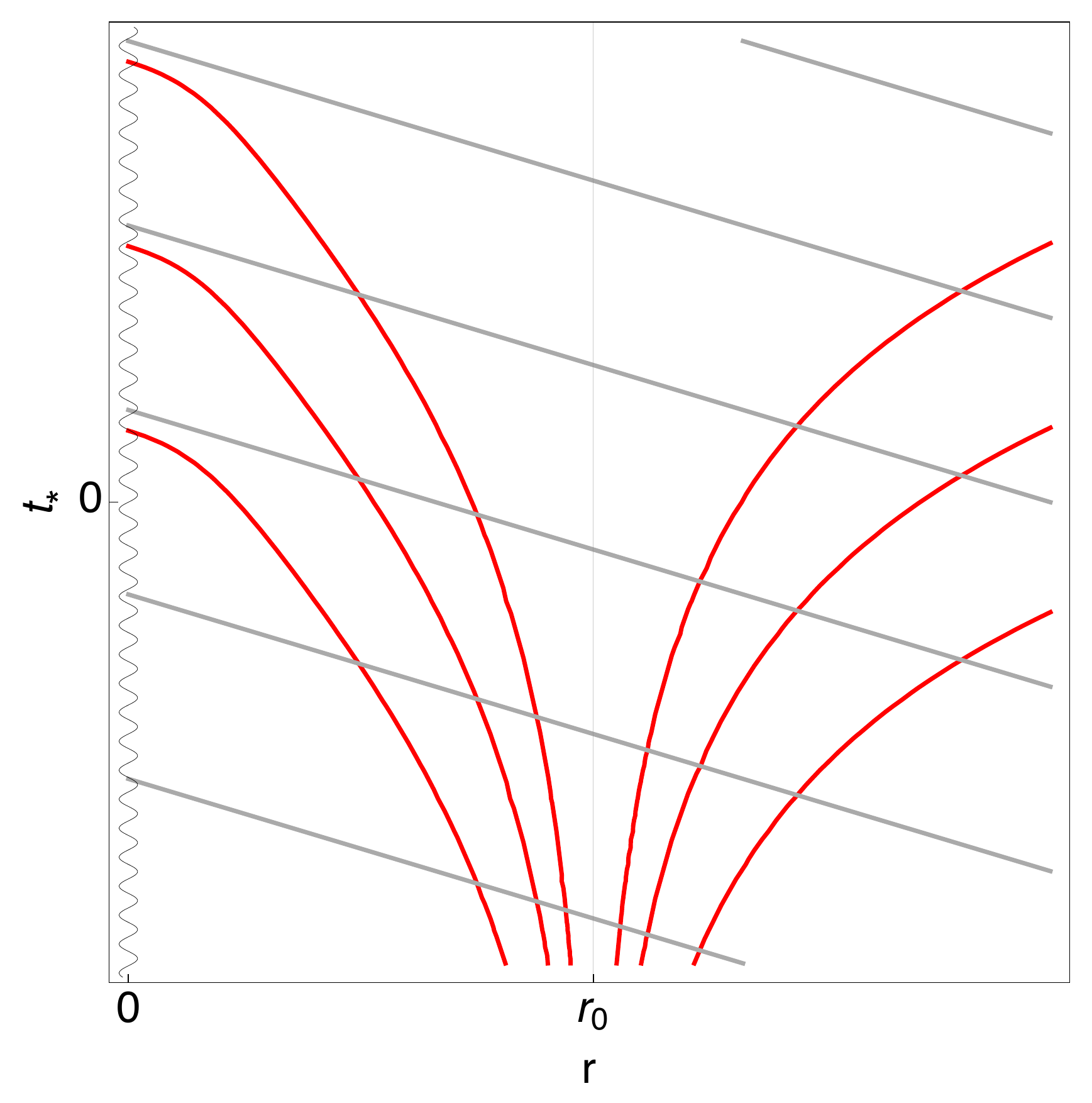}}
\caption{Light-cone structure in advanced coordinates for a Konoplya--Zhidenko black hole below the Kerr bound.\label{fig:lightcone}}
\end{figure*}

\begin{description}
\item[Below the Kerr bound] In region (I), there is only one real solution given by $r_2$ in \cref{horizons} which is always negative and hence the spacetime describes a naked singularity.

In region (II), the equation $\Delta=0$ admits three real solutions, and the event horizon is $r_0$.
The root $r_1$ is always positive while $r_2$ is negative (positive) for $\eta_-<\eta<0$ ($0<\eta<\eta_+$).
In particular for $\eta=\eta_-$, $r_0 = (1/3)(2M +\sqrt{4M^2-3a^2})$,\footnote{Notice, however, that $\partial r_0/\partial\eta$ diverges as $\eta\to\eta_-$.} while for $\eta=\eta_+$, $r_0 = (2/3)(M +\sqrt{4M^2-3a^2})$.

In region (III), $r_0$ is the only positive-definite real root.

\item[Highly spinning] For $a=M$ and $\eta>0$ the event horizon is $r_0 = (2M/3)\left[1 + \cos\left(\frac{1}{3} \arccos\left(27\eta/2M^3 - 1\right)\right)\right]$. The other solutions $r_1$ and $r_2$ are generically complex-valued but for $0<\eta<4M^3/27$ the imaginary part goes to zero and the real part is positive.

For $M<a<a_\ast$, $\eta_-$ is positive and in the subregion of region (I) such that $0<\eta<\eta_-$, the only real positive root is $r_2$.
In region (II) the three real roots are positive  and the event horizon is given by $r_0$, while in region (III) the only real solution is $r_0$.

Notice that for each value of $\eta$ in the range $0<\eta<8M^3/27$, there exists a value of $a$
\be
a_+ &= \frac{M}{\sqrt{3}}\left(1 + 2\sqrt{1+\frac{27\eta}{M^3}} \cos\beta_+\right)^{1/2},
\ee
with
\be
\beta_+ &= \frac{1}{3}\cos^{-1}\frac{8 M^6-540 \eta  M^3-729 \eta^2}{8 \left(M^4+27 \eta  M\right)^{3/2}}\,,\0
\ee
for which the largest root of $\Delta=0$ passes from $r_0$ to $r_2$ discontinuously.
Alternatively, for a fixed $a$, the largest root of $\Delta=0$ passes from $r_2$ to $r_0$ at $\eta=\eta_-$.
Depending on the specific values of the parameters the ratio $r_0/r_2$ can be of several orders of magnitude, and the compactness of the black hole changes accordingly.

\item[Ultra spinning] For the particular case $a = a_\ast$ with $\eta>0$, \mbox{$r_0= 2M/3 + \sqrt[3]{\eta - 8 M^3/27}$} and $r_1$ and $r_2$ are complex-valued unless $\eta=8M^3/27$, for which $r_0=r_1=r_2 = 2M/3$.
For \mbox{$a > a_\ast$}, $\eta_\pm$ in \cref{eq:hors} become complex-valued and independently on the value of $\eta$, $r_0$ and $r_2$ are complex-valued, while $r_1$ is positive for $\eta>0$.
\end{description}

The light-cone structure of these configurations can be richer and significantly different than that of a Kerr black hole.
As an example, consider a Konoplya--Zhidenko black hole below the Kerr bound.
Following a standard procedure, we define advanced coordinates and we plot null rays in \cref{fig:lightcone}, where $t_* = t+r-r_*$ being $r_*$ a tortoise coordinate defined by $\d r_*/\d r = (r^2+a^2)/\Delta$.

In the external regions, i.e.\ for $r>r_0$, we observe a peeling structure, typical of black-hole horizons.
In region (II), for $\eta_-<\eta<0$, the light-cone structure is nearly similar to that of a Kerr black hole, there are an outer and an inner horizon and a timelike singularity.
In region (II) but for \mbox{$0<\eta<\eta_+$}, a null trajectory encounters a black-hole horizon, a white-hole-like horizon and then again a black-hole-like horizon to eventually reach a spacelike singularity.
In region (III) there is only one horizon and the light-cone structure looks like the Schwarzschild one with a spacelike singularity.

For further convenience, the angular velocity $\Omega=-g_{t\varphi}/g_{\varphi\varphi}$ at the horizon reads
\be\label{Omega0}
\Omega_k = \frac{a}{r_k^2+a^2} = \frac{a}{2Mr_k+\eta/r_k}\,,
\ee
where the value of $k$ depends on the specific values of the black-hole parameters.

\subsection{Ergoregions}

An ergosurface is a static limit surface, \ie\ no static observer is allowed beyond this surface.
Ergosurfaces in these black-hole spacetimes are defined as the roots of the equation $g_{tt}=0$, or equivalently $r^2-2Mr+a^2\cos^2\theta-\eta/r=0$, which read
\be
r_k^\text{erg} &= \frac{2M}{3} + \frac{2}{3}\sqrt{4M^2 - 3a^2 \cos^2\theta}\, \cos \left(\beta^\text{\,erg} - \frac{2k\pi}{3}\right),\label{eq:k_inf}\\
\beta^\text{\,erg} &= \frac{1}{3}\cos^{-1}\frac{27 \eta +16 M^3-18 M a^2 \cos^2\theta}{2\left(4 M^2-3 a^2 \cos^2\theta\right)^{3/2}}\,,\quad k=0,1,2\,.\0
\ee

For configurations below the Kerr bound in regions (II) and (III), the location of the ergosurface is $r_0^\text{erg}$.

For highly spinning configurations, the ergosurface is again $r_0^\text{erg}$ in regions (II) and (III), but it is piecewise and non-continuous in region (I): it is given by $r_0^\text{erg}$ in the angular interval $[\theta_1,\theta_2]$ and $r_2^\text{erg}$ in the complementary interval, $[0,\theta_1)\cup(\theta_2,\pi]$ where $\theta_{1,2}$ ($\theta_2=\pi-\theta_1$) are the solutions of
\be\label{ergopiecewise}
\eta = \frac{2}{27} \left[9M a^2\cos^2\theta - 8M^3 -\left(4 M^2-3 a^2\cos^2\theta\right)^{3/2}\right],
\ee
once the values of $M$, $a$ and $\eta$ are fixed; the maximum value of $\theta_1$ is $\cos^{-1}\left(M/a\right)$, attained for $\eta\to0^+$.
This means that when passing from a configuration in region (I) to one in region (II), the volume between the ergosurface and the event horizon, the ergoregion, can change dramatically.

Notice that for configurations below the Kerr bound and highly spinning and values of $\eta$ in regions (II) and (III) the volume of the ergoregion is maximum for $\eta=\eta_-$ and it decreases for larger values.

For the particular case $a=a_\ast$, the location of the ergosurface is $r_0^\text{erg}$ as long as $\eta\geq8M^3/27$, but piecewise and discontinuous for $0<\eta<8M^3/27$ as described above.
For superspinning configurations, let $\theta_\ast$ the smallest root of $\cos^2\theta = a_\ast^2/a^2$.
For $0<\eta<8M^3/27$ the ergoregion is piecewise and discontinuous: it is given by $r_1^\text{erg}$ for $[0,\theta_\ast)\cup(\pi-\theta_\ast,\pi]$, $r_2^\text{erg}$ for $[\theta_\ast,\theta_1)\cup(\theta_2,\pi-\theta_\ast]$, and $r_0^\text{erg}$ for $[\theta_1,\theta_2]$ where $\theta_{1,2}$ are again the solutions of \cref{ergopiecewise}.
For $\eta\geq8M^3/27$ the ergoregion is still piecewise but no longer discontinuous: it is given by $r_0^\text{erg}$ in the interval $[\theta_1,\theta_2]$ and $r_1^\text{erg}$ in the complementary interval $[0,\theta_1)\cup(\theta_2,\pi]$.

The fact that superspinning configuration for some values of the deformation parameter can have a piecewise and non-continuous ergoregion, i.e.\ no longer an ergosurface, poses a serious problem on the viability of these particular configurations as black-hole mimickers. We expect these particular solutions to be dynamically unstable, but this analysis is beyond the scope of this paper and is left for future work.

\subsection{Photon orbits}

Photon orbits for the Konoplya--Zhidenko black hole can be studied starting from the geodesic equations derived in \cref{app:mattercontent}.
In particular, the radial null geodesic in the equatorial plane is
\be\label{eq:radial_equatorial_geod_eq}
\dot{r}^2 = E^2 + \frac{a^2 E^2-L^2}{r^2}+\frac{2M \left(L-a E\right)^2}{r^3}+\frac{\eta \left(L-a E\right)^2}{r^5}\,,
\ee
where a dot indicates derivative with respect to an affine parameter, while $E$ and $L$ are, respectively, the energy and the angular momentum of the photon, although it is more convenient to characterize the geodesic by the impact parameter $D\equiv L/E$.

The radius of photon orbits $r_c$ and its corresponding impact parameter $D_c$ are determined by \cref{eq:radial_equatorial_geod_eq} and its derivative evaluated at $r=r_c=\const$.
The problem is well-known for the Kerr black hole~\cite{Chandrasekhar}, but the term introduced by the deformation parameter $\eta$ makes the equation no longer amenable to analytical methods for all values of $L$ and $E$.
Therefore, we decide to adopt a small $\eta/M^3$ approximation and work below the Kerr bound. This guarantees some level of analyticity and exploits known results to be compared with.
In what follows, the sign of $a$ is important to distinguish between direct ($a>0$) and retrograde ($a<0$) orbits, so uniquely for the remainder of this subsection we allow $a\in[-M,M]$.

In practice, we expand the light ring radius $r_c$ and the impact parameter $D_c$ around the Kerr values in powers of $\eta/M^3$.
Here we report the leading-order corrections for the most familiar cases, i.e.\ $a=-M,\ 0,\ M$.
When $a=-M$ we find
\be
r_c \approx 4M+\frac{13\eta}{72M^2},\quad D_c \approx 7M+\frac{\eta}{6M^2}\,.
\ee
In the non-rotating limit, \ie\ for $a=0$, we get
\be
r_c \approx 3M+\frac{5\eta}{18M^2},\quad D_c \approx 3\sqrt{3}M + \frac{\sqrt{3}\eta}{6M^2}\,.
\ee
For $a=M$ the leading order correction is milder,
\be
r_c \approx M+\sqrt{\frac{4\eta}{3M}},\quad D_c \approx 2M+\sqrt{\frac{3\eta}{M}}\,.
\ee

For general values of the deformation parameter, and to allow the spin parameter above the Kerr bound, the radius of the photon orbits and the corresponding impact parameter can be determined numerically.
For $|\eta|/M^3\lesssim0.1$, $r_c$ and $D_c$ have maximum deviations from the Kerr values, respectively, of $\sim3\%$ and $\sim4\%$ for $0\leq a<0.9M$, which reduce to less than 1\% for $-M\leq a<0$.
We have also checked that the light ring is always outside the horizon for $\eta>\eta_-$ and $a\leq a_\ast$.

\subsection{The Konoplya--Zhidenko black hole as a solution of general relativity\label{s:KZinGR}}

Although these parametrized axially symmetric metrics are built \emph{not} to be exact solutions to any gravitational theory,\footnote{In Refs.~\cite{Suvorov:2020bvk,Suvorov:2021amy} it is shown that the Konoplya--Zhidenko metric is an exact solution of a (non-analytical) mixed scalar-$f(R)$ gravitational theory.} it is an interesting exercise to figure out what kind of matter distribution one would need in general relativity to obtain the Konoplya--Zhidenko black hole as an exact solution, and which energy conditions must be violated.

We start by defining the stress-energy tensor out of the Einstein tensor, \ie, $T_{\mu\nu}=G_{\mu\nu}/8\pi$, whose non-zero components are given in \cref{app:mattercontent}.

To characterize the would-be matter content of this spacetime, a first possibility is to compute the eigenvalues of $\tensor{T}{^\mu_\nu}$.
In particular, we identify the energy density with the opposite of the eigenvalue relative to the timelike eigenvector,%
\footnote{Being $v_t=\{a+r^2/a,0,0,1\}$, $v_r=\{0,1,0,0\}$, $v_\theta=\{0,0,1,0\}$, and $v_\varphi=\{a\sin^2\theta,0,0,1\}$ the eigenvectors of $\tensor{T}{^\mu_\nu}$, the timelike vector is $v_t$ for $\Delta>0$ and $v_r$ otherwise.}
\be\label{rhoeigenvalue}
\rho = -\frac{\eta }{4 \pi  r \left(r^2 + a^2 \cos^2\theta\right)^2}\,.
\ee
This matter distribution is concentrated close to the singularity and mainly along the equatorial plane, but it extends beyond the event horizon although it decays quite fast for large values of the radius.

Alternatively, the distribution of energy can be characterized in an observer-dependent way by analysing the contraction of the stress-energy tensor with the velocity of a physical observer, \ie, $\rho = T_{\mu\nu} u^\mu u^\nu$.
In view of the angular distribution of \cref{rhoeigenvalue}, for simplicity, consider a zero-angular-momentum observer (ZAMO) in the equatorial plane, whose four-velocity in given in \cref{app:mattercontent}.
It can be verified that
\be\label{rhofreefall}
\left.\rho_\text{ZAMO}\right|_{\theta=\pi/2} = -\frac{\eta \left(2 r^2 + 5 a^2\right)}{8\pi r^7}\,.
\ee

Inspection of \cref{rhofreefall,rhoeigenvalue} reveals that the sign of these energy densities is purely determined by the sign of $\eta$: negative (positive) values of $\eta$ correspond to a positive (negative) energy density; assuming $a<M$ and in the small $\eta/M^3$ regime, they also correspond to configurations more (less) compact than a Kerr black hole with the same spin --- cfr.\ \cref{compactness}.
These results further imply that, for positive values of $\eta$, this matter distribution violates --- at least --- the weak energy condition.

Within this effective description, it is possible to relate the above matter distribution to the flux contribution to the Komar mass~\cite{PoissonToolkit},
\be\label{Komar}
2\int_\Sigma \d^3y\,\sqrt{h}\left(T_{\mu\nu}-\frac{1}{2}\,T g_{\mu\nu}\right) n^\mu \xi^\nu,
\ee
where $\Sigma$ is a spacelike hypersurface that extends from the event horizon to infinity, $n^\mu$ the unit normal, $h$ the determinant of the induced metric on $\Sigma$, $T$ the trace of the stress-energy tensor, and $\xi^\nu$ the timelike Killing vector.
Explicit evaluation of \cref{Komar} indicates that this contribution can be of the same magnitude of $M$ for some specific values of the black-hole parameters, although for configurations below the Kerr bound it is typically of order $\pm20\%$ of $M$, where the sign depends on the sign of $\eta$.
It would be interesting to explore whether this amount of putative matter can be used to model ``dirty'' black holes as well.

Configurations on the edge of $\eta=\eta_-$, i.e.\ configurations between regions (I) and (II) --- which describe black holes for $a>M$ --- seem particularly unstable.
As the radius of the event horizon and the volume of the ergoregion can change abruptly and widely, one passes from small to enormous violations of the energy conditions.
Together with the odd piecewise and disconnected ergosurface for some values of the parameter space, this might suggest that not every configuration can mimic actual Kerr black holes.

Nonetheless, if we drop the assumption that general relativity is the correct gravitational theory, the discussion above might be extremely different.

\section{Superradiance from the Konoplya--Zhidenko black hole\label{s:results}}

In the Konoplya--Zhidenko background, the scalar ($s=0$) and electromagnetic ($s=\pm1$) wave equations are separable with the angular part described by the spin-weighted spheroidal harmonics equation and the radial part by
\be\label{radial}
\Delta^{-s}\frac{\d}{\d r}\left(\Delta^{s+1}\frac{\d R_s}{\d r}\right) +
\left(\frac{K^2 - 2\iu s\left(r-M+\frac{\eta}{2r^2}\right)K}{\Delta}\right.\0\\
\left.+ 4\iu s\omega r -\lambda - \frac{s(s+1)\eta}{r^3} \right)R_s = 0\,,
\ee
where $K\equiv (r^2+a^2)\omega - am$ and $\lambda\equiv A + a^2\omega^2 - 2ma\omega$, being $A$ the eigenvalue of the angular equation, $\omega$ the frequency of the perturbation and $m$ its azimuthal number. The angular eigenvalue is also characterized by the harmonic number $l$.
As discussed in \cref{app:maxwell}, the physical information contained in the solution with spin-weight $s$ is equivalent to that with spin-weight $-s$.
This property will be particularly important when computing the energy fluxes of electromagnetic waves at infinity.

\begin{figure*}[!thb]
\includegraphics[width=.42\textwidth]{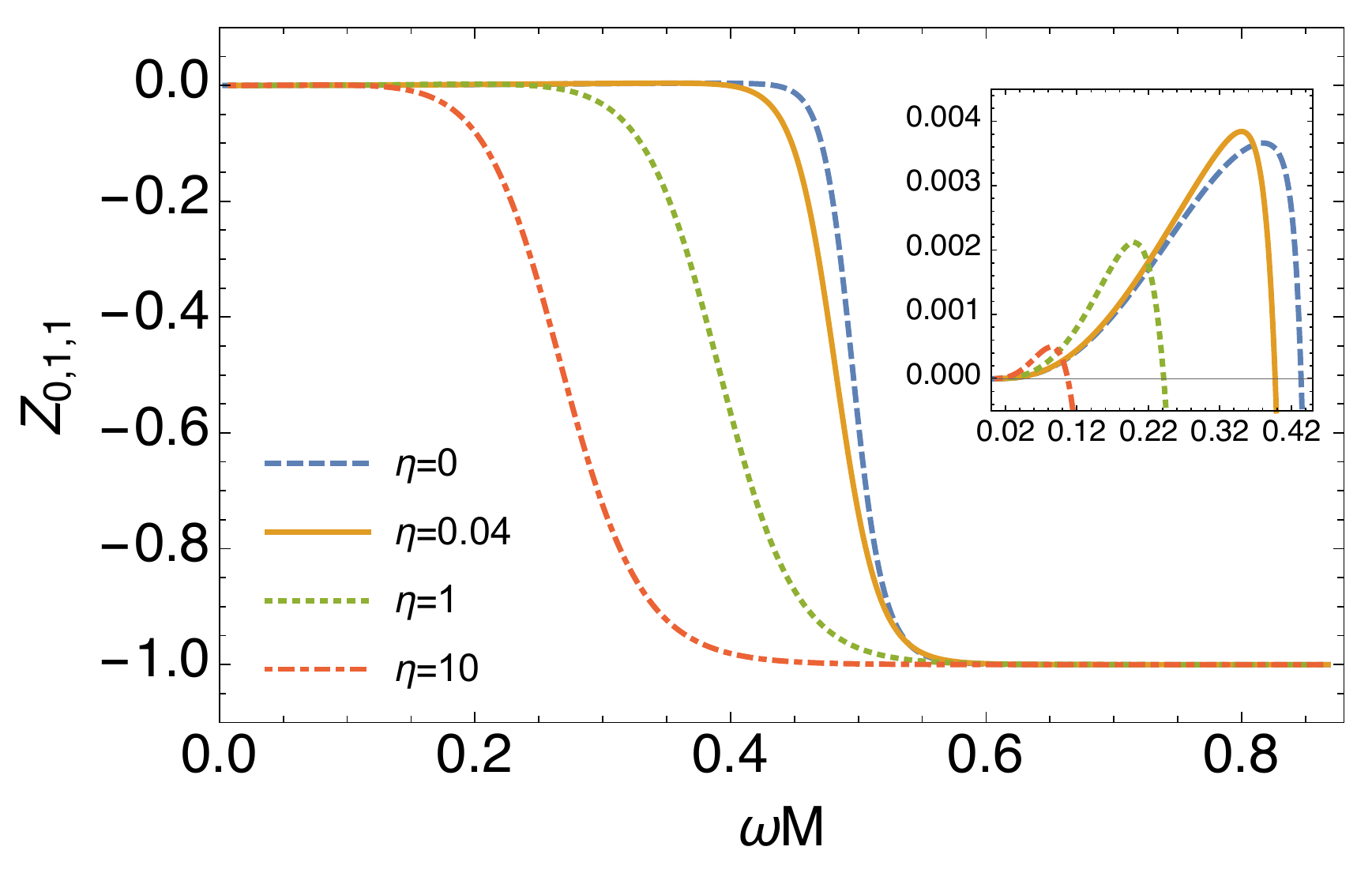}\quad
\includegraphics[width=.42\textwidth]{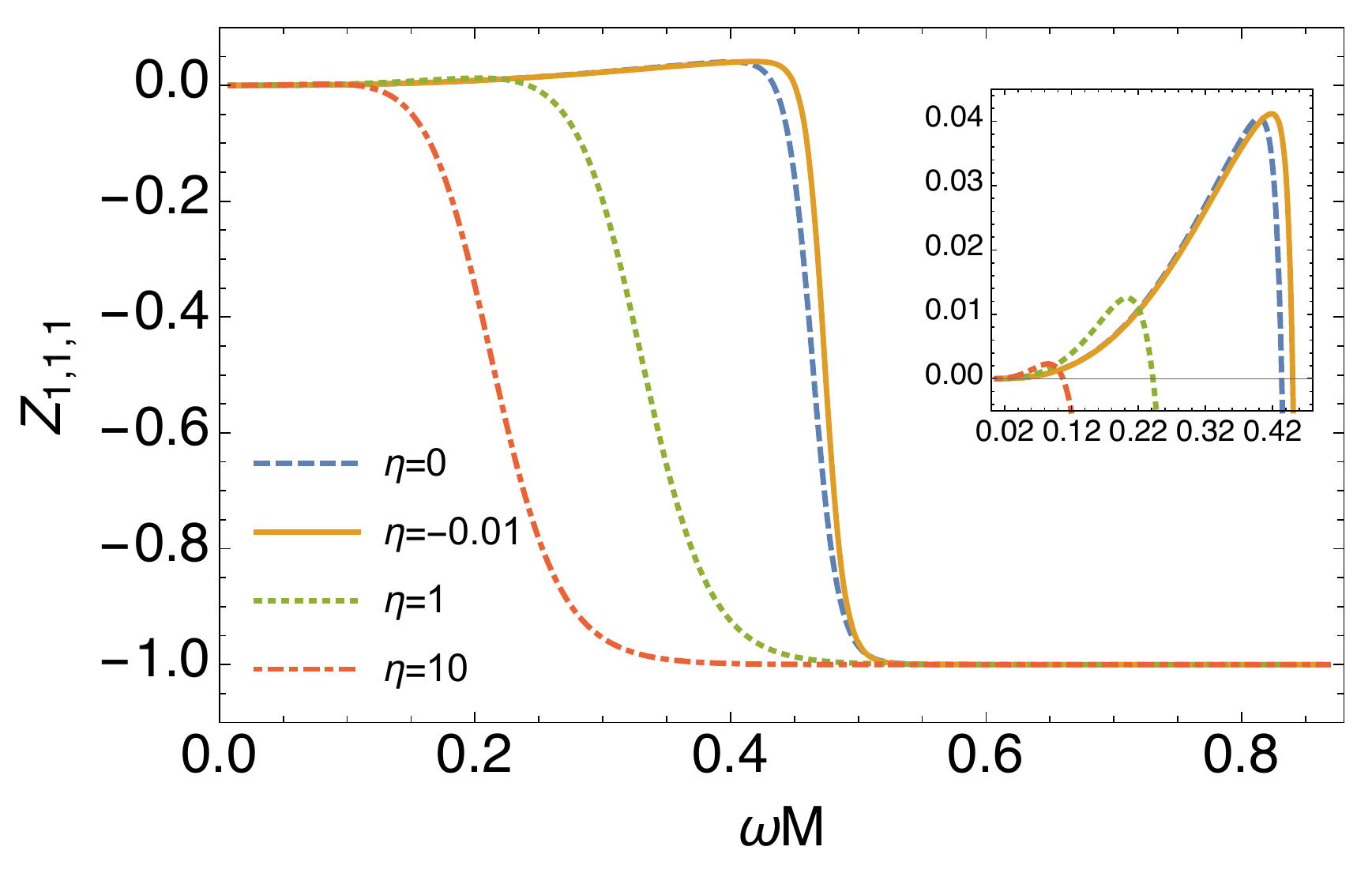}
\caption{Spectra of the amplification factor for a scalar (left panel) and electromagnetic (right panel) field with $l=m=1$ off a Konoplya--Zhidenko black hole with $a=0.99M$ for selected values of $\eta$ in units of $M^3$. Inset: Zoom in the superradiant region.}\label{fig:Z}
\end{figure*}

\subsection{Boundary conditions}

To integrate \cref{radial} we need to supply it with boundary conditions.
We first introduce a tortoise-like coordinate $\d r_*/\d r\equiv (r^2+a^2)/\Delta$ and a new radial function $Y_{\!s}(r) = \sqrt{r^2+a^2}\,\Delta^{s/2}\,R_s(r)$ such that the radial equation becomes
\be\label{radialstar}
\frac{\d^2 Y_{\!s}}{\d r_*^2} + \left(\frac{K^2 - 2 \iu s \left(r-M+\frac{\eta}{2 r^2}\right)K + (4 \iu r s \omega-\lambda)\Delta}{\left(r^2+a^2\right)^2}\right.\0\\
\left. - \frac{\d G}{\d r_*} - G^2 - \frac{s(s+1)\eta\Delta}{r^3 \left(r^2+a^2\right)^2}\right) Y_{\!s} = 0\,,
\ee
where $G = r\Delta/(r^2+a^2)^2 + s \Delta'/2(r^2+a^2)$.
Asymptotically ($r\to\infty$), \cref{radialstar} becomes
\be\label{radialsinf}
\frac{\d^2 Y_{\!s}}{\d r_*^2} + \left(\omega^2 + \frac{2\iu s\omega}{r}\right) Y_{\!s} \approx 0\,,
\ee
whose solutions are $Y_{\!s} \sim r^{\pm s}\e^{\mp\iu\omega r_*}$ where the plus (minus) sign refers to outgoing (ingoing) waves.

Near the event horizon $r_0$ ($r_*\to-\infty$), let $k\equiv\omega-m\Omega_0$, $\Omega_0$ being defined in \cref{Omega0}, then \cref{radialstar} becomes
\be\label{radialshor}
\frac{\d^2 Y_{\!s}}{\d r_*^2} + \left(k-\iu s\sigma\right)^2 Y_{\!s} \approx 0\,,\quad
\sigma = \frac{a^2+r_0 (3 r_0 - 4M)}{2 r_0 \left(r_0^2 + a^2\right)}\,,
\ee
and the purely ingoing solution at the horizon is $Y_{\!s} \sim \exp\left[-\iu\left(k - \iu s\sigma\right)r_*\right] \sim \Delta^{-s/2}\e^{-\iu k r_*}$.

\subsection{Amplification factors}

The asymptotic solutions to \cref{radialsinf} can be used to define the energy fluxes of bosonic fields at infinity.
Since the Konoplya--Zhidenko spacetime shares the same asymptotic behaviour and symmetries of the Kerr spacetime, the derivation of this section is very similar to what happens for Kerr~\cite{Teukolsky:1974yv}.

Consider an incident wave of amplitude $\mathcal{I}$ from infinity producing a reflected wave of amplitude $\mathcal{R}$, the asymptotic solution to \cref{radialsinf} can be written as
\be\label{asymptoticY}
Y_{\!s} \sim \mathcal{I}\,\e^{-\iu\omega r_*} r^s + \mathcal{R}\,\e^{\iu\omega r_*}/r^s\,.
\ee

The total energy flux at infinity per unit solid angle can be computed out of the stress-energy tensor of the test fields as
\be
\frac{\d^2E}{\d t\,\d\Omega} = \frac{\d^2}{\d t\,\d\Omega}\left(E_\text{in} + E_\text{out}\right)  = \lim_{r\to\infty} r^2 \tensor{T}{^r_t}\,,
\ee
where the ingoing and outgoing fluxes $\d E_\text{in/out}/\d t$ are proportional, respectively, to $|\mathcal{I}|^2$ and $|\mathcal{R}|^2$~\cite{Teukolsky:1974yv}.
When energy is extracted from the black hole, the flux of energy through the horizon is negative and energy conservation implies $\d E_\text{in}/\d t < \d E_\text{out}/\d t$.
It is then possible to define the quantity $Z_{s,l,m} = \d E_\text{out}/\d E_\text{in} - 1$ which gives the amplification or absorption factor for bosonic waves of spin-weight $s$ and quantum numbers $(l,m)$ off a black hole.

In our case of interest, the amplification factors are
\be\label{Zfactor}
Z_{0,l,m}    = \frac{|\mathcal{R}|^2}{|\mathcal{I}|^2} - 1\,,\quad
Z_{\pm1,l,m} = \frac{|\mathcal{R}|^2}{|\mathcal{I}|^2}\left(\frac{16\omega^4}{B^2}\right)^{\pm1} - 1\,,
\ee
where $B^2 = [\lambda+s(s+1)]^2 + 4ma\omega - 4a^2\omega^2$.
Notice that the expressions in \cref{Zfactor} are the same as for Kerr as the asymptotic behaviour and the symmetries of the Konoplya--Zhidenko black hole are the same.
However, the deformation parameter $\eta$ changes the geometry of the near-horizon region and it is responsible for a different amplification factor, as we show in the next section.

\subsection{Numerical results}

For general $\omega$ we need to numerical integrate the angular and radial equations. Our numerical routine works as follows.
For each value of the spin-weight $s$, the quantum numbers $(l,m)$ and $a\omega$, we first determine the angular eigenvalue using the Leaver method~\cite{Leaver:1985ax}.
Second, fixed a value for $\eta$, we integrate \cref{radialstar} from the horizon onwards until a sufficiently large radius.
Then we compare our numerical solution and its radial derivative to the asymptotic behaviour in \cref{asymptoticY} and its derivative to extract the coefficients $\mathcal{I}$ and $\mathcal{R}$.
Finally, we compute the amplification factor using \cref{Zfactor}.
To increase the accuracy of this numerical procedure, we consider a higher-order expansion near the horizon and in the asymptotic region which reduces to those reported in the previous section at the leading order.
The routine is repeated for several values of the frequency (typically) in the interval $0<\omega<2m\Omega_0$.
Modes with $m\leq0$ are not superradiant and as a consequence of the symmetries of the angular and radial equation, the amplification factor is symmetric under $Z_{s,l,m}(\omega)=Z_{s,l,-m}(-\omega)$ we can consider positive frequencies only.

We now define our working assumptions for what follows.
We allow the deformation parameter in the range $\eta\geq\eta_-$ and we mainly exclude superspinning configurations from our investigation, \ie, we focus on black holes below the Kerr bound and highly spinning in regions (II) and (III) introduced above. This has a practical advantage: the event horizon and the ergosurface are always given by $r_0$ and $r_0^\text{erg}$.
Despite the lack of observational evidence for rotating black holes beyond the Kerr bound~\cite{Reynolds:2019uxi}, it cannot be excluded that some highly spinning objects can be produced in high-energy astrophysical phenomena that dynamically evolve in less spinning configurations. Hence it makes sense to explore a bit this parameter region.

Some of our results are presented in \cref{fig:Z} and more results are available online~\cite{data}.
Both for scalar and electromagnetic fields with quantum numbers $l=m=1$, scattered off a black hole with spin $a=0.99M$, we observe in the insets of \cref{fig:Z} that the position of the maximum of the amplification factor is close to the superradiant threshold $\omega=m\Omega_0$ where the curve becomes very steep, as in the Kerr case.

\begin{figure}[!ht]
\includegraphics[width=.23\textwidth]{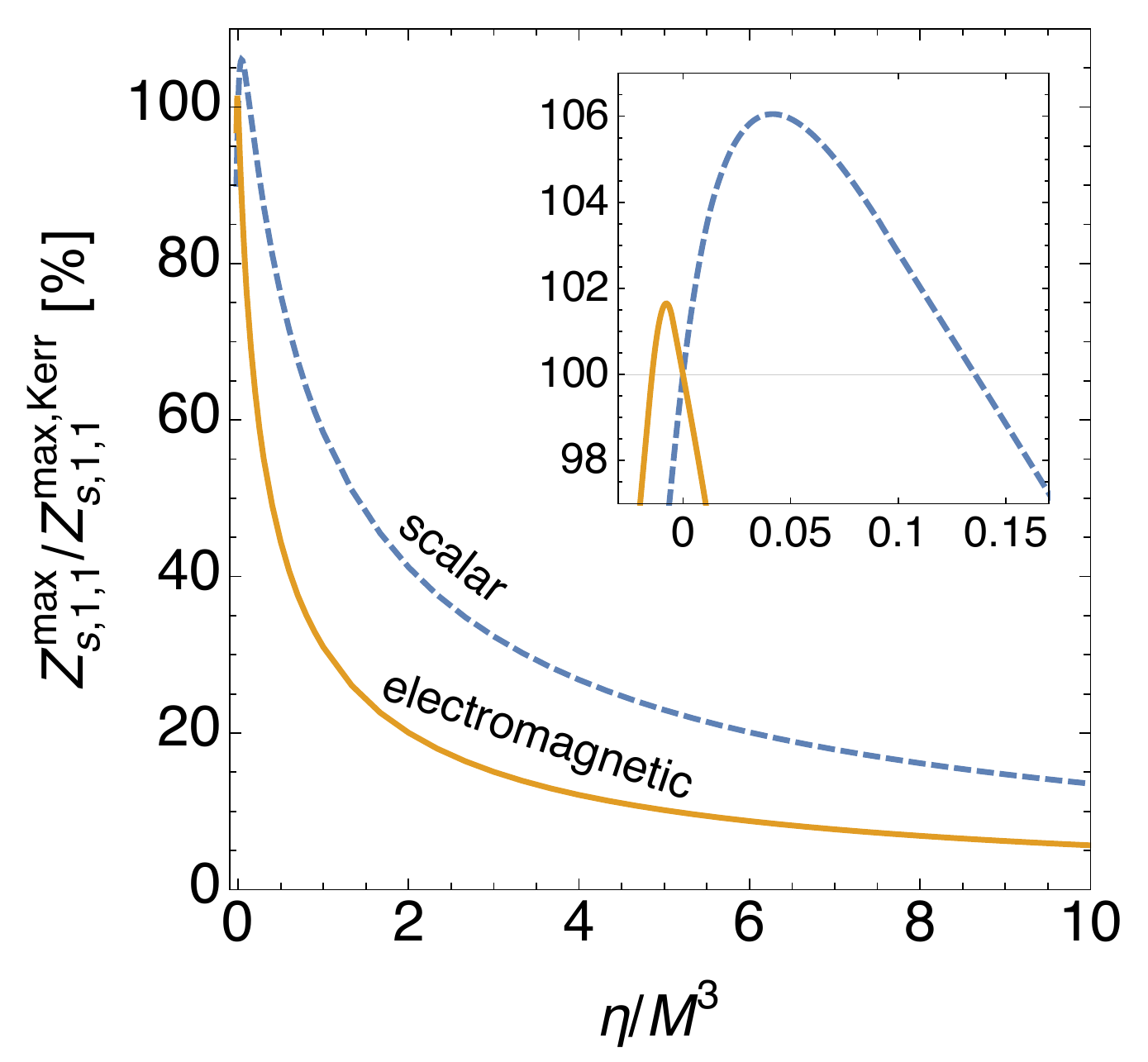}\quad
\includegraphics[width=.23\textwidth]{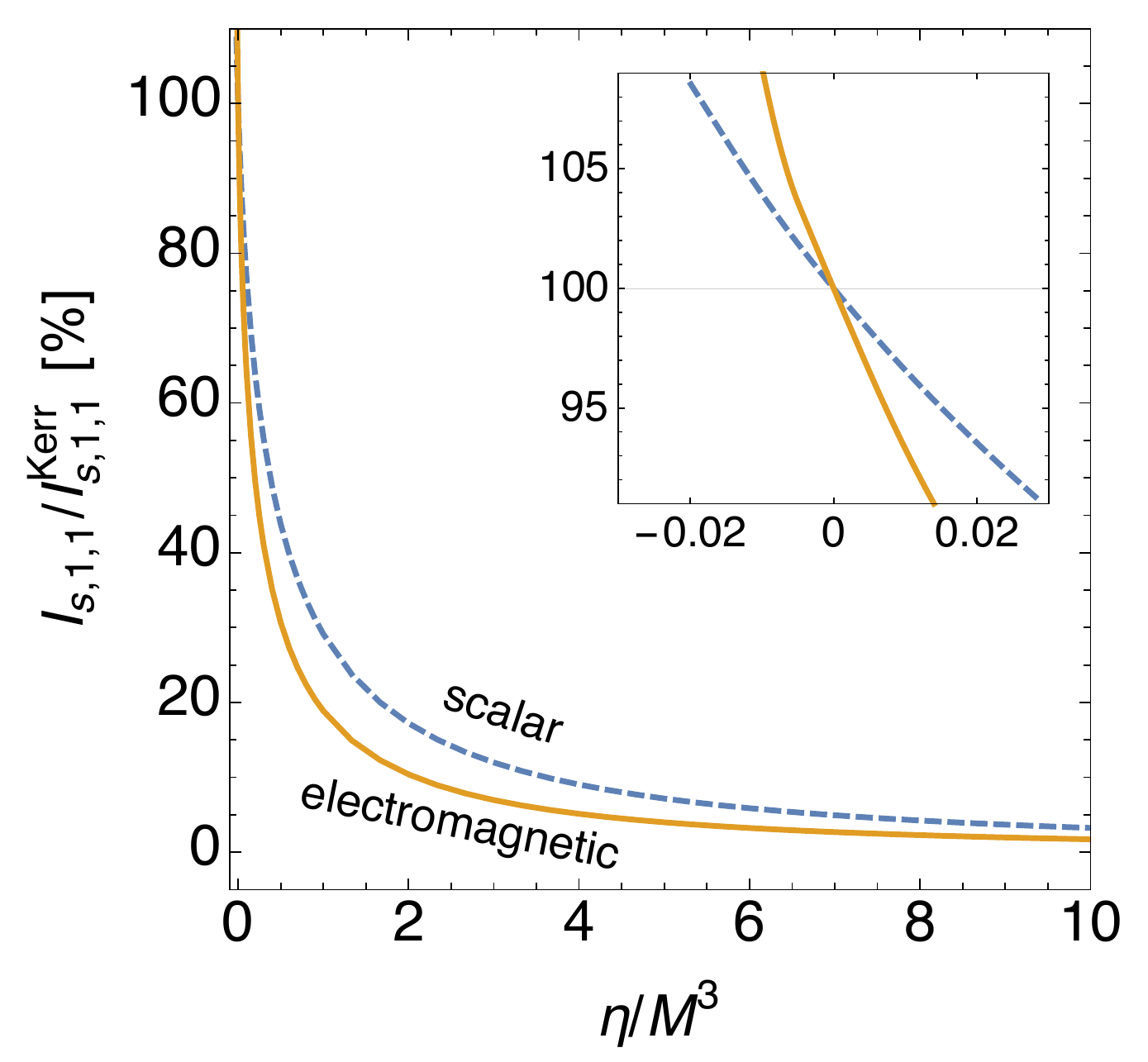}
\caption{Maximum value of the amplification factor $Z_{s,1,1}$ (left panel) and integral of the superradiant spectrum $I_{s,1,1}$ (right panel) for a scalar and electromagnetic field with $l=m=1$ as functions of~$\eta$, normalized to the maximum value in the Kerr case, \ie\ $\eta=0$, for $a=0.99M$.\label{fig:Zmax}}
\end{figure}

In absolute values, the maximum amplification factor is about 0.4\% and 4.4\% for scalar and electromagnetic waves, as for Kerr.
However, in the left panel of \cref{fig:Zmax} we notice that for scalar waves scattering off a Konoplya--Zhidenko black hole with $\eta/M^3\approx0.04$ the maximum amplification factor is about 6\% larger than in the Kerr case, while for electromagnetic waves, we observe a maximum amplification factor roughly 1\% larger than in the Kerr case for $\eta/M^3\approx-0.01$.

These values of $\eta/M^3$ are not universal, but depend on the value of $a/M$.
For smaller values of $a/M$, the maximum value of $Z_{s,l,m}$ gets smaller, the position of the peak moves towards smaller values of $\eta/M^3$ and the frequency range for which the amplification factor is positive shrinks.
For configurations with higher spin, say at the Thorne limit $a=0.998M$, the scalar (electromagnetic) amplification factor can be up to 15\% (1\%) larger than in the Kerr case.
This bigger amplification factor does not mean that these Kerr-like spacetimes are more superradiant than the Kerr spacetime, as the quantity
\be\label{integralZ}
I_{s,l,m} = \int_0^{m\Omega_0} \d\omega\,Z_{s,l,m}\,, 
\ee
is always smaller than in Kerr, for positive values of $\eta$, as shown in the right panel of \cref{fig:Zmax}.
However, a bosonic wave with frequency close to the superradiant threshold can be significantly more enhanced in a Konoplya--Zhidenko background.
For negative values of $\eta$, which correspond to more compact configurations, $I_{s,l,m}$ is typically bigger than in Kerr and maximal close to $\eta=\eta_-$.
For large enough positive values of the deformation parameter the maximum value of the amplification factor and the range of superradiant frequencies are always smaller than in the Kerr case.
The physical explanation to this result is that, typically, for values of $\eta/M^3\neq0$ the volume of the ergoregion is smaller and hence the energy that can be extracted.
In the non-rotating limit, \ie\ $a=0$, superradiance disappears and we recover the recent results on absorption in Schwarz\-schild-like backgrounds~\cite{Magalhaes:2020pyp,Magalhaes:2020sea}.

In the inset of the left panel of \cref{fig:Zmax}, we observe that the same maximum value of the amplification factor for a scalar field is obtained for Kerr ($\eta=0$) and for $\eta/M^3\approx0.12$.
This is nothing but an apparent degeneracy, as the spectra and the superradiant ranges of frequency are significantly different.

\begin{figure}[h]
\includegraphics[width=.42\textwidth]{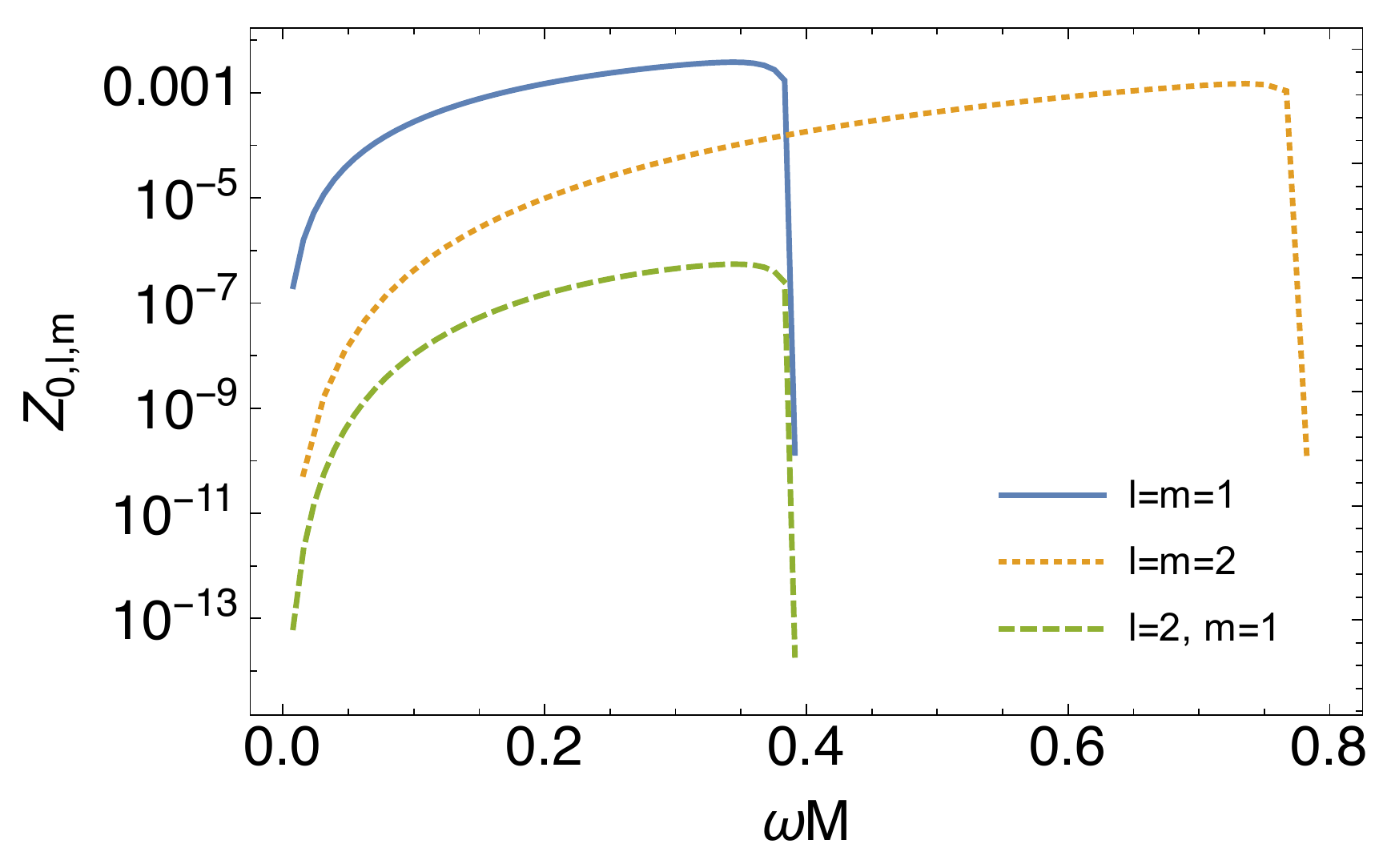}
\caption{Typical spectra of the amplification factor $Z_{0,l,m}$ for different superradiant scalar field modes off a Konoplya--Zhidenko black hole with $a=0.99M$ and $\eta/M^3=0.05$.}\label{fig:Zmodes}
\end{figure}
In \cref{fig:Zmodes} it is evident that the most superradiant mode corresponds to the minimum allowed value of $l=m$, as in the Kerr case.
Modes with different values of $(l,m>0)$ qualitatively share the same behaviour with the $l=m=1$ mode, though the maximum amplification factor is hierarchically smaller than the dominant one.
For example, in the range $0.5M\lesssim a < M$, for both scalar and electromagnetic fields we find $Z_{s,2,1}^{\max}/Z_{s,2,2}^{\max} \sim 10^{-3}$ while $Z^{\max}_{s,2,2}/Z^{\max}_{s,1,1}\sim 0.1$ for $a\gtrsim 0.8M$.
For the $l=m=2$ modes, $Z^{\max}_{s,2,2}$ and $I_{s,2,2}$ are always smaller than in the Kerr case for positive values of $\eta$ and $a<M$, but for negative values the amplification factor can be bigger than in Kerr. Again, this could be interpreted as a consequence of the fact that, for a given $a$, the ergoregion is larger than the Kerr ergoregion for negative values of $\eta$.
On the other hand, the $l=2$, $m=1$ modes can be more superradiant than in the Kerr case, in the sense of \cref{integralZ}, even for positive values of $\eta$ when $a\gtrsim0.8M$.
For the remaining modes, \ie\ with $m\leq0$, we have verified that the amplification factor is always negative, meaning that these modes are not superradiant.

As previously discussed, the Konoplya--Zhidenko black hole also admits superspinning configurations, \ie\ with spin parameter $a>M$.
If the rotation parameter is (slightly) above the Kerr bound, in principle, such energy extraction could rapidly spin down these configurations to produce a black hole with \mbox{$a<M$}.
\begin{figure}[!hb]
\includegraphics[width=.42\textwidth]{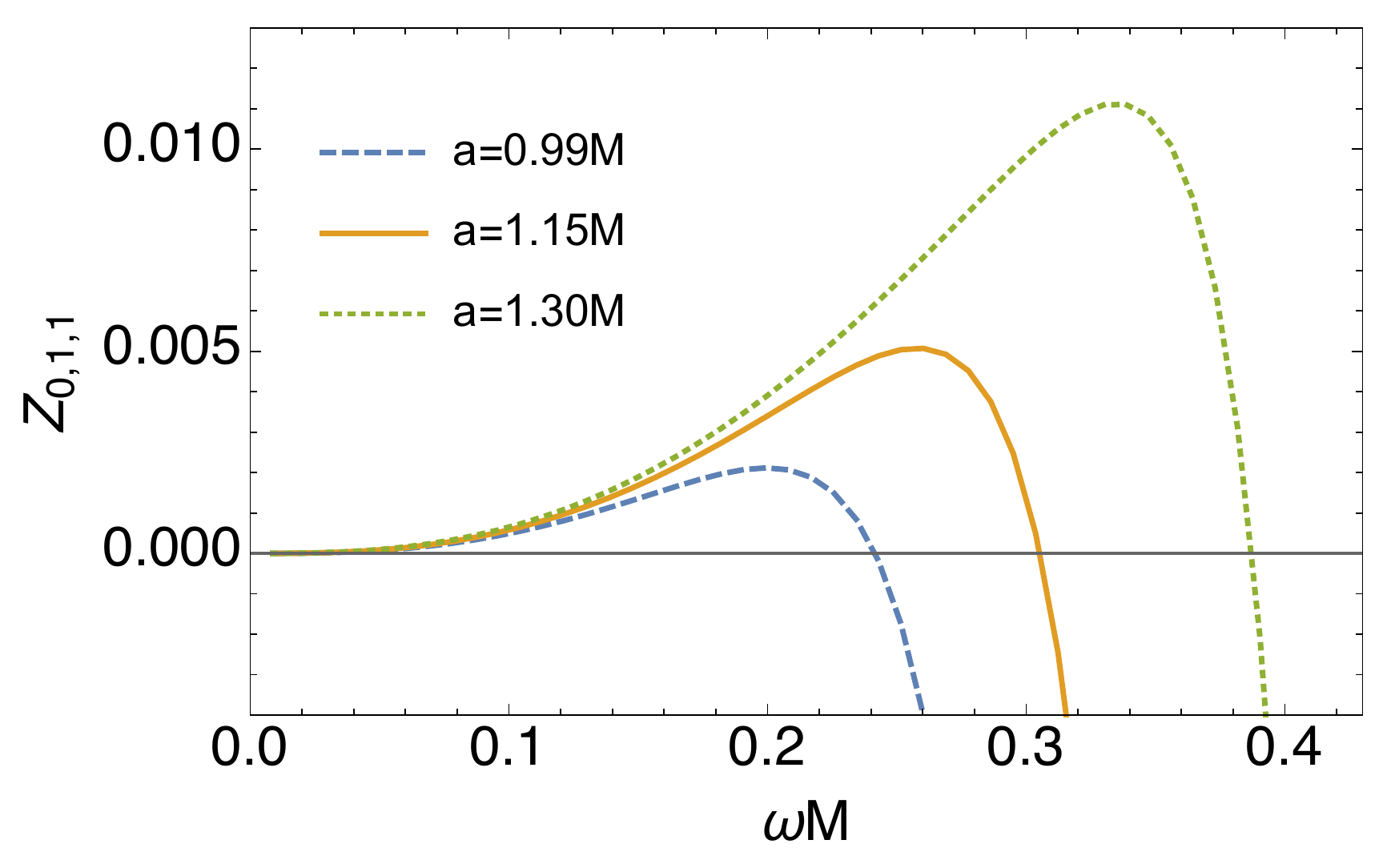}
\caption{Spectra of the amplification factor for a scalar field with $l=m=1$ off a superspinning Konoplya--Zhidenko black hole with $\eta/M^3=1$ for selected values of $a/M$.}\label{fig:superspinning}
\end{figure}

For completeness, we consider the scattering of a scalar field off a superspinning black hole.
We observe in \cref{fig:superspinning} that for $\eta/M^3=1$ and selected values of the black-hole spin, the maximum value of the amplification factor can grow (in principle indefinitely), as well as the range of frequency for which the process is superradiant.
But to obtain amplification factors larger than 100\% one needs configurations with very large spin parameter or very small positive deformation parameter, which are unlikely to describe astrophysical black holes.
Moreover, as discussed in \cref{KZblackhole:horizons}, one needs to be careful with these configurations, as in the range $0<\eta<8M^3/27$ the position of the event horizon is not always given by $r_0$ for all values of $a$, and perhaps even more gravely, the ergosurface can be piecewise and non-continuous.

\subsection{Massive scalar fields\label{s:ultralight}}

The extension to a massive scalar field with mass $\mu_s\hbar$ is quite simple: such mass term in the Klein--Gordon equation introduces, after separation, a quantity $-\mu_s^2 r^2 \Delta/\left(r^2+a^2\right)^2$ in the coefficient of $Y_0$ in \cref{radialstar} and shifts the frequency of the angular equation as $\omega^2\to\omega^2-\mu_s^2$.

The boundary conditions are slightly modified.
In particular, purely ingoing solutions at the horizon still require $Y_0 \sim \e^{-\iu k r_*}$, while the asymptotic behaviour at infinity is
\be\label{asymptoticYmass}
Y_0 \sim r^{-M \mu_s^2/\varpi}\,\e^{\varpi r_*}  \sim r^{M \left(\mu_s^2-2 \omega^2\right)/\varpi}\,\e^{\varpi r}\,,\quad
\varpi = \pm\sqrt{\mu_s^2 - \omega^2}\,.
\ee

Massive waves can be superradiant for frequencies in the range $\mu_s<\omega<m\Omega_0$, while they are trapped near the horizon and exponentially suppressed at infinity for $\omega<\mu_s$.

\begin{figure}[ht]
\includegraphics[width=.42\textwidth]{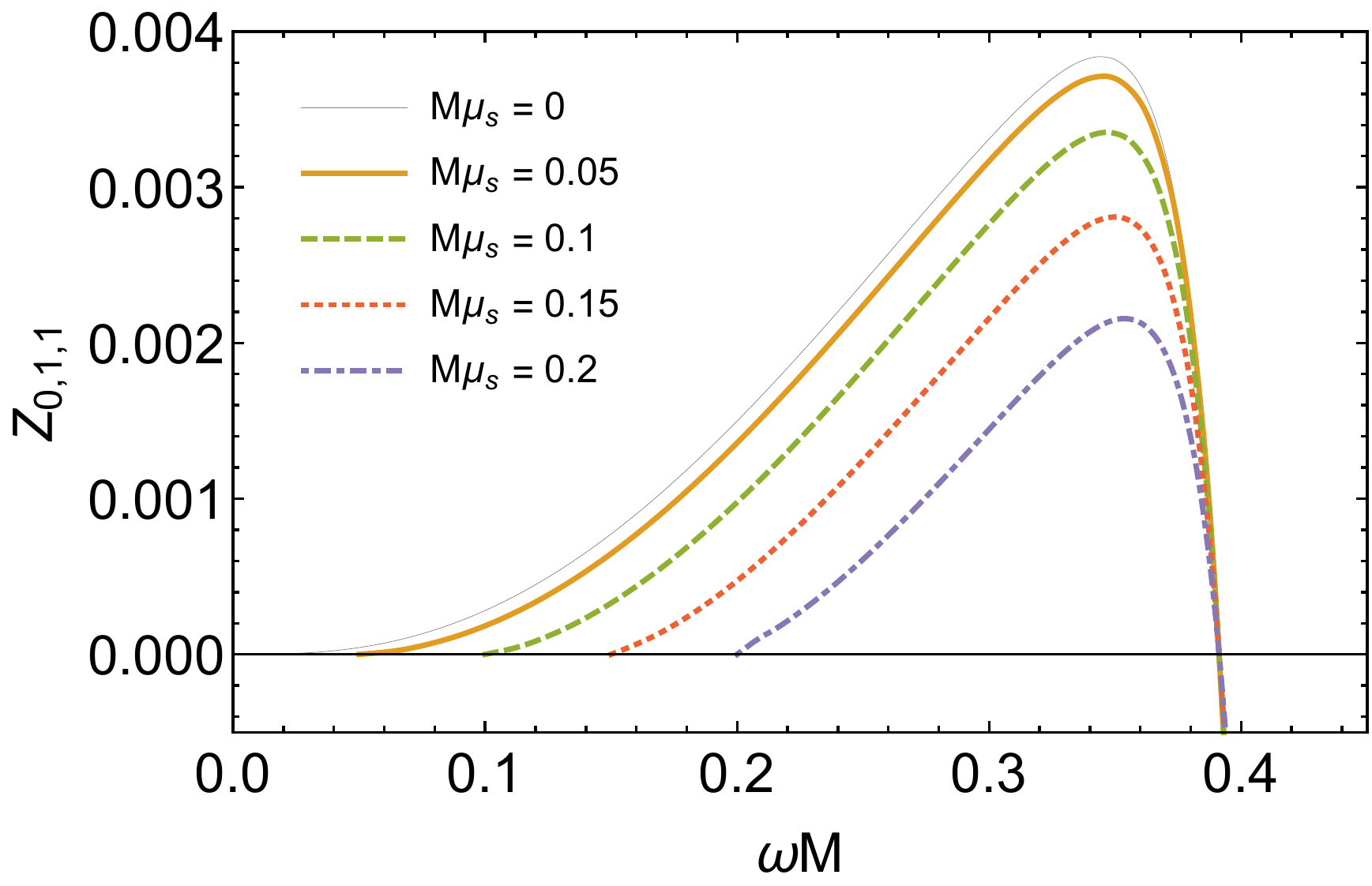}\newline
\includegraphics[width=.42\textwidth]{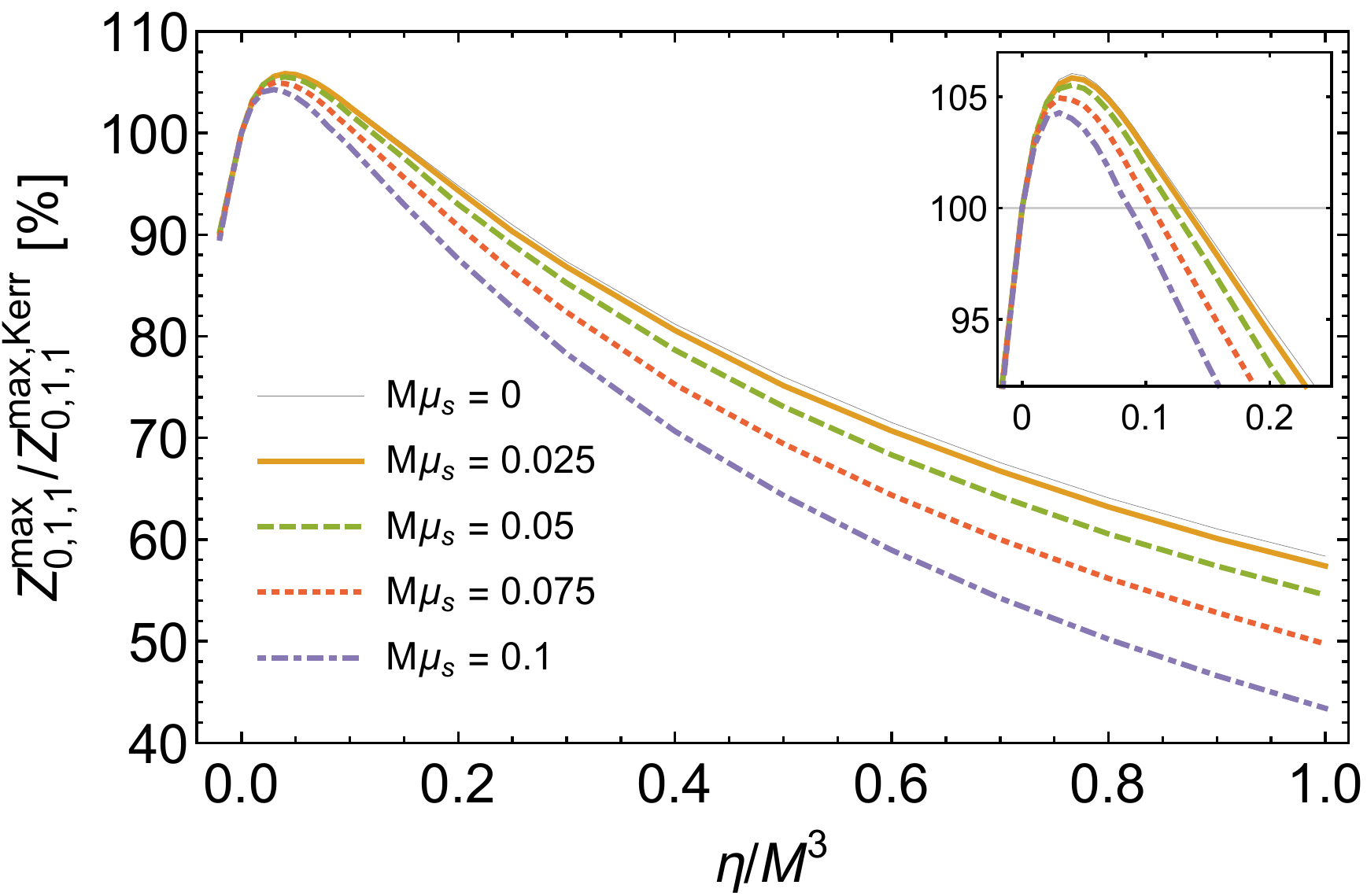}
\caption{(Top panel) Spectra of the amplification factor for massive scalar fields with $l=m=1$ off a Konoplya--Zhidenko black hole with $a=0.99M$ and $\eta/M^3=0.05$, for selected values of the mass parameter.
(Bottom panel) Maximum value of the amplification factor $Z_{0,1,1}$ for a massive scalar field with $l=m=1$ as function of $\eta$ normalized to the maximum value in the Kerr case, \ie\ $\eta=0$, for $a=0.99M$ and for selected values of the mass parameter.}\label{fig:ampl_factor_mass}
\end{figure}

The numerical routine for the computation of the amplification factor is adapted from that used for massless waves, correcting the asymptotic behaviours accordingly.
We limit this analysis to the $l=m=1$ mode for which, in analogy with the massless case, we expect the dominant contribution.
We repeat the routine for several values of the frequency in the interval $\mu_s<\omega<2\Omega_0-\mu_s$.
Our results, as those in the top panel of \cref{fig:ampl_factor_mass}, show that superradiance grows with the spin parameter $a$ and is less pronounced for more massive fields, as in the Kerr spacetime.
The bottom panel of \cref{fig:ampl_factor_mass} shows that massive waves can be more amplified than in a Kerr background with the same spin parameter for some values of the deformation parameter, analogously to what we found for massless fields, though waves with larger masses are still less enhanced.
Even in this case, for positive values of $\eta$, the Konoplya--Zhidenko black hole is less superradiant than Kerr in the sense of \cref{integralZ} with the interval of integration adapted to $[\mu_s,m\Omega_0]$.

Kerr black holes develop superradiant instabilities against massive fields~\cite{Dolan:2007mj} which can be used to constrain the existence and the mass of ultralight bosons, \ie\ using black holes as ``particle detectors''~\cite{Brito:2014wla}.
In addition, the bosonic cloud can produce long-lasting, monochromatic gravitational-wave signals observable, in principle, in the sensitive band of current detectors~\cite{Brito:2013wya,Arvanitaki:2014wva,Brito:2017wnc}.
We do not expect this picture to be considerably changed for Kerr-like black holes.
Small values of the deformation parameter unveiled an interesting feature in the massless case, and we also expect a good black-hole mimicker not to turn upside-down the Kerr metric.
This motivates us to investigate the stability of the Konoplya--Zhidenko spacetime against massive scalar fields in the small $\eta/M^3$ limit.
Remarkably, in this limit, in the low-frequency regime, \ie\ for $\omega M\ll1$ and $a\omega\ll1$ and in the small mass approximation $M\mu_s\ll1$, the problem can be tackled with analytical methods --- see \cref{app:instability} for details.

At leading order, the growth time of instability $\tau$ for a Kerr black hole perturbed by an axion with mass $m_\text{axion} =\mu_\text{axion}\hbar = \SI{e-20}{eV}$ is
\be
\tau = \left(\SI{1.58e6}{s}\right)\left(\frac{\mu_\text{axion}}{\mu_s}\right)\frac{M}{a}\,\frac{1}{(M\mu_s)^8}\,,
\ee
to which the deformation parameter adds the contribution (valid as long as $\eta/M^3$ is small)
\be
\delta\tau = -\left(\SI{7.89e3}{s}\right)\left(\frac{\mu_\text{axion}}{\mu_s}\right)\left(\frac{\eta/M^3}{0.01}\right)\left(\frac{M}{a}\right)^3\frac{1}{(M\mu_s)^8}\,.\label{deltatau}
\ee
\Cref{deltatau} implies that for positive (negative) values of the deformation parameter (within a perturbative regime), the growth time of instability is shorter (longer), i.e.\ the Konoplya--Zhidenko black hole is more (less) unstable than Kerr.
For an axion cloud around a supermassive black hole with $M=10^9 M_\odot$, $M\mu_\text{axion}\approx10^{-1}$, and the growth time of instability would be shorter but comparable with the age of the Universe.
Yet, this timescale should also be shorter than the decay time of the particle for the instability to be really effective.

As this preliminary result relies on several assumptions, it is to be confirmed by an exhaustive computation of quasi-normal modes and bound states, which is left for future work.
In fact, this result is valid for slowly rotating black holes hence we cannot conclude whether highly spinning configurations are more unstable or not.

\section{Discussion\label{s:conclusion}}

In this paper we have studied the superradiant scattering of scalar and electromagnetic test fields off a Kerr-like black hole.
In these spacetimes, the best that we can do is to study test fields propagating in a fixed background, but often test fields are a good proxy and the results for scalar and electromagnetic waves are similar to those for gravitational waves.
However, this is not always true and the case of superradiance in general relativity is illustrative: the maximum amplification factors are approximately 0.4\% for massless scalar fields with $l=m=1$, 4.4\% for electromagnetic waves with $l=m=1$ and 138\% for gravitational waves with $l=m=2$~\cite{Teukolsky:1974yv}.
To guarantee an analytical description of the problem, we have also limited our investigation to a very specific class of parametrized axially symmetric spacetimes, the Konoplya--Zhidenko black hole.
This does not mean that superradiance cannot be present in more general spacetimes, on the contrary, we do expect superradiance to occur in any spacetime rotating sufficiently fast, provided the presence of an ergoregion, but this would most likely require a full numerical simulation.
In this sense, our results represent a first step in the investigation of the phenomenon of superradiance in Kerr-like spacetimes.

Before exploring superradiant scattering around Konoplya--Zhidenko black holes, we have studied their structure thoroughly.
The simple Konoplya--Zhidenko metric, which shares with Kerr the same symmetries and asymptotic behaviour, translates into a complicated causal structure.
Depending on the values of the parameters, these configurations can have from zero up to three horizons. When the spin parameter is above the Kerr bound, the ergoregion can be piecewise and non-continuous. To consider this model as a valuable Kerr black-hole mimicker we might probably need to exclude some regions of the parameter space.
Moreover, when considered as non-vacuum general-relativistic solutions, these configurations require to be sustained by some exotic matter.
Yet, in the small-deformation limit and below the Kerr bound, we have shown that the horizon and the light ring radii are slightly modified with respect to the Kerr values of only a few percents.
Optimistically, future observations of e.g.\ black-hole shadows could set bounds on the deformation parameter~\cite{Gralla:2020srx,Volkel:2020xlc}.

Regarding superradiance, we have found maximum amplification for the minimum value allowed of $l=m$ (for scalar and electromagnetic fields $l=m=1$) and  highest values of the spin parameter, similarly to what happens for Kerr black holes.
Our numerical results show that for large values of the deformation parameter and considering the same spin, superradiance is highly suppressed with respect to the Kerr black hole.
This can be interpreted in terms of the volume of the ergoregion: for a Kerr and a Konoplya--Zhidenko black hole with same $a$, for positive values of the deformation parameter the ergoregion is smaller in the latter case as well as the amount of energy that can be extracted, and the effect of superradiance is damped.
This seems in agreement with the fact that the proper volume of the ergoregion of slowly rotating black holes in quadratic gravity decreases with respect to the general-relativistic case, suggesting a smaller amplification factor~\cite{Pani:2011gy}.
Our results for superspinning configurations shown in \cref{fig:superspinning} are compatible with the fact that the energy extraction by the Penrose process for the superspinning Johannsen--Psaltis~\cite{Johannsen:2011dh} and Konoplya--Zhidenko metric can be significantly larger than for a Kerr black hole~\cite{Liu:2012qe,Long:2017xqr}.
Analogously, for negative values of the deformation parameter, which correspond to more compact configurations, the volume of the ergoregion can be larger than that of a Kerr black hole with the same mass and spin, and as a consequence, the superradiant phenomenon can be enhanced.

The most interesting feature that we have found is the existence of an interval of small values of the deformation parameter for which the maximum of the amplification factor is larger than in Kerr. 
This interval contains positive values of $\eta$ for scalar fields when $a>0.97M$, while for electromagnetic fields it requires $a\approx M$, meaning that there are configurations less compact than a Kerr black hole with the same mass and angular momentum for which the superradiant scattering can be larger.
For different values of the spin parameter, to have more superradiance one needs more compact configurations, \ie\ with $M/r_0>1$. If this trend is respected by gravitational waves, then a higher amplification factor for less compact spacetimes would occur for values of $a$ extraordinary close to the extremal case.

Besides, we have also presented some initial results on massive scalar fields. Roughly, their behaviour is similar to the massless and Kerr cases.
Under some approximations --- small frequency, small spin parameter, small scalar mass and small deformation parameter --- the frequency eigenvalue can be determined analytically with asymptotic matching techniques.
For the expected most unstable mode $l=m=1$, positive values of the deformation parameter shorten the growth time of instability, meaning that these spacetimes are more unstable than Kerr against massive scalar fields.
The validity of this result is limited: it needs to be taken with great care and one should not infer too much information, as it is based on a large number of assumptions.
A complete numerical investigation is left for future work.

Within this context, knowledge of superradiant instabilities can be used to put bounds on the existence and mass of ultralight particles.
Nonetheless, if a black hole acts as a ``particle detector'', the presence of (dark) matter around it might modify the geometry and spoil superradiant effects.
In \cref{s:KZinGR} we interpreted the Konoplya--Zhidenko metric in terms of an exotic matter distribution and show that the matter flux contribution to the Komar mass of the spacetime can be a significant fraction of the black-hole mass $M$. To have this contribution less than, say, 10\% of $M$, and not to suppress superradiant effects, the deformation parameter should take values $|\eta|/M^3\lesssim0.1$, indicating once again that the most interesting phenomenology corresponds to small deviations from the Kerr geometry.

The spectra of the amplification factor can look very similar when comparing a Kerr black hole and a Konoplya--Zhidenko black hole with a small deformation parameter, as well as a massless scalar and a massive scalar with very small mass parameter.
In addition to this fact, there might be a similar ``degeneracy'' when comparing the spectra of a massless scalar off a slightly deformed Konoplya--Zhidenko black hole with the spectra of a little massive scalar off a Kerr black hole.
We have verified that this could actually happen in a number of cases. Our criterion for degenerate spectra is when both the maximum value of the amplification factor, its corresponding frequency, its integral as in \cref{integralZ} and the threshold frequency are the same within a tolerance of 5\%.
Since superradiance is suppressed for massive scalars, we expect this degeneracy to correspond to positive values of $\eta$. This is true for intermediate values of the spin parameter, but for $a\gtrsim0.9M$ the parameter space also include small negative values of $\eta/M^3$.
As an example, for $a=0.95M$ the spectra of a massive scalar field with $M\mu_s\approx0.025$ and $M\mu_s\approx0.05$ off a Kerr black hole resemble the spectra of a massless scalar field off a Konoplya--Zhidenko black hole with $-0.01\lesssim\eta/M^3\lesssim0.03$ and $0.02\lesssim\eta/M^3\lesssim0.03$.

There are several possible extensions of this work.
First notice that we considered rotating spacetimes with a horizon, meaning that they do not suffer ergoregion instability.
The presence of the horizon guarantees from the beginning the key ingredient for superradiance: dissipation.
In view of testing the Kerr hypothesis, it could be interesting to use this same parametrization and substitute the horizon with a partially reflective surface and see in which limits superradiance disappears.
Other possible developments might include considering non-minimally coupled scalar fields, or use the results in Ref.~\cite{Konoplya:2016jvv} to study superradiance of test fields off Kerr--Sen and Einstein--dilaton--Gauss--Bonnet black holes.

\begin{acknowledgments}
We are grateful to Vitor Cardoso for several valuable comments and a thorough reading of this manuscript, and to Ma\-ria\-no Cadoni for discussions.
EF acknowledges partial financial support by CNPq Brazil, process no.\ 301088/2020-9.
EF and SL acknowledge funding from the Italian Ministry of Education and Scientific Research (MIUR) under the grant PRIN MIUR 2017-MB8AEZ\@.
MO acknowledges partial financial support by the research project ``Theoretical and experimental investigations of accreting neutron stars and black holes'', CUP  F71I17000150002, funded by Fondazione di Sardegna.
The authors thankfully acknowledge Daniele Mura for assistance and computer resources provided by INFN, Sezione di Cagliari.
\end{acknowledgments}

\appendix

\section{The Klein--Gordon and Maxwell equations in Kerr-like backgrounds\label{app:maxwell}}

Being $s$ the spin weight of the test field, in linear perturbation theory the scalar ($s=0$) and Maxwell ($s=\pm1$) fields propagate in the background metric.
The Klein--Gordon equation for a massless scalar field $\Phi$ is easily obtained from $\Box\Phi=0$, where the D'Alambert operator is built out of the metric \eqref{eq:sep_functions}.
To derive the Maxwell equations in such spacetime, we follow the method proposed in Ref.~\cite{Teukolsky:1973ha}.

First we choose a suitable null tetrad $e_{(a)}^{\ \mu}=\left\{l^\mu,n^\mu,m^\mu,\bar{m}^\mu\right\}$ that easily reduces to the Kinnersley tetrad~\cite{Kinnersley:1969zza} in the Kerr spacetime, \ie
\begin{subequations}\label{eq:tetrad}
\be
l^\mu&=\frac{1}{\Delta}\left[r^2 R_\Sigma + a^2,\frac{\Delta}{R_B},0,a\right],\\
n^\mu&=\frac{1}{2\Sigma}\left[r^2 R_\Sigma + a^2,-\frac{\Delta }{R_B},0,a\right],\\
m^\mu&=\frac{1}{\sqrt{2} \bar{\rho}}\left[\iu a \sin\theta,0,1,\iu \csc\theta\right],
\ee
\end{subequations}
where $\Delta=r^2\,R_\Sigma-R_M\,r+a^2$, $\Sigma=r^2R_\Sigma+a^2\cos^2\theta$ and $\rho=r\sqrt{R_\Sigma}-\iu a \cos\theta$.
The tetrad vectors satisfy
\be
e_{(a)}^{\ \mu} e_{(b)\,\mu}^{\phantom{\mu}} =
\begin{pmatrix}
 0 & -1 & 0 & 0 \\
 -1 & 0 & 0 & 0 \\
 0 & 0 & 0 & 1 \\
 0 & 0 & 1 & 0
\end{pmatrix}\,.
\ee

The non-vanishing spin coefficients are
\begin{subequations}
\be
%
\varrho&=-\frac{\left(r^2R_\Sigma\right)'}{2 R_B \Sigma}-\frac{\iu a \cos\theta}{ \Sigma },\\
\epsilon&=-\frac{\iu a \cos\theta}{4 r R_B \Sigma }\left[2 r R_B-\frac{\left(r^2R_\Sigma\right)'}{\sqrt{R_\Sigma}}\right],\label{eps_sc}\\
\mu&=-\frac{\Delta}{4 R_B \Sigma^2}\left[\left(r^2R_\Sigma\right)'+2 \iu a R_B \cos\theta\right],\\
\gamma&=\frac{\Delta'}{4 R_B \Sigma}-\frac{\Delta}{8\Sigma^2} \left[\frac{\left(\bar{\rho}+r \sqrt{R_\Sigma}\right)}{r R_B \sqrt{R_\Sigma}}\left(r^2R_\Sigma\right)'+2 \iu a \cos\theta\right],\\
\tau&=\frac{a \sin\theta}{2 \sqrt{2} R_B \Sigma  \bar{\rho}}\left[2 a R_B \cos\theta-\iu \left(r^2R_\Sigma\right)'\right],\\
\alpha&=\frac{1}{8 \sqrt{2}\Sigma\rho}\Big[\cot\theta\left(\Sigma - 5a^2 -5r^2 R_\Sigma\right)\0\\
&\phantom{=}+\frac{2 \iu a \sin\theta}{R_B}\left(r^2R_\Sigma\right)'+\sin\theta\left(7 \iu a r \sqrt{R_\Sigma}-3\iu a\rho\right)\Big]\,,\\
\pi&=\frac{\iu a \sin\theta}{\sqrt{2}\Sigma\rho}\left[r\sqrt{R_\Sigma}+\frac{\left(r^2R_\Sigma\right)'}{2R_B} - \rho\right],\\
\beta&=\frac{1}{2\sqrt{2}\Sigma\bar{\rho}}\left\{\iu a\sin\theta\left[r\sqrt{R_\Sigma}-\frac{\left(r^2R_\Sigma\right)'}{2R_B}\right]+\Sigma\cot\theta\right\}.
\ee
\end{subequations}

The sourceless decoupled Newman--Penrose equations for the massless spin-$1$ field are given by~\cite{Teukolsky:1973ha}
\begin{subequations}\label{eq:Maxwell_eq_generic}
\be
\left[\left(\boldsymbol{D}-\epsilon+\bar\epsilon-2\varrho-\bar\varrho\right)\left(\boldsymbol{\Delta}+\mu-2\gamma\right)\right.\0\\
\left.-\left(\boldsymbol{\delta}-\beta-\bar\alpha-2\tau+\bar\pi\right)\left(\boldsymbol{\bar\delta}+\pi-2\alpha\right)\right]\phi_0 &= 0\,,\\
\left[\left(\boldsymbol{\Delta}+\gamma-\bar\gamma+2\mu+\bar\mu\right)\left(\boldsymbol{D}-\varrho+2\epsilon\right)\right.\0\\
-\left.\left(\boldsymbol{\bar\delta}+\alpha+\bar\beta+2\pi-\bar\tau\right)\left(\boldsymbol{\delta}-\tau+2\beta\right)\right]\phi_2 &= 0\,,
\ee
\end{subequations}
where $\boldsymbol{D} = l^\mu\,\nabla_\mu$, $\boldsymbol{\Delta} = n^\mu\,\nabla_\mu$ and $\boldsymbol{\delta} = m^\mu\,\nabla_\mu$,
and the complex fields are defined as $\phi_0 = F_{\mu\nu}\,l^\mu m^\nu$ and $\phi_2 = F_{\mu\nu}\,\bar{m}^\mu n^\nu$, being $F_{\mu\nu}$ the electromagnetic field tensor.

Differently from the Kerr case, the spin coefficient $\epsilon$ is generally non-zero and as a consequence \cref{eq:Maxwell_eq_generic} are not separable into a radial and angular part.
However, one can always perform a null rotation of the tetrad to set $\epsilon=0$~\cite{Janis:1965tx}.
Alternatively, we can restrict our metric performing a change of the radial coordinate such that $R_B=1$ and solving $\epsilon=0$ for $R_\Sigma$,
\be
R_\Sigma = \left(1+\frac{\xi}{r}\right)^2,
\ee
where $\xi$ is a constant parameter.

Under the above assumptions, decomposing the test fields as $\e^{-\iu\omega t}\,\e^{\iu m\varphi}\,S(\theta)\,R_s(r)$, the scalar and electromagnetic wave equations separate, with the angular part described by the spin-weighted spheroidal harmonics equation
\be\label{eq:general_angular_eq}
\frac{1}{\sin\theta}\frac{\d}{\d\theta}\left(\sin\theta\,\frac{\d S}{\d\theta}\right) + \left(a^2\omega^2\cos^2\theta-\frac{m^2}{\sin^2\theta}\right.\0\\
\left. -2a\omega s\cos\theta-\frac{2ms\cos\theta}{\sin^2\theta}-s^2\cot^2\theta+s+A\right)S = 0\,,
\ee
while the radial part by the following equation,
\be\label{eq:general_radial_eq}
\Delta^{-s}\frac{\d}{\d r}\left(\Delta^{s+1}\frac{\d R_s}{\d r}\right) + \left[\frac{K^2-\iu s \Delta' K}{\Delta} + 4\iu s rR_\Sigma\omega - \lambda\right.\0\\
\left.+\frac{s(s+1)\left(\Delta''-2\right)}{2}\right]R_s = 0\,,
\ee
where $K=\left(r^2R_\Sigma+a^2\right)\omega-am$ and $\lambda=A+a^2\omega^2-2am\omega$.
The radial functions $R_0$, $R_1$ and $R_{-1}$ correspond to $\Phi$, $\phi_0$ and $\phi_2/\rho^2$.

\Cref{eq:general_angular_eq} together with regular boundary conditions at $\theta=\{0,\ \pi\}$ is an eigenvalue problem for the separation constant $A$.
For each value of $s$, $m$ and $a\omega$, the eigenvalues are identified by a number $l$, whose smallest value is $\max\left(|m|,|s|\right)$.
The eigenfunctions form a complete and orthonormal set in $\theta\in[0,\pi]$.
For $a\omega=0$, \cref{eq:general_angular_eq} reduces to the spin-weighted spherical harmonics equation and $A=(l-s)(l+s+1)$~\cite{Goldberg:1966uu}; for $a\omega\ll1$, \cref{eq:general_angular_eq} can be solved perturbatively~\cite{Press:1973zz}, but in general it must be solved numerically~\cite{Berti:2005gp}.

To integrate \cref{eq:general_radial_eq} it is necessary to give boundary conditions at the horizon and at infinity. Therefore, we first introduce a tortoise-like coordinate given by $dr_\ast/dr\equiv (r^2\,R_\Sigma+a^2)/\Delta$ and the radial function $Y_{\!s}(r)=\sqrt{r^2\,R_\Sigma+a^2}\Delta^{s/2}R_s(r)$. With these substitutions, \cref{eq:general_radial_eq} becomes
\be\label{eq:r_tortoise_radial_eq}
\frac{\d^2Y_{\!s}}{\d r_\ast^2}+\left\{\frac{\Delta \left[s (s+1) \left(\Delta''-2\right)/2-\lambda +4 \iu s \omega(r+\xi)\right]}{\left[(r+\xi)^2 + a^2\right]^2}\right.\0\\
\left.+\frac{K^2-\iu K s \Delta'}{\left[(r+\xi)^2 + a^2\right]^2}-\frac{\d G}{\d r_\ast}-G^2\right\}Y_{\!s} = 0,
\ee
where $G=s\Delta'/2(r^2R_\Sigma+a^2)+r\sqrt{R_\Sigma}\Delta/(r^2R_\Sigma+a^2)^2$ and $r$ is an implicit function of $r_\ast$.

At infinity ($r_\ast\to\infty$), \cref{eq:r_tortoise_radial_eq} can be approximated as
\be
\frac{\d^2 Y_{\!s}}{\d r_\ast^2}+\left(\omega^2+\frac{2\iu s \omega}{r}\right)Y_{\!s}=0\,,
\ee
from which we see that $Y_{\!s}\sim r^{\pm s} \e^{\mp\iu\omega r_\ast}$, where the upper (lower) sign refers to outgoing (ingoing) waves.

Near the event horizon $r_0$ ($r_\ast\to-\infty$), \cref{eq:r_tortoise_radial_eq} becomes
\be
\frac{\d^2 Y_{\!s}}{\d r_\ast^2}+\left(k- \iu s \sigma\right)^2 Y_{\!s} = 0\,,
\ee
where
\be
k &= \omega\left(1 + \frac{\xi(2r_0+\xi)}{(r_0+\xi)^2 + a^2}\right) - m\Omega_0\,,\\
\sigma &= \frac{r_0^2\left(1 - R_M'(r_0)\right) - \xi^2 - a^2}{2 r_0 \left[(r_0+\xi)^2 + a^2\right]}\,.
\ee
The purely ingoing solution at the horizon is given by $Y_{\!s}\sim\exp\left[\iu(k-\iu\sigma)r_\ast\right]\sim\Delta^{-s/2}\e^{-\iu k r_\ast}$.

Teukolsky and Press showed that one solution of the Teukolsky equation with spin-weight $s$ contains the same physical information of that with spin-weight $-s$~\cite{Teukolsky:1974yv}.
This result is a consequence of the fact that the Kerr spacetime is stationary and axisymmetric. This fact holds for this class of metrics too, in fact, repeating the same derivation for \cref{eq:general_radial_eq} but starting with the tetrad
\be
\tilde{l}^\mu=-\frac{2\Sigma}{\Delta}\,n^\mu\,,\quad
\tilde{n}^\mu=-\frac{\Delta}{2\Sigma}\,l^\mu\,,\quad
\tilde{m}^\mu=\frac{r\sqrt{R_\Sigma}-\iu a\cos\theta}{r\sqrt{R_\Sigma}+\iu a\cos\theta}\,\bar{m}^\mu\,,
\ee
related to \cref{eq:tetrad} by the simultaneous transformation \mbox{$\varphi\to-\varphi$}, $t\to-t$, one finds that, after the separation of the radial and angular variables, the radial function $\tilde{R}_s$ satisfies \cref{eq:general_radial_eq} with $s\to-s$ and it is related to $R_{-s}$ through
\be
\tilde{R}_s = \left(\frac{2}{\Delta}\right)^sR_{-s}\,.
\ee

\section{Einstein tensor, geodesic equations and zero angular momentum observers for the Konoplya--Zhidenko spacetime\label{app:mattercontent}}

The non-zero components of the Einstein tensor for the Konoplya--Zhidenko metric read
\begin{widetext}
\begin{subequations}\be
G_{tt} &= \eta\,\frac{r^2 \left(3 \cos^2\theta-5\right) a^2+2 r \left(-r^3+2 M r^2+\eta \right)-a^4\cos^2\theta \sin^2\theta}{r^3 \left(r^2+a^2 \cos^2\theta\right)^3}\,,\\
G_{t\varphi} &= a \eta\sin^2\theta\,\frac{a^2 \left(r^2+a^2\right) \cos^2\theta + r \left(5 r^3-4 M r^2+5 a^2 r-2 \eta \right)}{r^3 \left(r^2+a^2 \cos^2\theta\right)^3}\,,\\
G_{rr} &= \frac{2\eta}{r \Delta \left(r^2+a^2 \cos^2\theta\right)}\,,\\
G_{\theta\theta} & = -\frac{\eta  \left(3 r^2+a^2 \cos^2\theta\right)}{r^3 \left(r^2+a^2 \cos^2\theta\right)}\,,\\
G_{\varphi\varphi} &= -\eta\sin^2\theta\,\frac{a^2 \left(a^4-r^4+4 M r^3+2 r \eta \right) \cos^2\theta+r \left[3 r^5+5 a^4 r-2 a^2 \left((2M-4r) r^2+\eta \right)\right]}{r^3 \left(r^2+a^2 \cos^2\theta\right)^3}\,.
\ee\end{subequations}
\end{widetext}

The geodesic equations can be obtained via the Euler--Lagrange equations from the Lagrangian $\mathcal{L}=\frac{1}{2}g_{\mu\nu}\dot{x}^\mu \dot{x}^\nu$, where a dot indicates differentiation with respect to an affine parameter $\lambda$.
However, it is simpler to use the integrals of motion, two of which are related to the obvious symmetries of the metric, \ie\ stationarity and axisymmetry, that can be expressed respectively by
\be
p_t \equiv g_{tt}\dot{t}+g_{t\varphi}\dot{\varphi}=-E\,,\quad
p_\varphi \equiv g_{\varphi\varphi}\dot{\varphi}+g_{t\varphi}\dot{t}=L_z\,,
\ee
where $E$ and $L_z$ represent the energy and the angular momentum along the $\varphi$ axis of the particle.
Another constant of motion can be obtained observing that the Hamiltonian $\mathcal{H}=\frac{1}{2}g_{\mu\nu}p^\mu p^\nu$, where $p^\mu=\partial\mathcal{L}/\partial \dot{x^\mu}$, is independent of the affine parameter.
Therefore we can write $\mathcal{H}=-\frac{1}{2}\epsilon^2$, where $\epsilon^2$ is a constant parameter that can be $+1,0,-1$, respectively, for timelike, null and spacelike geodesics.
The last integral of motion is less obvious and it is related to the separability of the Hamilton--Jacobi equation
\be\label{eq:HJ}
\dot{S}=\frac{1}{2}g^{\mu\nu}\frac{\partial S}{\partial x^\mu}\frac{\partial S}{\partial x^\nu}\,,
\ee
where $S$ is a function of $\lambda$ and the coordinates. In fact, with the ansatz $S=-\frac{\epsilon^2}{2}\lambda-Et+S_{\!\theta} (\theta)+S_{\!r}(r)+L_z\varphi$, \cref{eq:HJ} separates into an angular and a radial part.
The (generalized) Carter constant $Q=K-(aE-L_z)^2$ is related to the separation constant $K$ associated to the hidden symmetry of the metric generated by a second-order Killing tensor $K^{\mu\nu}$ that satisfies $\nabla_{(\rho}K_{\mu\nu)}=0$, where the round parentheses denote symmetrization with respect to the indices.
The explicit form of $K^{\mu\nu}$ is
\be
K^{\mu\nu}=2\Sigma\,l^{(\mu}n^{\nu)}+r^2g^{\mu\nu}\,,
\ee
where $\Sigma=r^2+a^2\cos^2\theta$ while $l^\mu$ and $n^\mu$ are the vectors defined in \cref{eq:tetrad} with $R_\Sigma=R_B=1$ and $R_M=2M+\eta/r^2$.

Using these four integrals of motion it is possible to write the geodesic equations as
\begin{subequations}
\be
\dot{t} &= E + \frac{\left(2 M r^2 + \eta\right) \left((r^2+a^2) E - a L_z\right)}{r \Delta \Sigma}\,,\\
\dot{\varphi} &= \frac{1}{r\Sigma}\left(\frac{a\left(2 M r^2+ \eta \right) E -a^2 L_z r}{\Delta}+\frac{r\,L_z}{\sin^2\theta}\right),\\
\Sigma^2\dot{r}^2 &= \left[a L_z-E \left(a^2+r^2\right)\right]^2 -\Delta \left[(a E-L_z)^2+Q+r^2 \epsilon ^2\right],\\
\Sigma^2\dot{\theta}^2 &= a^2 \cos ^2\theta \,(E^2-\epsilon^2) -L_z^2 \cot ^2\theta +Q\,.
\ee
\end{subequations}

The four-velocity of a zero-angular-momentum observer in the equatorial plane is readily obtained,
\begin{subequations}\be
u^t & = \frac{r^5 + a^2 \left(r^3+2 M r^2+\eta\right)}{r^3 \Delta}\,,\\
u^r &= -\sqrt{\frac{\left(a^2+r^2\right) \left(\eta +2 M r^2\right)}{r^5}}\,,\\
u^\varphi &= \frac{a \left(2 M r^2 + \eta\right)}{r^3 \Delta}\,.
\ee\end{subequations}

\section{Frequency eigeinvalues in the low-frequency, small-mass and small-deformation limit\label{app:instability}}

In the low-frequency regime, \ie, $\omega M\ll1$ and $a\omega\ll1$, the amplification factor for waves scattered off a Kerr black hole can be computed analytically~\cite{Starobinskii:1973,Starobinskii:1974,Page:1976df}. The angular equation reduces to the scalar spherical harmonics equation and the angular eigenvalue $\lambda$ can be approximated as $l(l+1)$. For massive scalar field a similar technique can be applied in the small mass limit $M\mu_s\ll1$~\cite{Detweiler:1980uk}.
We extend this result to the Konoplya--Zhidenko black hole in the limit $\eta/M^3\ll1$.

The asymptotic matching technique consists in solving the radial equation in the asymptotic and near-horizon regions and relies on the existence of an overlap region in which the two solutions can be matched.


In the large $r$ limit the radial equation for a massive scalar field in the Konoplya--Zhidenko background becomes
\be
R_0''(r) + \frac{2}{r}\,R_0'(r) + \left(-\frac{l (l+1)}{r^2} + \frac{2 M \mu_s^2}{r} + \omega^2 - \mu_s^2\right)R_0(r) = 0\,.
\ee
Defining $k^2 = \mu_s^2-\omega^2$, $\nu=M\mu_s^2/k$, and $x=2kr$ the above equation reads
\be\label{eqRxinf}
x R_0''(x) + 2 R_0'(x) + \left(-\frac{l (l+1)}{x}+\nu -\frac{x}{4}\right)R_0(x) = 0\,,
\ee
i.e., the same equation which governs an electron in the hydrogen atom. For large $x$ the two independent solutions of \cref{eqRxinf} behave as $R_0(x)\sim x^{\pm(\nu+1)}\,\e^{\mp x/2}$. Since we are interested in the unstable modes we take the solution with the upper signs, and the complete solution to \cref{eqRxinf} with such asymptotic behaviour is
\be\label{RxinfSol}
R_0(x) = \e^{-x/2} x^l U(l-\nu +1,2 l+2,x)
\ee
being $U$ the confluent hypergeometric function.

The regularity of the electron wave-function in $x=0$ implies that the bound states of the hydrogen atom corresponds to integer values of $\nu$ as $\nu=l+1+n$ with $n$ positive.
As the boundary conditions in this case are slightly different from the quantum mechanics problem (ingoing waves at the horizon) we guess $\nu=l+1+n+\delta\nu$ where $\delta\nu$ is a small complex number.

In the small $x$ limit, \cref{RxinfSol} is
\be
R_0(x) \approx \frac{\Gamma (-2 l-1)}{\Gamma (-l-\nu )}\,x^l + \frac{\Gamma (2 l+1)}{\Gamma (l-\nu +1)}\,x^{-l-1}\,.
\ee
In terms of the coordinate $r$ and in the small $\delta\nu$ limit
\be
R_0(r) \approx (-1)^n\frac{(2 l+n+1)!}{(2 l+1)!}\,(2 k r)^l + (-1)^{n+1} \delta\nu (2 l)! n! (2 k r)^{-l-1}\,.\label{Rinfasym}
\ee


In the near-horizon region we write $R_0(r)=\mathring{R}_0(r)+\eta\,\delta R_0(r)$ and we solve order by order in $\eta/M^3$.
We define a new dimensionless coordinate $x \equiv (r-r_+)/(r_+-r_-)$ and the quantity $q \equiv (a m-2 M r_+ \omega)/(r_+-r_-)$
where $r_\pm=M\pm\sqrt{M^2-a^2}$ are the radial location of the Kerr event and Cauchy horizon.


At zeroth order, the radial equation reduces to
\be\label{eqR0hor}
x^2 (x+1)^2 \mathring{R}_0''(x) + x (2 x+1) (x+1) \mathring{R}_0'(x) \0\\+ \left(q^2-l (l+1) x (x+1)\right)\mathring{R}_0(x) = 0\,,
\ee
whose general solution is a combination of the associated Legendre functions $c_1 P^l_{2\iu q}(1+2x) + c_2 Q^l_{2\iu q}(1+2x)$ which represent, respectively, the ingoing and outgoing waves at the horizon.

Now assume there exists an intermediate region in which the two solutions calculated asymptotically and close to the horizon overlap. Then the small $x$ limit of the asymptotic solution \eqref{Rinfasym} must be equal to the large $x$ limit of the near-horizon solution, supplied with the requirement of no outgoing waves at the horizon ($c_2=0$).
We have
\be
P^l_{2\iu q}(1+2x) \sim \frac{(2l)!\,x^l}{l! \Gamma (l + 1 - 2\iu q)} + \frac{(-1)^{-1-l} l!\,x^{-l-1}}{(2l+1)!\Gamma (-l-2\iu q)}\,.\label{LegendrePlargex}
\ee

The constant $c_1$ can be determined by comparing the $r^l$ terms,
\be\label{c1sol}
c_1 = \frac{(2k)^l (r_+-r_-)^l (-1)^n l! (2l+n+1)! \Gamma (l+1-2\iu q)}{(2l+1)! (2l)!}\,,
\ee
while by comparing the $r^{-l-1}$ terms we get
\be\label{deltanu}
\delta\nu = 2\iu q \left[2k (r_+-r_-)\right]^{2 l+1} \left(\frac{l!}{(2 l)! (2 l+1)!}\right)^2 \times\0\\
\frac{(2l+n+1)! }{n!}\,\prod_{j=1}^l \left(j^2 + 4q^2\right).
\ee

Finally, the relation among $n$, $\delta\nu$ and $\omega=\sigma+\iu\gamma$ gives $\sigma\approx\mu_s$ from the real part, while from the imaginary part
\be\label{igamma}
\iu\gamma = \left(\frac{M\mu_s}{l+1+n}\right)^3\,\frac{\delta\nu}{M}\,.
\ee

Now we are able to give an estimate for the growth time of the instability.
At zeroth order, combining \cref{deltanu,igamma} we notice that for $m>0$ the imaginary part of the frequency is positive and hence the mode is unstable.
In particular, for the most unstable mode, $l=m=1$ and $n=0$, at leading order
\be
\gamma = \mu_s \frac{a}{M}\,\frac{(M\mu_s)^8}{24}\,,
\ee
and the growth time, for an axion with mass $m_\text{axion} =\mu_\text{axion}\hbar = \SI{e-20}{eV}$,
\be
\tau \equiv 1/\gamma = \left(\SI{1.58e6}{s}\right)\left(\frac{\mu_\text{axion}}{\mu_s}\right)\frac{M}{a}\,\frac{1}{(M\mu_s)^8}\,.
\ee


At first order, the zeroth-order solution enters as a ``source term'',
\be\label{eqdeltaR0hor}
x^2 (x+1)^2 \delta R_0''(x) + x (2 x+1) (x+1) \delta R_0'(x) \0\\+ \left(q^2-l (l+1) x (x+1)\right)\delta R_0(x) = T(x)\,,
\ee
where
\be
&r_+ (r_+-r_-)^2 T(x) = - \mathring{R}_0'(x)\0\\
&-\left(\frac{2 q^2}{x} + \frac{q^2 \left(r_-^2-5 r_- r_++2 r_+^2\right)}{M r_+} - l(l+1)\right)\mathring{R}_0(x)\,,
\ee
with $R_0 = c_1 P^l_{2\iu q}(1+2x)$ and $c_1$ given by \cref{c1sol}.

The homogenous problem associated to \cref{eqdeltaR0hor} for $\delta R_0$ is the same as in \cref{eqR0hor} for $\mathring{R}_0$, meaning that its general solution is again a combination of the associated Legendre functions, $c_3 P^l_{2\iu q}(1+2x) + c_4 Q^l_{2\iu q}(1+2x)$.  Again, $c_4$ can be set to zero by the request of no outgoing waves at the horizon.
The particular solution can be obtained with the method of variation of constants,
\be
\delta R_{0,\text{p}} = &- \delta R_{0,1} \int\d z\,\frac{T(z)\,\delta R_{0,2}(z)}{z^2(1+z)^2 W(z)}\0\\
&+ \delta R_{0,2} \int\d z\,\frac{T(z)\,\delta R_{0,1}(z)}{z^2(1+z)^2 W(z)}\,,
\ee
where $W(x)$ is the Wronskian associated with $\delta R_{0,1}(x) = P^l_{2\iu q}(1+2x)$ and $\delta R_{0,2}(x) = Q^l_{2\iu q}(1+2x)$.

As in the zeroth-order calculation, assume that there exists an intermediate overlapping region in which the solution in \cref{Rinfasym} is glued with the large $r$ behaviour of the near-horizon solution.

At this stage, we focus on the $l=m=1$ and $n=0$ mode which is, at zeroth order, the most unstable.
Using \cref{LegendrePlargex} with $l=1$ and
\be
\delta R_{0,\text{p}} \sim - \frac{c_1 x\left[\mathfrak{R}_1 M r_+ + \mathfrak{R}_2 r_-^2 + \mathfrak{R}_3 r_- r_+ + \mathfrak{R}_4 r_+^2\right]}{2M r_+^2 (r_+-r_-)^2 q^2 (1+2\iu q) \left(1-2\iu q\right)^3 \Gamma (1-2\iu q)}\,,
\ee
where
\begin{subequations}\be
\mathfrak{R}_1 & = 8\iu q (1-2\iu q) \left(2q^2 + 1\right) \left[{\psi\left(-2\iu q\right)} + \gamma_\text{E}\right],\\
\mathfrak{R}_2 & = q^2 (1-2\iu q) \left(28q^2 - 4\iu q +1\right),\\
\mathfrak{R}_3 & = 280\iu q^5 - 120q^4 + 70\iu q^3 - 32q^2 + 5\iu q - 2\,,\\
\mathfrak{R}_4 & = -112\iu q^5 + 20q^4 + 28\iu q^3 - 25q^2 + 5\iu q - 2\,,
\ee\end{subequations}
being $\psi(z)$ the digamma function, $\gamma_\text{E}$ the Euler--Mascheroni constant and $q$ is now meant to be computed for $m=1$, we repeat what we have done for the zeroth-order solution, but matching \cref{Rinfasym} with $c_1 \mathring{R}_0 + \eta \left(c_3 \delta R_{0,1} + \delta R_{0,\text{p}}\right)$.
We first solve for $c_3$ and find that $\delta\nu$ gains a correction proportional to $\eta$, whose imaginary part sums up to $\gamma$ computed at zeroth order,
\be
\delta\gamma = \frac{\eta k^3 M \mu_s^3 (r_+-r_-)}{48r_+^2 q \left(4q^2+1\right)}
\left[\mathfrak{g}_1 M r_+ + \mathfrak{g}_2 r_-^2 + \mathfrak{g}_3 r_- r_+ + \mathfrak{g}_4 r_+^2\right],
\ee
where
\begin{subequations}\be
\mathfrak{g}_1 & = 8 q \left(2 q^2+1\right) \left(4 q^2+1\right) \Im\psi(-2\iu q)\,,\\
\mathfrak{g}_2 & = -q^2 \left(4 q^2+1\right) \left(28 q^2+1\right),\\
\mathfrak{g}_3 & = 2 \left(280q^6 + 130q^4 + 21q^2 + 1\right),\\
\mathfrak{g}_4 & = -\left(224q^6 - 36q^4 - 35q^2 - 2\right).
\ee\end{subequations}

We can now evaluate how this correction contributes to the growth time of the instability. At leading order, for an axion,
\be
\delta\tau = -\left(\SI{7.89e3}{s}\right)\left(\frac{\mu_\text{axion}}{\mu_s}\right)\left(\frac{\eta/M^3}{0.01}\right)\left(\frac{M}{a}\right)^3\frac{1}{(M\mu_s)^8}\,.
\ee

\bibliography{refs}

\begin{thebibliography}{84}%
\makeatletter
\providecommand \@ifxundefined [1]{%
 \@ifx{#1\undefined}
}%
\providecommand \@ifnum [1]{%
 \ifnum #1\expandafter \@firstoftwo
 \else \expandafter \@secondoftwo
 \fi
}%
\providecommand \@ifx [1]{%
 \ifx #1\expandafter \@firstoftwo
 \else \expandafter \@secondoftwo
 \fi
}%
\providecommand \natexlab [1]{#1}%
\providecommand \enquote  [1]{``#1''}%
\providecommand \bibnamefont  [1]{#1}%
\providecommand \bibfnamefont [1]{#1}%
\providecommand \citenamefont [1]{#1}%
\providecommand \href@noop [0]{\@secondoftwo}%
\providecommand \href [0]{\begingroup \@sanitize@url \@href}%
\providecommand \@href[1]{\@@startlink{#1}\@@href}%
\providecommand \@@href[1]{\endgroup#1\@@endlink}%
\providecommand \@sanitize@url [0]{\catcode `\\12\catcode `\$12\catcode
  `\&12\catcode `\#12\catcode `\^12\catcode `\_12\catcode `\%12\relax}%
\providecommand \@@startlink[1]{}%
\providecommand \@@endlink[0]{}%
\providecommand \url  [0]{\begingroup\@sanitize@url \@url }%
\providecommand \@url [1]{\endgroup\@href {#1}{\urlprefix }}%
\providecommand \urlprefix  [0]{URL }%
\providecommand \Eprint [0]{\href }%
\providecommand \doibase [0]{https://doi.org/}%
\providecommand \selectlanguage [0]{\@gobble}%
\providecommand \bibinfo  [0]{\@secondoftwo}%
\providecommand \bibfield  [0]{\@secondoftwo}%
\providecommand \translation [1]{[#1]}%
\providecommand \BibitemOpen [0]{}%
\providecommand \bibitemStop [0]{}%
\providecommand \bibitemNoStop [0]{.\EOS\space}%
\providecommand \EOS [0]{\spacefactor3000\relax}%
\providecommand \BibitemShut  [1]{\csname bibitem#1\endcsname}%
\let\auto@bib@innerbib\@empty
\bibitem [{\citenamefont {Will}(2014)}]{Will:2014kxa}%
  \BibitemOpen
  \bibfield  {author} {\bibinfo {author} {\bibfnamefont {C.~M.}\ \bibnamefont
  {Will}},\ }\bibfield  {title} {\bibinfo {title} {{The Confrontation between
  General Relativity and Experiment}},\ }\href
  {https://doi.org/10.12942/lrr-2014-4} {\bibfield  {journal} {\bibinfo
  {journal} {Living Rev. Relativ.}\ }\textbf {\bibinfo {volume} {17}},\
  \bibinfo {pages} {4} (\bibinfo {year} {2014})},\ \Eprint
  {https://arxiv.org/abs/1403.7377} {arXiv:1403.7377 [gr-qc]} \BibitemShut
  {NoStop}%
\bibitem [{\citenamefont {Yagi}\ and\ \citenamefont
  {Stein}(2016)}]{Yagi:2016jml}%
  \BibitemOpen
  \bibfield  {author} {\bibinfo {author} {\bibfnamefont {K.}~\bibnamefont
  {Yagi}}\ and\ \bibinfo {author} {\bibfnamefont {L.~C.}\ \bibnamefont
  {Stein}},\ }\bibfield  {title} {\bibinfo {title} {{Black Hole Based Tests of
  General Relativity}},\ }\href {https://doi.org/10.1088/0264-9381/33/5/054001}
  {\bibfield  {journal} {\bibinfo  {journal} {Class. Quantum Grav.}\ }\textbf
  {\bibinfo {volume} {33}},\ \bibinfo {pages} {054001} (\bibinfo {year}
  {2016})},\ \Eprint {https://arxiv.org/abs/1602.02413} {arXiv:1602.02413
  [gr-qc]} \BibitemShut {NoStop}%
\bibitem [{\citenamefont {Abbott}\ \emph
  {et~al.}(2016{\natexlab{a}})\citenamefont {Abbott} \emph
  {et~al.}}]{Abbott:2016blz}%
  \BibitemOpen
  \bibfield  {author} {\bibinfo {author} {\bibfnamefont {B.~P.}\ \bibnamefont
  {Abbott}} \emph {et~al.} (\bibinfo {collaboration} {LIGO Scientific,
  Virgo}),\ }\bibfield  {title} {\bibinfo {title} {{Observation of
  Gravitational Waves from a Binary Black Hole Merger}},\ }\href
  {https://doi.org/10.1103/PhysRevLett.116.061102} {\bibfield  {journal}
  {\bibinfo  {journal} {Phys. Rev. Lett.}\ }\textbf {\bibinfo {volume} {116}},\
  \bibinfo {pages} {061102} (\bibinfo {year} {2016}{\natexlab{a}})},\ \Eprint
  {https://arxiv.org/abs/1602.03837} {arXiv:1602.03837 [gr-qc]} \BibitemShut
  {NoStop}%
\bibitem [{\citenamefont {Akiyama}\ \emph {et~al.}(2019)\citenamefont {Akiyama}
  \emph {et~al.}}]{Akiyama:2019cqa}%
  \BibitemOpen
  \bibfield  {author} {\bibinfo {author} {\bibfnamefont {K.}~\bibnamefont
  {Akiyama}} \emph {et~al.} (\bibinfo {collaboration} {Event Horizon
  Telescope}),\ }\bibfield  {title} {\bibinfo {title} {{First M87 Event Horizon
  Telescope Results. I. The Shadow of the Supermassive Black Hole}},\ }\href
  {https://doi.org/10.3847/2041-8213/ab0ec7} {\bibfield  {journal} {\bibinfo
  {journal} {Astrophys. J.}\ }\textbf {\bibinfo {volume} {875}},\ \bibinfo
  {pages} {L1} (\bibinfo {year} {2019})},\ \Eprint
  {https://arxiv.org/abs/1906.11238} {arXiv:1906.11238 [astro-ph.GA]}
  \BibitemShut {NoStop}%
\bibitem [{\citenamefont {Celotti}\ \emph {et~al.}(1999)\citenamefont
  {Celotti}, \citenamefont {Miller},\ and\ \citenamefont
  {Sciama}}]{Celotti:1999tg}%
  \BibitemOpen
  \bibfield  {author} {\bibinfo {author} {\bibfnamefont {A.}~\bibnamefont
  {Celotti}}, \bibinfo {author} {\bibfnamefont {J.~C.}\ \bibnamefont
  {Miller}},\ and\ \bibinfo {author} {\bibfnamefont {D.~W.}\ \bibnamefont
  {Sciama}},\ }\bibfield  {title} {\bibinfo {title} {{Astrophysical evidence
  for the existence of black holes}},\ }\href
  {https://doi.org/10.1088/0264-9381/16/12A/301} {\bibfield  {journal}
  {\bibinfo  {journal} {Class. Quantum Grav.}\ }\textbf {\bibinfo {volume}
  {16}},\ \bibinfo {pages} {A3} (\bibinfo {year} {1999})},\ \Eprint
  {https://arxiv.org/abs/astro-ph/9912186} {arXiv:astro-ph/9912186}
  \BibitemShut {NoStop}%
\bibitem [{\citenamefont {Bambi}(2017)}]{Bambi:2015kza}%
  \BibitemOpen
  \bibfield  {author} {\bibinfo {author} {\bibfnamefont {C.}~\bibnamefont
  {Bambi}},\ }\bibfield  {title} {\bibinfo {title} {{Testing black hole
  candidates with electromagnetic radiation}},\ }\href
  {https://doi.org/10.1103/RevModPhys.89.025001} {\bibfield  {journal}
  {\bibinfo  {journal} {Rev. Mod. Phys.}\ }\textbf {\bibinfo {volume} {89}},\
  \bibinfo {pages} {025001} (\bibinfo {year} {2017})},\ \Eprint
  {https://arxiv.org/abs/1509.03884} {arXiv:1509.03884 [gr-qc]} \BibitemShut
  {NoStop}%
\bibitem [{\citenamefont {Kerr}(1963)}]{Kerr:1963ud}%
  \BibitemOpen
  \bibfield  {author} {\bibinfo {author} {\bibfnamefont {R.~P.}\ \bibnamefont
  {Kerr}},\ }\bibfield  {title} {\bibinfo {title} {{Gravitational field of a
  spinning mass as an example of algebraically special metrics}},\ }\href
  {https://doi.org/10.1103/PhysRevLett.11.237} {\bibfield  {journal} {\bibinfo
  {journal} {Phys. Rev. Lett.}\ }\textbf {\bibinfo {volume} {11}},\ \bibinfo
  {pages} {237} (\bibinfo {year} {1963})}\BibitemShut {NoStop}%
\bibitem [{\citenamefont {Visser}(2014)}]{Visser:2014zqa}%
  \BibitemOpen
  \bibfield  {author} {\bibinfo {author} {\bibfnamefont {M.}~\bibnamefont
  {Visser}},\ }\bibfield  {title} {\bibinfo {title} {{Physical observability of
  horizons}},\ }\href {https://doi.org/10.1103/PhysRevD.90.127502} {\bibfield
  {journal} {\bibinfo  {journal} {Phys. Rev. D}\ }\textbf {\bibinfo {volume}
  {90}},\ \bibinfo {pages} {127502} (\bibinfo {year} {2014})},\ \Eprint
  {https://arxiv.org/abs/1407.7295} {arXiv:1407.7295 [gr-qc]} \BibitemShut
  {NoStop}%
\bibitem [{\citenamefont {Abramowicz}\ \emph {et~al.}(2016)\citenamefont
  {Abramowicz}, \citenamefont {Bulik}, \citenamefont {Ellis}, \citenamefont
  {Meissner},\ and\ \citenamefont {Wielgus}}]{Abramowicz:2016lja}%
  \BibitemOpen
  \bibfield  {author} {\bibinfo {author} {\bibfnamefont {M.~A.}\ \bibnamefont
  {Abramowicz}}, \bibinfo {author} {\bibfnamefont {T.}~\bibnamefont {Bulik}},
  \bibinfo {author} {\bibfnamefont {G.~F.~R.}\ \bibnamefont {Ellis}}, \bibinfo
  {author} {\bibfnamefont {K.~A.}\ \bibnamefont {Meissner}},\ and\ \bibinfo
  {author} {\bibfnamefont {M.}~\bibnamefont {Wielgus}},\ }\bibfield  {title}
  {\bibinfo {title} {{The electromagnetic afterglows of gravitational waves as
  a test for Quantum Gravity}},\ }\href@noop {} {\  (\bibinfo {year} {2016})},\
  \Eprint {https://arxiv.org/abs/1603.07830} {arXiv:1603.07830 [gr-qc]}
  \BibitemShut {NoStop}%
\bibitem [{\citenamefont {Cardoso}\ \emph
  {et~al.}(2016{\natexlab{a}})\citenamefont {Cardoso}, \citenamefont
  {Franzin},\ and\ \citenamefont {Pani}}]{Cardoso:2016rao}%
  \BibitemOpen
  \bibfield  {author} {\bibinfo {author} {\bibfnamefont {V.}~\bibnamefont
  {Cardoso}}, \bibinfo {author} {\bibfnamefont {E.}~\bibnamefont {Franzin}},\
  and\ \bibinfo {author} {\bibfnamefont {P.}~\bibnamefont {Pani}},\ }\bibfield
  {title} {\bibinfo {title} {{Is the gravitational-wave ringdown a probe of the
  event horizon?}},\ }\href {https://doi.org/10.1103/PhysRevLett.116.171101}
  {\bibfield  {journal} {\bibinfo  {journal} {Phys. Rev. Lett.}\ }\textbf
  {\bibinfo {volume} {116}},\ \bibinfo {pages} {171101} (\bibinfo {year}
  {2016}{\natexlab{a}})},\ \bibinfo {note} {[Erratum: Phys. Rev. Lett.
  {\bf117}, 089902 (2016)]},\ \Eprint {https://arxiv.org/abs/1602.07309}
  {arXiv:1602.07309 [gr-qc]} \BibitemShut {NoStop}%
\bibitem [{\citenamefont {Cardoso}\ and\ \citenamefont
  {Pani}(2019)}]{Cardoso:2019rvt}%
  \BibitemOpen
  \bibfield  {author} {\bibinfo {author} {\bibfnamefont {V.}~\bibnamefont
  {Cardoso}}\ and\ \bibinfo {author} {\bibfnamefont {P.}~\bibnamefont {Pani}},\
  }\bibfield  {title} {\bibinfo {title} {{Testing the nature of dark compact
  objects:\ a status report}},\ }\href
  {https://doi.org/10.1007/s41114-019-0020-4} {\bibfield  {journal} {\bibinfo
  {journal} {Living Rev. Relativ.}\ }\textbf {\bibinfo {volume} {22}},\
  \bibinfo {pages} {4} (\bibinfo {year} {2019})},\ \Eprint
  {https://arxiv.org/abs/1904.05363} {arXiv:1904.05363 [gr-qc]} \BibitemShut
  {NoStop}%
\bibitem [{\citenamefont {Abbott}\ \emph
  {et~al.}(2016{\natexlab{b}})\citenamefont {Abbott} \emph
  {et~al.}}]{TheLIGOScientific:2016src}%
  \BibitemOpen
  \bibfield  {author} {\bibinfo {author} {\bibfnamefont {B.~P.}\ \bibnamefont
  {Abbott}} \emph {et~al.} (\bibinfo {collaboration} {LIGO Scientific,
  Virgo}),\ }\bibfield  {title} {\bibinfo {title} {{Tests of general relativity
  with GW150914}},\ }\href {https://doi.org/10.1103/PhysRevLett.116.221101}
  {\bibfield  {journal} {\bibinfo  {journal} {Phys. Rev. Lett.}\ }\textbf
  {\bibinfo {volume} {116}},\ \bibinfo {pages} {221101} (\bibinfo {year}
  {2016}{\natexlab{b}})},\ \bibinfo {note} {[Erratum: Phys. Rev. Lett.
  {\bf121}, 129902 (2018)]},\ \Eprint {https://arxiv.org/abs/1602.03841}
  {arXiv:1602.03841 [gr-qc]} \BibitemShut {NoStop}%
\bibitem [{\citenamefont {Johannsen}(2016{\natexlab{a}})}]{Johannsen:2015mdd}%
  \BibitemOpen
  \bibfield  {author} {\bibinfo {author} {\bibfnamefont {T.}~\bibnamefont
  {Johannsen}},\ }\bibfield  {title} {\bibinfo {title} {{Sgr A* and General
  Relativity}},\ }\href {https://doi.org/10.1088/0264-9381/33/11/113001}
  {\bibfield  {journal} {\bibinfo  {journal} {Class. Quantum Grav.}\ }\textbf
  {\bibinfo {volume} {33}},\ \bibinfo {pages} {113001} (\bibinfo {year}
  {2016}{\natexlab{a}})},\ \Eprint {https://arxiv.org/abs/1512.03818}
  {arXiv:1512.03818 [astro-ph.GA]} \BibitemShut {NoStop}%
\bibitem [{\citenamefont {Yunes}\ \emph {et~al.}(2016)\citenamefont {Yunes},
  \citenamefont {Yagi},\ and\ \citenamefont {Pretorius}}]{Yunes:2016jcc}%
  \BibitemOpen
  \bibfield  {author} {\bibinfo {author} {\bibfnamefont {N.}~\bibnamefont
  {Yunes}}, \bibinfo {author} {\bibfnamefont {K.}~\bibnamefont {Yagi}},\ and\
  \bibinfo {author} {\bibfnamefont {F.}~\bibnamefont {Pretorius}},\ }\bibfield
  {title} {\bibinfo {title} {{Theoretical Physics Implications of the Binary
  Black-Hole Mergers GW150914 and GW151226}},\ }\href
  {https://doi.org/10.1103/PhysRevD.94.084002} {\bibfield  {journal} {\bibinfo
  {journal} {Phys. Rev. D}\ }\textbf {\bibinfo {volume} {94}},\ \bibinfo
  {pages} {084002} (\bibinfo {year} {2016})},\ \Eprint
  {https://arxiv.org/abs/1603.08955} {arXiv:1603.08955 [gr-qc]} \BibitemShut
  {NoStop}%
\bibitem [{\citenamefont {Cardoso}\ and\ \citenamefont
  {Gualtieri}(2016)}]{Cardoso:2016ryw}%
  \BibitemOpen
  \bibfield  {author} {\bibinfo {author} {\bibfnamefont {V.}~\bibnamefont
  {Cardoso}}\ and\ \bibinfo {author} {\bibfnamefont {L.}~\bibnamefont
  {Gualtieri}},\ }\bibfield  {title} {\bibinfo {title} {{Testing the black hole
  `no-hair' hypothesis}},\ }\href
  {https://doi.org/10.1088/0264-9381/33/17/174001} {\bibfield  {journal}
  {\bibinfo  {journal} {Class. Quantum Grav.}\ }\textbf {\bibinfo {volume}
  {33}},\ \bibinfo {pages} {174001} (\bibinfo {year} {2016})},\ \Eprint
  {https://arxiv.org/abs/1607.03133} {arXiv:1607.03133 [gr-qc]} \BibitemShut
  {NoStop}%
\bibitem [{\citenamefont {Johannsen}(2016{\natexlab{b}})}]{Johannsen:2016uoh}%
  \BibitemOpen
  \bibfield  {author} {\bibinfo {author} {\bibfnamefont {T.}~\bibnamefont
  {Johannsen}},\ }\bibfield  {title} {\bibinfo {title} {{Testing the No-Hair
  Theorem with Observations of Black Holes in the Electromagnetic Spectrum}},\
  }\href {https://doi.org/10.1088/0264-9381/33/12/124001} {\bibfield  {journal}
  {\bibinfo  {journal} {Class. Quantum Grav.}\ }\textbf {\bibinfo {volume}
  {33}},\ \bibinfo {pages} {124001} (\bibinfo {year} {2016}{\natexlab{b}})},\
  \Eprint {https://arxiv.org/abs/1602.07694} {arXiv:1602.07694 [astro-ph.HE]}
  \BibitemShut {NoStop}%
\bibitem [{\citenamefont {Carballo-Rubio}\ \emph {et~al.}(2018)\citenamefont
  {Carballo-Rubio}, \citenamefont {Di~Filippo}, \citenamefont {Liberati},\ and\
  \citenamefont {Visser}}]{Carballo-Rubio:2018jzw}%
  \BibitemOpen
  \bibfield  {author} {\bibinfo {author} {\bibfnamefont {R.}~\bibnamefont
  {Carballo-Rubio}}, \bibinfo {author} {\bibfnamefont {F.}~\bibnamefont
  {Di~Filippo}}, \bibinfo {author} {\bibfnamefont {S.}~\bibnamefont
  {Liberati}},\ and\ \bibinfo {author} {\bibfnamefont {M.}~\bibnamefont
  {Visser}},\ }\bibfield  {title} {\bibinfo {title} {{Phenomenological aspects
  of black holes beyond general relativity}},\ }\href
  {https://doi.org/10.1103/PhysRevD.98.124009} {\bibfield  {journal} {\bibinfo
  {journal} {Phys. Rev. D}\ }\textbf {\bibinfo {volume} {98}},\ \bibinfo
  {pages} {124009} (\bibinfo {year} {2018})},\ \Eprint
  {https://arxiv.org/abs/1809.08238} {arXiv:1809.08238 [gr-qc]} \BibitemShut
  {NoStop}%
\bibitem [{\citenamefont {Penrose}(1965)}]{Penrose:1964wq}%
  \BibitemOpen
  \bibfield  {author} {\bibinfo {author} {\bibfnamefont {R.}~\bibnamefont
  {Penrose}},\ }\bibfield  {title} {\bibinfo {title} {{Gravitational collapse
  and space-time singularities}},\ }\href
  {https://doi.org/10.1103/PhysRevLett.14.57} {\bibfield  {journal} {\bibinfo
  {journal} {Phys. Rev. Lett.}\ }\textbf {\bibinfo {volume} {14}},\ \bibinfo
  {pages} {57} (\bibinfo {year} {1965})}\BibitemShut {NoStop}%
\bibitem [{\citenamefont {Ferrari}\ and\ \citenamefont
  {Kokkotas}(2000)}]{Ferrari:2000sr}%
  \BibitemOpen
  \bibfield  {author} {\bibinfo {author} {\bibfnamefont {V.}~\bibnamefont
  {Ferrari}}\ and\ \bibinfo {author} {\bibfnamefont {K.~D.}\ \bibnamefont
  {Kokkotas}},\ }\bibfield  {title} {\bibinfo {title} {{Scattering of particles
  by neutron stars:\ Time evolutions for axial perturbations}},\ }\href
  {https://doi.org/10.1103/PhysRevD.62.107504} {\bibfield  {journal} {\bibinfo
  {journal} {Phys. Rev. D}\ }\textbf {\bibinfo {volume} {62}},\ \bibinfo
  {pages} {107504} (\bibinfo {year} {2000})},\ \Eprint
  {https://arxiv.org/abs/gr-qc/0008057} {arXiv:gr-qc/0008057} \BibitemShut
  {NoStop}%
\bibitem [{\citenamefont {Abedi}\ \emph {et~al.}(2020)\citenamefont {Abedi},
  \citenamefont {Afshordi}, \citenamefont {Oshita},\ and\ \citenamefont
  {Wang}}]{Abedi:2020ujo}%
  \BibitemOpen
  \bibfield  {author} {\bibinfo {author} {\bibfnamefont {J.}~\bibnamefont
  {Abedi}}, \bibinfo {author} {\bibfnamefont {N.}~\bibnamefont {Afshordi}},
  \bibinfo {author} {\bibfnamefont {N.}~\bibnamefont {Oshita}},\ and\ \bibinfo
  {author} {\bibfnamefont {Q.}~\bibnamefont {Wang}},\ }\bibfield  {title}
  {\bibinfo {title} {{Quantum Black Holes in the Sky}},\ }\href
  {https://doi.org/10.3390/universe6030043} {\bibfield  {journal} {\bibinfo
  {journal} {Universe}\ }\textbf {\bibinfo {volume} {6}},\ \bibinfo {pages}
  {43} (\bibinfo {year} {2020})},\ \Eprint {https://arxiv.org/abs/2001.09553}
  {arXiv:2001.09553 [gr-qc]} \BibitemShut {NoStop}%
\bibitem [{\citenamefont {Cardoso}\ \emph
  {et~al.}(2016{\natexlab{b}})\citenamefont {Cardoso}, \citenamefont {Hopper},
  \citenamefont {Macedo}, \citenamefont {Palenzuela},\ and\ \citenamefont
  {Pani}}]{Cardoso:2016oxy}%
  \BibitemOpen
  \bibfield  {author} {\bibinfo {author} {\bibfnamefont {V.}~\bibnamefont
  {Cardoso}}, \bibinfo {author} {\bibfnamefont {S.}~\bibnamefont {Hopper}},
  \bibinfo {author} {\bibfnamefont {C.~F.~B.}\ \bibnamefont {Macedo}}, \bibinfo
  {author} {\bibfnamefont {C.}~\bibnamefont {Palenzuela}},\ and\ \bibinfo
  {author} {\bibfnamefont {P.}~\bibnamefont {Pani}},\ }\bibfield  {title}
  {\bibinfo {title} {{Gravitational-wave signatures of exotic compact objects
  and of quantum corrections at the horizon scale}},\ }\href
  {https://doi.org/10.1103/PhysRevD.94.084031} {\bibfield  {journal} {\bibinfo
  {journal} {Phys. Rev. D}\ }\textbf {\bibinfo {volume} {94}},\ \bibinfo
  {pages} {084031} (\bibinfo {year} {2016}{\natexlab{b}})},\ \Eprint
  {https://arxiv.org/abs/1608.08637} {arXiv:1608.08637 [gr-qc]} \BibitemShut
  {NoStop}%
\bibitem [{\citenamefont {Cardoso}\ and\ \citenamefont
  {Pani}(2017)}]{Cardoso:2017cqb}%
  \BibitemOpen
  \bibfield  {author} {\bibinfo {author} {\bibfnamefont {V.}~\bibnamefont
  {Cardoso}}\ and\ \bibinfo {author} {\bibfnamefont {P.}~\bibnamefont {Pani}},\
  }\bibfield  {title} {\bibinfo {title} {{Tests for the existence of black
  holes through gravitational wave echoes}},\ }\href
  {https://doi.org/10.1038/s41550-017-0225-y} {\bibfield  {journal} {\bibinfo
  {journal} {Nature Astron.}\ }\textbf {\bibinfo {volume} {1}},\ \bibinfo
  {pages} {586} (\bibinfo {year} {2017})},\ \Eprint
  {https://arxiv.org/abs/1709.01525} {arXiv:1709.01525 [gr-qc]} \BibitemShut
  {NoStop}%
\bibitem [{\citenamefont {Berti}\ \emph {et~al.}(2018)\citenamefont {Berti},
  \citenamefont {Yagi}, \citenamefont {Yang},\ and\ \citenamefont
  {Yunes}}]{Berti:2018vdi}%
  \BibitemOpen
  \bibfield  {author} {\bibinfo {author} {\bibfnamefont {E.}~\bibnamefont
  {Berti}}, \bibinfo {author} {\bibfnamefont {K.}~\bibnamefont {Yagi}},
  \bibinfo {author} {\bibfnamefont {H.}~\bibnamefont {Yang}},\ and\ \bibinfo
  {author} {\bibfnamefont {N.}~\bibnamefont {Yunes}},\ }\bibfield  {title}
  {\bibinfo {title} {{Extreme Gravity Tests with Gravitational Waves from
  Compact Binary Coalescences:\ (II) Ringdown}},\ }\href
  {https://doi.org/10.1007/s10714-018-2372-6} {\bibfield  {journal} {\bibinfo
  {journal} {Gen. Relativ. Gravit.}\ }\textbf {\bibinfo {volume} {50}},\
  \bibinfo {pages} {49} (\bibinfo {year} {2018})},\ \Eprint
  {https://arxiv.org/abs/1801.03587} {arXiv:1801.03587 [gr-qc]} \BibitemShut
  {NoStop}%
\bibitem [{\citenamefont {Johannsen}\ and\ \citenamefont
  {Psaltis}(2011)}]{Johannsen:2011dh}%
  \BibitemOpen
  \bibfield  {author} {\bibinfo {author} {\bibfnamefont {T.}~\bibnamefont
  {Johannsen}}\ and\ \bibinfo {author} {\bibfnamefont {D.}~\bibnamefont
  {Psaltis}},\ }\bibfield  {title} {\bibinfo {title} {{A Metric for Rapidly
  Spinning Black Holes Suitable for Strong-Field Tests of the No-Hair
  Theorem}},\ }\href {https://doi.org/10.1103/PhysRevD.83.124015} {\bibfield
  {journal} {\bibinfo  {journal} {Phys. Rev. D}\ }\textbf {\bibinfo {volume}
  {83}},\ \bibinfo {pages} {124015} (\bibinfo {year} {2011})},\ \Eprint
  {https://arxiv.org/abs/1105.3191} {arXiv:1105.3191 [gr-qc]} \BibitemShut
  {NoStop}%
\bibitem [{\citenamefont {Johannsen}(2013)}]{Johannsen:2015pca}%
  \BibitemOpen
  \bibfield  {author} {\bibinfo {author} {\bibfnamefont {T.}~\bibnamefont
  {Johannsen}},\ }\bibfield  {title} {\bibinfo {title} {{Regular Black Hole
  Metric with Three Constants of Motion}},\ }\href
  {https://doi.org/10.1103/PhysRevD.88.044002} {\bibfield  {journal} {\bibinfo
  {journal} {Phys. Rev. D}\ }\textbf {\bibinfo {volume} {88}},\ \bibinfo
  {pages} {044002} (\bibinfo {year} {2013})},\ \Eprint
  {https://arxiv.org/abs/1501.02809} {arXiv:1501.02809 [gr-qc]} \BibitemShut
  {NoStop}%
\bibitem [{\citenamefont {Johannsen}\ \emph {et~al.}(2016)\citenamefont
  {Johannsen}, \citenamefont {Broderick}, \citenamefont {Plewa}, \citenamefont
  {Chatzopoulos}, \citenamefont {Doeleman}, \citenamefont {Eisenhauer},
  \citenamefont {Fish}, \citenamefont {Genzel}, \citenamefont {Gerhard},\ and\
  \citenamefont {Johnson}}]{Johannsen:2015hib}%
  \BibitemOpen
  \bibfield  {author} {\bibinfo {author} {\bibfnamefont {T.}~\bibnamefont
  {Johannsen}}, \bibinfo {author} {\bibfnamefont {A.~E.}\ \bibnamefont
  {Broderick}}, \bibinfo {author} {\bibfnamefont {P.~M.}\ \bibnamefont
  {Plewa}}, \bibinfo {author} {\bibfnamefont {S.}~\bibnamefont {Chatzopoulos}},
  \bibinfo {author} {\bibfnamefont {S.~S.}\ \bibnamefont {Doeleman}}, \bibinfo
  {author} {\bibfnamefont {F.}~\bibnamefont {Eisenhauer}}, \bibinfo {author}
  {\bibfnamefont {V.~L.}\ \bibnamefont {Fish}}, \bibinfo {author}
  {\bibfnamefont {R.}~\bibnamefont {Genzel}}, \bibinfo {author} {\bibfnamefont
  {O.}~\bibnamefont {Gerhard}},\ and\ \bibinfo {author} {\bibfnamefont {M.~D.}\
  \bibnamefont {Johnson}},\ }\bibfield  {title} {\bibinfo {title} {{Testing
  General Relativity with the Shadow Size of Sgr A*}},\ }\href
  {https://doi.org/10.1103/PhysRevLett.116.031101} {\bibfield  {journal}
  {\bibinfo  {journal} {Phys. Rev. Lett.}\ }\textbf {\bibinfo {volume} {116}},\
  \bibinfo {pages} {031101} (\bibinfo {year} {2016})},\ \Eprint
  {https://arxiv.org/abs/1512.02640} {arXiv:1512.02640 [astro-ph.GA]}
  \BibitemShut {NoStop}%
\bibitem [{\citenamefont {Cardoso}\ \emph {et~al.}(2014)\citenamefont
  {Cardoso}, \citenamefont {Pani},\ and\ \citenamefont
  {Rico}}]{Cardoso:2014rha}%
  \BibitemOpen
  \bibfield  {author} {\bibinfo {author} {\bibfnamefont {V.}~\bibnamefont
  {Cardoso}}, \bibinfo {author} {\bibfnamefont {P.}~\bibnamefont {Pani}},\ and\
  \bibinfo {author} {\bibfnamefont {J.}~\bibnamefont {Rico}},\ }\bibfield
  {title} {\bibinfo {title} {{On generic parametrizations of spinning
  black-hole geometries}},\ }\href {https://doi.org/10.1103/PhysRevD.89.064007}
  {\bibfield  {journal} {\bibinfo  {journal} {Phys. Rev. D}\ }\textbf {\bibinfo
  {volume} {89}},\ \bibinfo {pages} {064007} (\bibinfo {year} {2014})},\
  \Eprint {https://arxiv.org/abs/1401.0528} {arXiv:1401.0528 [gr-qc]}
  \BibitemShut {NoStop}%
\bibitem [{\citenamefont {Rezzolla}\ and\ \citenamefont
  {Zhidenko}(2014)}]{Rezzolla:2014mua}%
  \BibitemOpen
  \bibfield  {author} {\bibinfo {author} {\bibfnamefont {L.}~\bibnamefont
  {Rezzolla}}\ and\ \bibinfo {author} {\bibfnamefont {A.}~\bibnamefont
  {Zhidenko}},\ }\bibfield  {title} {\bibinfo {title} {{New parametrization for
  spherically symmetric black holes in metric theories of gravity}},\ }\href
  {https://doi.org/10.1103/PhysRevD.90.084009} {\bibfield  {journal} {\bibinfo
  {journal} {Phys. Rev. D}\ }\textbf {\bibinfo {volume} {90}},\ \bibinfo
  {pages} {084009} (\bibinfo {year} {2014})},\ \Eprint
  {https://arxiv.org/abs/1407.3086} {arXiv:1407.3086 [gr-qc]} \BibitemShut
  {NoStop}%
\bibitem [{\citenamefont {Konoplya}\ \emph {et~al.}(2016)\citenamefont
  {Konoplya}, \citenamefont {Rezzolla},\ and\ \citenamefont
  {Zhidenko}}]{Konoplya:2016jvv}%
  \BibitemOpen
  \bibfield  {author} {\bibinfo {author} {\bibfnamefont {R.}~\bibnamefont
  {Konoplya}}, \bibinfo {author} {\bibfnamefont {L.}~\bibnamefont {Rezzolla}},\
  and\ \bibinfo {author} {\bibfnamefont {A.}~\bibnamefont {Zhidenko}},\
  }\bibfield  {title} {\bibinfo {title} {{General parametrization of
  axisymmetric black holes in metric theories of gravity}},\ }\href
  {https://doi.org/10.1103/PhysRevD.93.064015} {\bibfield  {journal} {\bibinfo
  {journal} {Phys. Rev. D}\ }\textbf {\bibinfo {volume} {93}},\ \bibinfo
  {pages} {064015} (\bibinfo {year} {2016})},\ \Eprint
  {https://arxiv.org/abs/1602.02378} {arXiv:1602.02378 [gr-qc]} \BibitemShut
  {NoStop}%
\bibitem [{\citenamefont {Ni}\ \emph {et~al.}(2016)\citenamefont {Ni},
  \citenamefont {Jiang},\ and\ \citenamefont {Bambi}}]{Ni:2016uik}%
  \BibitemOpen
  \bibfield  {author} {\bibinfo {author} {\bibfnamefont {Y.}~\bibnamefont
  {Ni}}, \bibinfo {author} {\bibfnamefont {J.}~\bibnamefont {Jiang}},\ and\
  \bibinfo {author} {\bibfnamefont {C.}~\bibnamefont {Bambi}},\ }\bibfield
  {title} {\bibinfo {title} {{Testing the Kerr metric with the iron line and
  the KRZ parametrization}},\ }\href
  {https://doi.org/10.1088/1475-7516/2016/09/014} {\bibfield  {journal}
  {\bibinfo  {journal} {J. Cosmol. Astropart. Phys.}\ }\textbf {\bibinfo
  {volume} {\normalfont09 (2016) 014}}},\ \Eprint
  {https://arxiv.org/abs/1607.04893} {arXiv:1607.04893 [gr-qc]} \BibitemShut
  {NoStop}%
\bibitem [{\citenamefont {C\'ardenas-Avenda\~no}\ \emph
  {et~al.}(2016)\citenamefont {C\'ardenas-Avenda\~no}, \citenamefont {Jiang},\
  and\ \citenamefont {Bambi}}]{Cardenas-Avendano:2016zml}%
  \BibitemOpen
  \bibfield  {author} {\bibinfo {author} {\bibfnamefont {A.}~\bibnamefont
  {C\'ardenas-Avenda\~no}}, \bibinfo {author} {\bibfnamefont {J.}~\bibnamefont
  {Jiang}},\ and\ \bibinfo {author} {\bibfnamefont {C.}~\bibnamefont {Bambi}},\
  }\bibfield  {title} {\bibinfo {title} {{Testing the Kerr black hole
  hypothesis:\ comparison between the gravitational wave and the iron line
  approaches}},\ }\href {https://doi.org/10.1016/j.physletb.2016.06.075}
  {\bibfield  {journal} {\bibinfo  {journal} {Phys. Lett. B}\ }\textbf
  {\bibinfo {volume} {760}},\ \bibinfo {pages} {254} (\bibinfo {year}
  {2016})},\ \Eprint {https://arxiv.org/abs/1603.04720} {arXiv:1603.04720
  [gr-qc]} \BibitemShut {NoStop}%
\bibitem [{\citenamefont {Nampalliwar}\ \emph {et~al.}(2020)\citenamefont
  {Nampalliwar}, \citenamefont {Xin}, \citenamefont {Srivastava}, \citenamefont
  {Abdikamalov}, \citenamefont {Ayzenberg}, \citenamefont {Bambi},
  \citenamefont {Dauser}, \citenamefont {Garcia},\ and\ \citenamefont
  {Tripathi}}]{Nampalliwar:2019iti}%
  \BibitemOpen
  \bibfield  {author} {\bibinfo {author} {\bibfnamefont {S.}~\bibnamefont
  {Nampalliwar}}, \bibinfo {author} {\bibfnamefont {S.}~\bibnamefont {Xin}},
  \bibinfo {author} {\bibfnamefont {S.}~\bibnamefont {Srivastava}}, \bibinfo
  {author} {\bibfnamefont {A.~B.}\ \bibnamefont {Abdikamalov}}, \bibinfo
  {author} {\bibfnamefont {D.}~\bibnamefont {Ayzenberg}}, \bibinfo {author}
  {\bibfnamefont {C.}~\bibnamefont {Bambi}}, \bibinfo {author} {\bibfnamefont
  {T.}~\bibnamefont {Dauser}}, \bibinfo {author} {\bibfnamefont {J.~A.}\
  \bibnamefont {Garcia}},\ and\ \bibinfo {author} {\bibfnamefont
  {A.}~\bibnamefont {Tripathi}},\ }\bibfield  {title} {\bibinfo {title}
  {{Testing General Relativity with X-ray reflection spectroscopy:\ The
  Konoplya-Rezzolla-Zhidenko parametrization}},\ }\href
  {https://doi.org/10.1103/PhysRevD.102.124071} {\bibfield  {journal} {\bibinfo
   {journal} {Phys. Rev. D}\ }\textbf {\bibinfo {volume} {102}},\ \bibinfo
  {pages} {124071} (\bibinfo {year} {2020})},\ \Eprint
  {https://arxiv.org/abs/1903.12119} {arXiv:1903.12119 [gr-qc]} \BibitemShut
  {NoStop}%
\bibitem [{\citenamefont {Younsi}\ \emph {et~al.}(2016)\citenamefont {Younsi},
  \citenamefont {Zhidenko}, \citenamefont {Rezzolla}, \citenamefont
  {Konoplya},\ and\ \citenamefont {Mizuno}}]{Younsi:2016azx}%
  \BibitemOpen
  \bibfield  {author} {\bibinfo {author} {\bibfnamefont {Z.}~\bibnamefont
  {Younsi}}, \bibinfo {author} {\bibfnamefont {A.}~\bibnamefont {Zhidenko}},
  \bibinfo {author} {\bibfnamefont {L.}~\bibnamefont {Rezzolla}}, \bibinfo
  {author} {\bibfnamefont {R.}~\bibnamefont {Konoplya}},\ and\ \bibinfo
  {author} {\bibfnamefont {Y.}~\bibnamefont {Mizuno}},\ }\bibfield  {title}
  {\bibinfo {title} {{New method for shadow calculations:\ Application to
  parametrized axisymmetric black holes}},\ }\href
  {https://doi.org/10.1103/PhysRevD.94.084025} {\bibfield  {journal} {\bibinfo
  {journal} {Phys. Rev. D}\ }\textbf {\bibinfo {volume} {94}},\ \bibinfo
  {pages} {084025} (\bibinfo {year} {2016})},\ \Eprint
  {https://arxiv.org/abs/1607.05767} {arXiv:1607.05767 [gr-qc]} \BibitemShut
  {NoStop}%
\bibitem [{\citenamefont {Mizuno}\ \emph {et~al.}(2018)\citenamefont {Mizuno},
  \citenamefont {Younsi}, \citenamefont {Fromm}, \citenamefont {Porth},
  \citenamefont {De~Laurentis}, \citenamefont {Olivares}, \citenamefont
  {Falcke}, \citenamefont {Kramer},\ and\ \citenamefont
  {Rezzolla}}]{Mizuno:2018lxz}%
  \BibitemOpen
  \bibfield  {author} {\bibinfo {author} {\bibfnamefont {Y.}~\bibnamefont
  {Mizuno}}, \bibinfo {author} {\bibfnamefont {Z.}~\bibnamefont {Younsi}},
  \bibinfo {author} {\bibfnamefont {C.~M.}\ \bibnamefont {Fromm}}, \bibinfo
  {author} {\bibfnamefont {O.}~\bibnamefont {Porth}}, \bibinfo {author}
  {\bibfnamefont {M.}~\bibnamefont {De~Laurentis}}, \bibinfo {author}
  {\bibfnamefont {H.}~\bibnamefont {Olivares}}, \bibinfo {author}
  {\bibfnamefont {H.}~\bibnamefont {Falcke}}, \bibinfo {author} {\bibfnamefont
  {M.}~\bibnamefont {Kramer}},\ and\ \bibinfo {author} {\bibfnamefont
  {L.}~\bibnamefont {Rezzolla}},\ }\bibfield  {title} {\bibinfo {title} {{The
  Current Ability to Test Theories of Gravity with Black Hole Shadows}},\
  }\href {https://doi.org/10.1038/s41550-018-0449-5} {\bibfield  {journal}
  {\bibinfo  {journal} {Nature Astron.}\ }\textbf {\bibinfo {volume} {2}},\
  \bibinfo {pages} {585} (\bibinfo {year} {2018})},\ \Eprint
  {https://arxiv.org/abs/1804.05812} {arXiv:1804.05812 [astro-ph.GA]}
  \BibitemShut {NoStop}%
\bibitem [{\citenamefont {Ghasemi-Nodehi}\ and\ \citenamefont
  {Bambi}(2016)}]{Ghasemi-Nodehi:2016wao}%
  \BibitemOpen
  \bibfield  {author} {\bibinfo {author} {\bibfnamefont {M.}~\bibnamefont
  {Ghasemi-Nodehi}}\ and\ \bibinfo {author} {\bibfnamefont {C.}~\bibnamefont
  {Bambi}},\ }\bibfield  {title} {\bibinfo {title} {{Note on a new
  parametrization for testing the Kerr metric}},\ }\href
  {https://doi.org/10.1140/epjc/s10052-016-4137-2} {\bibfield  {journal}
  {\bibinfo  {journal} {Eur. Phys. J. C}\ }\textbf {\bibinfo {volume} {76}},\
  \bibinfo {pages} {290} (\bibinfo {year} {2016})},\ \Eprint
  {https://arxiv.org/abs/1604.07032} {arXiv:1604.07032 [gr-qc]} \BibitemShut
  {NoStop}%
\bibitem [{\citenamefont {Carson}\ and\ \citenamefont
  {Yagi}(2020)}]{Carson:2020dez}%
  \BibitemOpen
  \bibfield  {author} {\bibinfo {author} {\bibfnamefont {Z.}~\bibnamefont
  {Carson}}\ and\ \bibinfo {author} {\bibfnamefont {K.}~\bibnamefont {Yagi}},\
  }\bibfield  {title} {\bibinfo {title} {{Asymptotically flat, parametrized
  black hole metric preserving Kerr symmetries}},\ }\href
  {https://doi.org/10.1103/PhysRevD.101.084030} {\bibfield  {journal} {\bibinfo
   {journal} {Phys. Rev. D}\ }\textbf {\bibinfo {volume} {101}},\ \bibinfo
  {pages} {084030} (\bibinfo {year} {2020})},\ \Eprint
  {https://arxiv.org/abs/2002.01028} {arXiv:2002.01028 [gr-qc]} \BibitemShut
  {NoStop}%
\bibitem [{\citenamefont {Konoplya}\ and\ \citenamefont
  {Zhidenko}(2020)}]{Konoplya:2020hyk}%
  \BibitemOpen
  \bibfield  {author} {\bibinfo {author} {\bibfnamefont {R.~A.}\ \bibnamefont
  {Konoplya}}\ and\ \bibinfo {author} {\bibfnamefont {A.}~\bibnamefont
  {Zhidenko}},\ }\bibfield  {title} {\bibinfo {title} {{General parametrization
  of black holes:\ the only parameters that matter}},\ }\href
  {https://doi.org/10.1103/PhysRevD.101.124004} {\bibfield  {journal} {\bibinfo
   {journal} {Phys. Rev. D}\ }\textbf {\bibinfo {volume} {101}},\ \bibinfo
  {pages} {124004} (\bibinfo {year} {2020})},\ \Eprint
  {https://arxiv.org/abs/2001.06100} {arXiv:2001.06100 [gr-qc]} \BibitemShut
  {NoStop}%
\bibitem [{\citenamefont {Zel'dovich}(1971)}]{Zeldovich:1971}%
  \BibitemOpen
  \bibfield  {author} {\bibinfo {author} {\bibfnamefont {Y.~B.}\ \bibnamefont
  {Zel'dovich}},\ }\bibfield  {title} {\bibinfo {title} {{Generation of Waves
  by a Rotating Body}},\ }\href@noop {} {\bibfield  {journal} {\bibinfo
  {journal} {J. Exp. Theor. Phys. Letters}\ }\textbf {\bibinfo {volume} {14}},\
  \bibinfo {pages} {180} (\bibinfo {year} {1971})}\BibitemShut {NoStop}%
\bibitem [{\citenamefont {Bekenstein}\ and\ \citenamefont
  {Schiffer}(1998)}]{Bekenstein:1998nt}%
  \BibitemOpen
  \bibfield  {author} {\bibinfo {author} {\bibfnamefont {J.~D.}\ \bibnamefont
  {Bekenstein}}\ and\ \bibinfo {author} {\bibfnamefont {M.}~\bibnamefont
  {Schiffer}},\ }\bibfield  {title} {\bibinfo {title} {{The many faces of
  superradiance}},\ }\href {https://doi.org/10.1103/PhysRevD.58.064014}
  {\bibfield  {journal} {\bibinfo  {journal} {Phys. Rev. D}\ }\textbf {\bibinfo
  {volume} {58}},\ \bibinfo {pages} {064014} (\bibinfo {year} {1998})},\
  \Eprint {https://arxiv.org/abs/gr-qc/9803033} {arXiv:gr-qc/9803033}
  \BibitemShut {NoStop}%
\bibitem [{\citenamefont {Brito}\ \emph {et~al.}(2020)\citenamefont {Brito},
  \citenamefont {Cardoso},\ and\ \citenamefont {Pani}}]{Brito:2015oca}%
  \BibitemOpen
  \bibfield  {author} {\bibinfo {author} {\bibfnamefont {R.}~\bibnamefont
  {Brito}}, \bibinfo {author} {\bibfnamefont {V.}~\bibnamefont {Cardoso}},\
  and\ \bibinfo {author} {\bibfnamefont {P.}~\bibnamefont {Pani}},\ }\href
  {https://doi.org/10.1007/978-3-030-46622-0} {\emph {\bibinfo {title}
  {{Superradiance:\ Energy Extraction, Black-Hole Bombs and Implications for
  Astrophysics and Particle Physics}}}},\ \bibinfo {edition} {2nd}\ ed.,\
  \bibinfo {series} {Lect. Notes Phys.}, Vol.\ \bibinfo {volume} {971}\
  (\bibinfo  {publisher} {Springer},\ \bibinfo {year} {2020})\ pp.\ \bibinfo
  {pages} {1--293},\ \Eprint {https://arxiv.org/abs/1501.06570}
  {arXiv:1501.06570 [gr-qc]} \BibitemShut {NoStop}%
\bibitem [{\citenamefont {Press}\ and\ \citenamefont
  {Teukolsky}(1972)}]{Press:1972zz}%
  \BibitemOpen
  \bibfield  {author} {\bibinfo {author} {\bibfnamefont {W.~H.}\ \bibnamefont
  {Press}}\ and\ \bibinfo {author} {\bibfnamefont {S.~A.}\ \bibnamefont
  {Teukolsky}},\ }\bibfield  {title} {\bibinfo {title} {{Floating Orbits,
  Superradiant Scattering and the Black-hole Bomb}},\ }\href
  {https://doi.org/10.1038/238211a0} {\bibfield  {journal} {\bibinfo  {journal}
  {Nature}\ }\textbf {\bibinfo {volume} {238}},\ \bibinfo {pages} {211}
  (\bibinfo {year} {1972})}\BibitemShut {NoStop}%
\bibitem [{\citenamefont {Cardoso}\ \emph {et~al.}(2004)\citenamefont
  {Cardoso}, \citenamefont {Dias}, \citenamefont {Lemos},\ and\ \citenamefont
  {Yoshida}}]{Cardoso:2004nk}%
  \BibitemOpen
  \bibfield  {author} {\bibinfo {author} {\bibfnamefont {V.}~\bibnamefont
  {Cardoso}}, \bibinfo {author} {\bibfnamefont {O.~J.~C.}\ \bibnamefont
  {Dias}}, \bibinfo {author} {\bibfnamefont {J.~P.~S.}\ \bibnamefont {Lemos}},\
  and\ \bibinfo {author} {\bibfnamefont {S.}~\bibnamefont {Yoshida}},\
  }\bibfield  {title} {\bibinfo {title} {{The black hole bomb and superradiant
  instabilities}},\ }\href {https://doi.org/10.1103/PhysRevD.70.044039}
  {\bibfield  {journal} {\bibinfo  {journal} {Phys. Rev. D}\ }\textbf {\bibinfo
  {volume} {70}},\ \bibinfo {pages} {044039} (\bibinfo {year} {2004})},\
  \bibinfo {note} {[Erratum: Phys. Rev. D {\bf70}, 049903 (2004)]},\ \Eprint
  {https://arxiv.org/abs/hep-th/0404096} {arXiv:hep-th/0404096} \BibitemShut
  {NoStop}%
\bibitem [{\citenamefont {Damour}\ \emph {et~al.}(1976)\citenamefont {Damour},
  \citenamefont {Deruelle},\ and\ \citenamefont {Ruffini}}]{Damour:1976kh}%
  \BibitemOpen
  \bibfield  {author} {\bibinfo {author} {\bibfnamefont {T.}~\bibnamefont
  {Damour}}, \bibinfo {author} {\bibfnamefont {N.}~\bibnamefont {Deruelle}},\
  and\ \bibinfo {author} {\bibfnamefont {R.}~\bibnamefont {Ruffini}},\
  }\bibfield  {title} {\bibinfo {title} {{On Quantum Resonances in Stationary
  Geometries}},\ }\href {https://doi.org/10.1007/BF02725534} {\bibfield
  {journal} {\bibinfo  {journal} {Lett. Nuovo Cim.}\ }\textbf {\bibinfo
  {volume} {15}},\ \bibinfo {pages} {257} (\bibinfo {year} {1976})}\BibitemShut
  {NoStop}%
\bibitem [{\citenamefont {Detweiler}(1980)}]{Detweiler:1980uk}%
  \BibitemOpen
  \bibfield  {author} {\bibinfo {author} {\bibfnamefont {S.~L.}\ \bibnamefont
  {Detweiler}},\ }\bibfield  {title} {\bibinfo {title} {{Klein-Gordon equation
  and rotating black holes}},\ }\href
  {https://doi.org/10.1103/PhysRevD.22.2323} {\bibfield  {journal} {\bibinfo
  {journal} {Phys. Rev. D}\ }\textbf {\bibinfo {volume} {22}},\ \bibinfo
  {pages} {2323} (\bibinfo {year} {1980})}\BibitemShut {NoStop}%
\bibitem [{\citenamefont {Cardoso}\ \emph {et~al.}(2011)\citenamefont
  {Cardoso}, \citenamefont {Chakrabarti}, \citenamefont {Pani}, \citenamefont
  {Berti},\ and\ \citenamefont {Gualtieri}}]{Cardoso:2011xi}%
  \BibitemOpen
  \bibfield  {author} {\bibinfo {author} {\bibfnamefont {V.}~\bibnamefont
  {Cardoso}}, \bibinfo {author} {\bibfnamefont {S.}~\bibnamefont
  {Chakrabarti}}, \bibinfo {author} {\bibfnamefont {P.}~\bibnamefont {Pani}},
  \bibinfo {author} {\bibfnamefont {E.}~\bibnamefont {Berti}},\ and\ \bibinfo
  {author} {\bibfnamefont {L.}~\bibnamefont {Gualtieri}},\ }\bibfield  {title}
  {\bibinfo {title} {{Floating and sinking:\ The imprint of massive scalars
  around rotating black holes}},\ }\href
  {https://doi.org/10.1103/PhysRevLett.107.241101} {\bibfield  {journal}
  {\bibinfo  {journal} {Phys. Rev. Lett.}\ }\textbf {\bibinfo {volume} {107}},\
  \bibinfo {pages} {241101} (\bibinfo {year} {2011})},\ \Eprint
  {https://arxiv.org/abs/1109.6021} {arXiv:1109.6021 [gr-qc]} \BibitemShut
  {NoStop}%
\bibitem [{\citenamefont {Witek}\ \emph {et~al.}(2013)\citenamefont {Witek},
  \citenamefont {Cardoso}, \citenamefont {Ishibashi},\ and\ \citenamefont
  {Sperhake}}]{Witek:2012tr}%
  \BibitemOpen
  \bibfield  {author} {\bibinfo {author} {\bibfnamefont {H.}~\bibnamefont
  {Witek}}, \bibinfo {author} {\bibfnamefont {V.}~\bibnamefont {Cardoso}},
  \bibinfo {author} {\bibfnamefont {A.}~\bibnamefont {Ishibashi}},\ and\
  \bibinfo {author} {\bibfnamefont {U.}~\bibnamefont {Sperhake}},\ }\bibfield
  {title} {\bibinfo {title} {{Superradiant instabilities in astrophysical
  systems}},\ }\href {https://doi.org/10.1103/PhysRevD.87.043513} {\bibfield
  {journal} {\bibinfo  {journal} {Phys. Rev. D}\ }\textbf {\bibinfo {volume}
  {87}},\ \bibinfo {pages} {043513} (\bibinfo {year} {2013})},\ \Eprint
  {https://arxiv.org/abs/1212.0551} {arXiv:1212.0551 [gr-qc]} \BibitemShut
  {NoStop}%
\bibitem [{\citenamefont {Brito}\ \emph {et~al.}(2013)\citenamefont {Brito},
  \citenamefont {Cardoso},\ and\ \citenamefont {Pani}}]{Brito:2013wya}%
  \BibitemOpen
  \bibfield  {author} {\bibinfo {author} {\bibfnamefont {R.}~\bibnamefont
  {Brito}}, \bibinfo {author} {\bibfnamefont {V.}~\bibnamefont {Cardoso}},\
  and\ \bibinfo {author} {\bibfnamefont {P.}~\bibnamefont {Pani}},\ }\bibfield
  {title} {\bibinfo {title} {{Massive spin-2 fields on black hole spacetimes:\
  Instability of the Schwarzschild and Kerr solutions and bounds on the
  graviton mass}},\ }\href {https://doi.org/10.1103/PhysRevD.88.023514}
  {\bibfield  {journal} {\bibinfo  {journal} {Phys. Rev. D}\ }\textbf {\bibinfo
  {volume} {88}},\ \bibinfo {pages} {023514} (\bibinfo {year} {2013})},\
  \Eprint {https://arxiv.org/abs/1304.6725} {arXiv:1304.6725 [gr-qc]}
  \BibitemShut {NoStop}%
\bibitem [{\citenamefont {Brito}\ \emph
  {et~al.}(2017{\natexlab{a}})\citenamefont {Brito}, \citenamefont {Ghosh},
  \citenamefont {Barausse}, \citenamefont {Berti}, \citenamefont {Cardoso},
  \citenamefont {Dvorkin}, \citenamefont {Klein},\ and\ \citenamefont
  {Pani}}]{Brito:2017wnc}%
  \BibitemOpen
  \bibfield  {author} {\bibinfo {author} {\bibfnamefont {R.}~\bibnamefont
  {Brito}}, \bibinfo {author} {\bibfnamefont {S.}~\bibnamefont {Ghosh}},
  \bibinfo {author} {\bibfnamefont {E.}~\bibnamefont {Barausse}}, \bibinfo
  {author} {\bibfnamefont {E.}~\bibnamefont {Berti}}, \bibinfo {author}
  {\bibfnamefont {V.}~\bibnamefont {Cardoso}}, \bibinfo {author} {\bibfnamefont
  {I.}~\bibnamefont {Dvorkin}}, \bibinfo {author} {\bibfnamefont
  {A.}~\bibnamefont {Klein}},\ and\ \bibinfo {author} {\bibfnamefont
  {P.}~\bibnamefont {Pani}},\ }\bibfield  {title} {\bibinfo {title}
  {{Stochastic and resolvable gravitational waves from ultralight bosons}},\
  }\href {https://doi.org/10.1103/PhysRevLett.119.131101} {\bibfield  {journal}
  {\bibinfo  {journal} {Phys. Rev. Lett.}\ }\textbf {\bibinfo {volume} {119}},\
  \bibinfo {pages} {131101} (\bibinfo {year} {2017}{\natexlab{a}})},\ \Eprint
  {https://arxiv.org/abs/1706.05097} {arXiv:1706.05097 [gr-qc]} \BibitemShut
  {NoStop}%
\bibitem [{\citenamefont {Brito}\ \emph
  {et~al.}(2017{\natexlab{b}})\citenamefont {Brito}, \citenamefont {Ghosh},
  \citenamefont {Barausse}, \citenamefont {Berti}, \citenamefont {Cardoso},
  \citenamefont {Dvorkin}, \citenamefont {Klein},\ and\ \citenamefont
  {Pani}}]{Brito:2017zvb}%
  \BibitemOpen
  \bibfield  {author} {\bibinfo {author} {\bibfnamefont {R.}~\bibnamefont
  {Brito}}, \bibinfo {author} {\bibfnamefont {S.}~\bibnamefont {Ghosh}},
  \bibinfo {author} {\bibfnamefont {E.}~\bibnamefont {Barausse}}, \bibinfo
  {author} {\bibfnamefont {E.}~\bibnamefont {Berti}}, \bibinfo {author}
  {\bibfnamefont {V.}~\bibnamefont {Cardoso}}, \bibinfo {author} {\bibfnamefont
  {I.}~\bibnamefont {Dvorkin}}, \bibinfo {author} {\bibfnamefont
  {A.}~\bibnamefont {Klein}},\ and\ \bibinfo {author} {\bibfnamefont
  {P.}~\bibnamefont {Pani}},\ }\bibfield  {title} {\bibinfo {title}
  {{Gravitational wave searches for ultralight bosons with LIGO and LISA}},\
  }\href {https://doi.org/10.1103/PhysRevD.96.064050} {\bibfield  {journal}
  {\bibinfo  {journal} {Phys. Rev. D}\ }\textbf {\bibinfo {volume} {96}},\
  \bibinfo {pages} {064050} (\bibinfo {year} {2017}{\natexlab{b}})},\ \Eprint
  {https://arxiv.org/abs/1706.06311} {arXiv:1706.06311 [gr-qc]} \BibitemShut
  {NoStop}%
\bibitem [{\citenamefont {Cardoso}\ \emph {et~al.}(2018)\citenamefont
  {Cardoso}, \citenamefont {Dias}, \citenamefont {Hartnett}, \citenamefont
  {Middleton}, \citenamefont {Pani},\ and\ \citenamefont
  {Santos}}]{Cardoso:2018tly}%
  \BibitemOpen
  \bibfield  {author} {\bibinfo {author} {\bibfnamefont {V.}~\bibnamefont
  {Cardoso}}, \bibinfo {author} {\bibfnamefont {O.~J.~C.}\ \bibnamefont
  {Dias}}, \bibinfo {author} {\bibfnamefont {G.~S.}\ \bibnamefont {Hartnett}},
  \bibinfo {author} {\bibfnamefont {M.}~\bibnamefont {Middleton}}, \bibinfo
  {author} {\bibfnamefont {P.}~\bibnamefont {Pani}},\ and\ \bibinfo {author}
  {\bibfnamefont {J.~E.}\ \bibnamefont {Santos}},\ }\bibfield  {title}
  {\bibinfo {title} {{Constraining the mass of dark photons and axion-like
  particles through black-hole superradiance}},\ }\href
  {https://doi.org/10.1088/1475-7516/2018/03/043} {\bibfield  {journal}
  {\bibinfo  {journal} {J. Cosmol. Astropart. Phys.}\ }\textbf {\bibinfo
  {volume} {\normalfont03 (2018) 043}}},\ \Eprint
  {https://arxiv.org/abs/1801.01420} {arXiv:1801.01420 [gr-qc]} \BibitemShut
  {NoStop}%
\bibitem [{\citenamefont {Yunes}\ and\ \citenamefont
  {Siemens}(2013)}]{Yunes:2013dva}%
  \BibitemOpen
  \bibfield  {author} {\bibinfo {author} {\bibfnamefont {N.}~\bibnamefont
  {Yunes}}\ and\ \bibinfo {author} {\bibfnamefont {X.}~\bibnamefont
  {Siemens}},\ }\bibfield  {title} {\bibinfo {title} {{Gravitational-Wave Tests
  of General Relativity with Ground-Based Detectors and Pulsar
  Timing-Arrays}},\ }\href {https://doi.org/10.12942/lrr-2013-9} {\bibfield
  {journal} {\bibinfo  {journal} {Living Rev. Relativ.}\ }\textbf {\bibinfo
  {volume} {16}},\ \bibinfo {pages} {9} (\bibinfo {year} {2013})},\ \Eprint
  {https://arxiv.org/abs/1304.3473} {arXiv:1304.3473 [gr-qc]} \BibitemShut
  {NoStop}%
\bibitem [{\citenamefont {Berti}\ \emph {et~al.}(2015)\citenamefont {Berti}
  \emph {et~al.}}]{Berti:2015itd}%
  \BibitemOpen
  \bibfield  {author} {\bibinfo {author} {\bibfnamefont {E.}~\bibnamefont
  {Berti}} \emph {et~al.},\ }\bibfield  {title} {\bibinfo {title} {{Testing
  General Relativity with Present and Future Astrophysical Observations}},\
  }\href {https://doi.org/10.1088/0264-9381/32/24/243001} {\bibfield  {journal}
  {\bibinfo  {journal} {Class. Quantum Grav.}\ }\textbf {\bibinfo {volume}
  {32}},\ \bibinfo {pages} {243001} (\bibinfo {year} {2015})},\ \Eprint
  {https://arxiv.org/abs/1501.07274} {arXiv:1501.07274 [gr-qc]} \BibitemShut
  {NoStop}%
\bibitem [{\citenamefont {Konoplya}\ \emph {et~al.}(2018)\citenamefont
  {Konoplya}, \citenamefont {Stuchl{\'i}k},\ and\ \citenamefont
  {Zhidenko}}]{Konoplya:2018arm}%
  \BibitemOpen
  \bibfield  {author} {\bibinfo {author} {\bibfnamefont {R.~A.}\ \bibnamefont
  {Konoplya}}, \bibinfo {author} {\bibfnamefont {Z.}~\bibnamefont
  {Stuchl{\'i}k}},\ and\ \bibinfo {author} {\bibfnamefont {A.}~\bibnamefont
  {Zhidenko}},\ }\bibfield  {title} {\bibinfo {title} {{Axisymmetric black
  holes allowing for separation of variables in the Klein-Gordon and
  Hamilton-Jacobi equations}},\ }\href
  {https://doi.org/10.1103/PhysRevD.97.084044} {\bibfield  {journal} {\bibinfo
  {journal} {Phys. Rev. D}\ }\textbf {\bibinfo {volume} {97}},\ \bibinfo
  {pages} {084044} (\bibinfo {year} {2018})},\ \Eprint
  {https://arxiv.org/abs/1801.07195} {arXiv:1801.07195 [gr-qc]} \BibitemShut
  {NoStop}%
\bibitem [{\citenamefont {Konoplya}\ and\ \citenamefont
  {Zhidenko}(2016)}]{Konoplya:2016pmh}%
  \BibitemOpen
  \bibfield  {author} {\bibinfo {author} {\bibfnamefont {R.}~\bibnamefont
  {Konoplya}}\ and\ \bibinfo {author} {\bibfnamefont {A.}~\bibnamefont
  {Zhidenko}},\ }\bibfield  {title} {\bibinfo {title} {{Detection of
  gravitational waves from black holes:\ Is there a window for alternative
  theories?}},\ }\href {https://doi.org/10.1016/j.physletb.2016.03.044}
  {\bibfield  {journal} {\bibinfo  {journal} {Phys. Lett. B}\ }\textbf
  {\bibinfo {volume} {756}},\ \bibinfo {pages} {350} (\bibinfo {year}
  {2016})},\ \Eprint {https://arxiv.org/abs/1602.04738} {arXiv:1602.04738
  [gr-qc]} \BibitemShut {NoStop}%
\bibitem [{\citenamefont {Wang}\ \emph {et~al.}(2016)\citenamefont {Wang},
  \citenamefont {Chen},\ and\ \citenamefont {Jing}}]{Wang:2016paq}%
  \BibitemOpen
  \bibfield  {author} {\bibinfo {author} {\bibfnamefont {S.}~\bibnamefont
  {Wang}}, \bibinfo {author} {\bibfnamefont {S.}~\bibnamefont {Chen}},\ and\
  \bibinfo {author} {\bibfnamefont {J.}~\bibnamefont {Jing}},\ }\bibfield
  {title} {\bibinfo {title} {{Strong gravitational lensing by a
  Konoplya-Zhidenko rotating non-Kerr compact object}},\ }\href
  {https://doi.org/10.1088/1475-7516/2016/11/020} {\bibfield  {journal}
  {\bibinfo  {journal} {J. Cosmol. Astropart. Phys.}\ }\textbf {\bibinfo
  {volume} {\normalfont11 (2016) 020}}},\ \Eprint
  {https://arxiv.org/abs/1609.00802} {arXiv:1609.00802 [gr-qc]} \BibitemShut
  {NoStop}%
\bibitem [{\citenamefont {Wang}\ \emph {et~al.}(2017)\citenamefont {Wang},
  \citenamefont {Chen},\ and\ \citenamefont {Jing}}]{Wang:2017hjl}%
  \BibitemOpen
  \bibfield  {author} {\bibinfo {author} {\bibfnamefont {M.}~\bibnamefont
  {Wang}}, \bibinfo {author} {\bibfnamefont {S.}~\bibnamefont {Chen}},\ and\
  \bibinfo {author} {\bibfnamefont {J.}~\bibnamefont {Jing}},\ }\bibfield
  {title} {\bibinfo {title} {{Shadow casted by a Konoplya-Zhidenko rotating
  non-Kerr black hole}},\ }\href
  {https://doi.org/10.1088/1475-7516/2017/10/051} {\bibfield  {journal}
  {\bibinfo  {journal} {J. Cosmol. Astropart. Phys.}\ }\textbf {\bibinfo
  {volume} {\normalfont10 (2017) 051}}},\ \Eprint
  {https://arxiv.org/abs/1707.09451} {arXiv:1707.09451 [gr-qc]} \BibitemShut
  {NoStop}%
\bibitem [{\citenamefont {Konoplya}(2020)}]{Konoplya:2019xmn}%
  \BibitemOpen
  \bibfield  {author} {\bibinfo {author} {\bibfnamefont {R.~A.}\ \bibnamefont
  {Konoplya}},\ }\bibfield  {title} {\bibinfo {title} {{Quantum corrected black
  holes:\ quasinormal modes, scattering, shadows}},\ }\href
  {https://doi.org/10.1016/j.physletb.2020.135363} {\bibfield  {journal}
  {\bibinfo  {journal} {Phys. Lett. B}\ }\textbf {\bibinfo {volume} {804}},\
  \bibinfo {pages} {135363} (\bibinfo {year} {2020})},\ \Eprint
  {https://arxiv.org/abs/1912.10582} {arXiv:1912.10582 [gr-qc]} \BibitemShut
  {NoStop}%
\bibitem [{\citenamefont {Chandrasekhar}(1983)}]{Chandrasekhar}%
  \BibitemOpen
  \bibfield  {author} {\bibinfo {author} {\bibfnamefont {S.}~\bibnamefont
  {Chandrasekhar}},\ }\href@noop {} {\emph {\bibinfo {title} {{The Mathematical
  Theory of Black Holes}}}}\ (\bibinfo  {publisher} {Oxford University Press},\
  \bibinfo {address} {Oxford, UK},\ \bibinfo {year} {1983})\BibitemShut
  {NoStop}%
\bibitem [{\citenamefont {Suvorov}(2021)}]{Suvorov:2020bvk}%
  \BibitemOpen
  \bibfield  {author} {\bibinfo {author} {\bibfnamefont {A.~G.}\ \bibnamefont
  {Suvorov}},\ }\bibfield  {title} {\bibinfo {title} {{A family of solutions to
  the inverse problem in gravitation:\ building a theory around a metric}},\
  }\href {https://doi.org/10.1007/s10714-020-02779-8} {\bibfield  {journal}
  {\bibinfo  {journal} {Gen. Relativ. Gravit.}\ }\textbf {\bibinfo {volume}
  {53}},\ \bibinfo {pages} {6} (\bibinfo {year} {2021})},\ \Eprint
  {https://arxiv.org/abs/2008.02510} {arXiv:2008.02510 [gr-qc]} \BibitemShut
  {NoStop}%
\bibitem [{\citenamefont {Suvorov}\ and\ \citenamefont
  {V\"olkel}(2021)}]{Suvorov:2021amy}%
  \BibitemOpen
  \bibfield  {author} {\bibinfo {author} {\bibfnamefont {A.~G.}\ \bibnamefont
  {Suvorov}}\ and\ \bibinfo {author} {\bibfnamefont {S.~H.}\ \bibnamefont
  {V\"olkel}},\ }\bibfield  {title} {\bibinfo {title} {{Exact theory for the
  Rezzolla-Zhidenko metric and self-consistent calculation of quasinormal
  modes}},\ }\href {https://doi.org/10.1103/PhysRevD.103.044027} {\bibfield
  {journal} {\bibinfo  {journal} {Phys. Rev. D}\ }\textbf {\bibinfo {volume}
  {103}},\ \bibinfo {pages} {044027} (\bibinfo {year} {2021})},\ \Eprint
  {https://arxiv.org/abs/2101.09697} {arXiv:2101.09697 [gr-qc]} \BibitemShut
  {NoStop}%
\bibitem [{\citenamefont {Poisson}(2004)}]{PoissonToolkit}%
  \BibitemOpen
  \bibfield  {author} {\bibinfo {author} {\bibfnamefont {E.}~\bibnamefont
  {Poisson}},\ }\href@noop {} {\emph {\bibinfo {title} {{A Relativist's
  Toolkit}}}}\ (\bibinfo  {publisher} {Cambridge University Press},\ \bibinfo
  {address} {Cambridge, UK},\ \bibinfo {year} {2004})\BibitemShut {NoStop}%
\bibitem [{\citenamefont {Teukolsky}\ and\ \citenamefont
  {Press}(1974)}]{Teukolsky:1974yv}%
  \BibitemOpen
  \bibfield  {author} {\bibinfo {author} {\bibfnamefont {S.~A.}\ \bibnamefont
  {Teukolsky}}\ and\ \bibinfo {author} {\bibfnamefont {W.~H.}\ \bibnamefont
  {Press}},\ }\bibfield  {title} {\bibinfo {title} {{Perturbations of a
  rotating black hole. III\@. Interaction of the hole with gravitational and
  electromagnetic radiation}},\ }\href {https://doi.org/10.1086/153180}
  {\bibfield  {journal} {\bibinfo  {journal} {Astrophys. J.}\ }\textbf
  {\bibinfo {volume} {193}},\ \bibinfo {pages} {443} (\bibinfo {year}
  {1974})}\BibitemShut {NoStop}%
\bibitem [{\citenamefont {Leaver}(1985)}]{Leaver:1985ax}%
  \BibitemOpen
  \bibfield  {author} {\bibinfo {author} {\bibfnamefont {E.~W.}\ \bibnamefont
  {Leaver}},\ }\bibfield  {title} {\bibinfo {title} {{An analytic
  representation for the quasi normal modes of Kerr black holes}},\ }\href
  {https://doi.org/10.1098/rspa.1985.0119} {\bibfield  {journal} {\bibinfo
  {journal} {Proc. Roy. Soc. Lond. A}\ }\textbf {\bibinfo {volume} {402}},\
  \bibinfo {pages} {285} (\bibinfo {year} {1985})}\BibitemShut {NoStop}%
\bibitem [{\citenamefont {Reynolds}(2019)}]{Reynolds:2019uxi}%
  \BibitemOpen
  \bibfield  {author} {\bibinfo {author} {\bibfnamefont {C.~S.}\ \bibnamefont
  {Reynolds}},\ }\bibfield  {title} {\bibinfo {title} {{Observing black holes
  spin}},\ }\href {https://doi.org/10.1038/s41550-018-0665-z} {\bibfield
  {journal} {\bibinfo  {journal} {Nature Astron.}\ }\textbf {\bibinfo {volume}
  {3}},\ \bibinfo {pages} {41} (\bibinfo {year} {2019})},\ \Eprint
  {https://arxiv.org/abs/1903.11704} {arXiv:1903.11704 [astro-ph.HE]}
  \BibitemShut {NoStop}%
\bibitem [{dat()}]{data}%
  \BibitemOpen
  \href@noop {} {\bibinfo {title} {{\texttt{SuperradianceKZ} GitHub
  repository}}},\ \bibinfo {howpublished}
  {\url{https://github.com/efranzin/SuperradianceKZ/}}\BibitemShut {NoStop}%
\bibitem [{\citenamefont {Magalh\~aes}\ \emph
  {et~al.}(2020{\natexlab{a}})\citenamefont {Magalh\~aes}, \citenamefont
  {Leite},\ and\ \citenamefont {Crispino}}]{Magalhaes:2020pyp}%
  \BibitemOpen
  \bibfield  {author} {\bibinfo {author} {\bibfnamefont {R.~B.}\ \bibnamefont
  {Magalh\~aes}}, \bibinfo {author} {\bibfnamefont {L.~C.~S.}\ \bibnamefont
  {Leite}},\ and\ \bibinfo {author} {\bibfnamefont {L.~C.~B.}\ \bibnamefont
  {Crispino}},\ }\bibfield  {title} {\bibinfo {title} {{Absorption by deformed
  black holes}},\ }\href {https://doi.org/10.1016/j.physletb.2020.135418}
  {\bibfield  {journal} {\bibinfo  {journal} {Phys. Lett. B}\ }\textbf
  {\bibinfo {volume} {805}},\ \bibinfo {pages} {135418} (\bibinfo {year}
  {2020}{\natexlab{a}})},\ \Eprint {https://arxiv.org/abs/2004.08438}
  {arXiv:2004.08438 [gr-qc]} \BibitemShut {NoStop}%
\bibitem [{\citenamefont {Magalh\~aes}\ \emph
  {et~al.}(2020{\natexlab{b}})\citenamefont {Magalh\~aes}, \citenamefont
  {Leite},\ and\ \citenamefont {Crispino}}]{Magalhaes:2020sea}%
  \BibitemOpen
  \bibfield  {author} {\bibinfo {author} {\bibfnamefont {R.~B.}\ \bibnamefont
  {Magalh\~aes}}, \bibinfo {author} {\bibfnamefont {L.~C.~S.}\ \bibnamefont
  {Leite}},\ and\ \bibinfo {author} {\bibfnamefont {L.~C.~B.}\ \bibnamefont
  {Crispino}},\ }\bibfield  {title} {\bibinfo {title} {{Schwarzschild-like
  black holes:\ Light-like trajectories and massless scalar absorption}},\
  }\href {https://doi.org/10.1140/epjc/s10052-020-7909-7} {\bibfield  {journal}
  {\bibinfo  {journal} {Eur. Phys. J. C}\ }\textbf {\bibinfo {volume} {80}},\
  \bibinfo {pages} {386} (\bibinfo {year} {2020}{\natexlab{b}})},\ \Eprint
  {https://arxiv.org/abs/2005.04515} {arXiv:2005.04515 [gr-qc]} \BibitemShut
  {NoStop}%
\bibitem [{\citenamefont {Dolan}(2007)}]{Dolan:2007mj}%
  \BibitemOpen
  \bibfield  {author} {\bibinfo {author} {\bibfnamefont {S.~R.}\ \bibnamefont
  {Dolan}},\ }\bibfield  {title} {\bibinfo {title} {{Instability of the massive
  Klein-Gordon field on the Kerr spacetime}},\ }\href
  {https://doi.org/10.1103/PhysRevD.76.084001} {\bibfield  {journal} {\bibinfo
  {journal} {Phys. Rev. D}\ }\textbf {\bibinfo {volume} {76}},\ \bibinfo
  {pages} {084001} (\bibinfo {year} {2007})},\ \Eprint
  {https://arxiv.org/abs/0705.2880} {arXiv:0705.2880 [gr-qc]} \BibitemShut
  {NoStop}%
\bibitem [{\citenamefont {Brito}\ \emph {et~al.}(2015)\citenamefont {Brito},
  \citenamefont {Cardoso},\ and\ \citenamefont {Pani}}]{Brito:2014wla}%
  \BibitemOpen
  \bibfield  {author} {\bibinfo {author} {\bibfnamefont {R.}~\bibnamefont
  {Brito}}, \bibinfo {author} {\bibfnamefont {V.}~\bibnamefont {Cardoso}},\
  and\ \bibinfo {author} {\bibfnamefont {P.}~\bibnamefont {Pani}},\ }\bibfield
  {title} {\bibinfo {title} {{Black holes as particle detectors:\ evolution of
  superradiant instabilities}},\ }\href
  {https://doi.org/10.1088/0264-9381/32/13/134001} {\bibfield  {journal}
  {\bibinfo  {journal} {Class. Quantum Grav.}\ }\textbf {\bibinfo {volume}
  {32}},\ \bibinfo {pages} {134001} (\bibinfo {year} {2015})},\ \Eprint
  {https://arxiv.org/abs/1411.0686} {arXiv:1411.0686 [gr-qc]} \BibitemShut
  {NoStop}%
\bibitem [{\citenamefont {Arvanitaki}\ \emph {et~al.}(2015)\citenamefont
  {Arvanitaki}, \citenamefont {Baryakhtar},\ and\ \citenamefont
  {Huang}}]{Arvanitaki:2014wva}%
  \BibitemOpen
  \bibfield  {author} {\bibinfo {author} {\bibfnamefont {A.}~\bibnamefont
  {Arvanitaki}}, \bibinfo {author} {\bibfnamefont {M.}~\bibnamefont
  {Baryakhtar}},\ and\ \bibinfo {author} {\bibfnamefont {X.}~\bibnamefont
  {Huang}},\ }\bibfield  {title} {\bibinfo {title} {{Discovering the QCD Axion
  with Black Holes and Gravitational Waves}},\ }\href
  {https://doi.org/10.1103/PhysRevD.91.084011} {\bibfield  {journal} {\bibinfo
  {journal} {Phys. Rev. D}\ }\textbf {\bibinfo {volume} {91}},\ \bibinfo
  {pages} {084011} (\bibinfo {year} {2015})},\ \Eprint
  {https://arxiv.org/abs/1411.2263} {arXiv:1411.2263 [hep-ph]} \BibitemShut
  {NoStop}%
\bibitem [{\citenamefont {Gralla}\ \emph {et~al.}(2020)\citenamefont {Gralla},
  \citenamefont {Lupsasca},\ and\ \citenamefont {Marrone}}]{Gralla:2020srx}%
  \BibitemOpen
  \bibfield  {author} {\bibinfo {author} {\bibfnamefont {S.~E.}\ \bibnamefont
  {Gralla}}, \bibinfo {author} {\bibfnamefont {A.}~\bibnamefont {Lupsasca}},\
  and\ \bibinfo {author} {\bibfnamefont {D.~P.}\ \bibnamefont {Marrone}},\
  }\bibfield  {title} {\bibinfo {title} {{The shape of the black hole photon
  ring:\ A precise test of strong-field general relativity}},\ }\href
  {https://doi.org/10.1103/PhysRevD.102.124004} {\bibfield  {journal} {\bibinfo
   {journal} {Phys. Rev. D}\ }\textbf {\bibinfo {volume} {102}},\ \bibinfo
  {pages} {124004} (\bibinfo {year} {2020})},\ \Eprint
  {https://arxiv.org/abs/2008.03879} {arXiv:2008.03879 [gr-qc]} \BibitemShut
  {NoStop}%
\bibitem [{\citenamefont {V\"olkel}\ \emph {et~al.}(2020)\citenamefont
  {V\"olkel}, \citenamefont {Barausse}, \citenamefont {Franchini},\ and\
  \citenamefont {Broderick}}]{Volkel:2020xlc}%
  \BibitemOpen
  \bibfield  {author} {\bibinfo {author} {\bibfnamefont {S.~H.}\ \bibnamefont
  {V\"olkel}}, \bibinfo {author} {\bibfnamefont {E.}~\bibnamefont {Barausse}},
  \bibinfo {author} {\bibfnamefont {N.}~\bibnamefont {Franchini}},\ and\
  \bibinfo {author} {\bibfnamefont {A.~E.}\ \bibnamefont {Broderick}},\
  }\bibfield  {title} {\bibinfo {title} {{EHT tests of the strong-field regime
  of General Relativity}},\ }\href@noop {} {\  (\bibinfo {year} {2020})},\
  \Eprint {https://arxiv.org/abs/2011.06812} {arXiv:2011.06812 [gr-qc]}
  \BibitemShut {NoStop}%
\bibitem [{\citenamefont {Pani}\ \emph {et~al.}(2011)\citenamefont {Pani},
  \citenamefont {Macedo}, \citenamefont {Crispino},\ and\ \citenamefont
  {Cardoso}}]{Pani:2011gy}%
  \BibitemOpen
  \bibfield  {author} {\bibinfo {author} {\bibfnamefont {P.}~\bibnamefont
  {Pani}}, \bibinfo {author} {\bibfnamefont {C.~F.~B.}\ \bibnamefont {Macedo}},
  \bibinfo {author} {\bibfnamefont {L.~C.~B.}\ \bibnamefont {Crispino}},\ and\
  \bibinfo {author} {\bibfnamefont {V.}~\bibnamefont {Cardoso}},\ }\bibfield
  {title} {\bibinfo {title} {{Slowly rotating black holes in alternative
  theories of gravity}},\ }\href {https://doi.org/10.1103/PhysRevD.84.087501}
  {\bibfield  {journal} {\bibinfo  {journal} {Phys. Rev. D}\ }\textbf {\bibinfo
  {volume} {84}},\ \bibinfo {pages} {087501} (\bibinfo {year} {2011})},\
  \Eprint {https://arxiv.org/abs/1109.3996} {arXiv:1109.3996 [gr-qc]}
  \BibitemShut {NoStop}%
\bibitem [{\citenamefont {Liu}\ \emph {et~al.}(2012)\citenamefont {Liu},
  \citenamefont {Chen},\ and\ \citenamefont {Jing}}]{Liu:2012qe}%
  \BibitemOpen
  \bibfield  {author} {\bibinfo {author} {\bibfnamefont {C.}~\bibnamefont
  {Liu}}, \bibinfo {author} {\bibfnamefont {S.}~\bibnamefont {Chen}},\ and\
  \bibinfo {author} {\bibfnamefont {J.}~\bibnamefont {Jing}},\ }\bibfield
  {title} {\bibinfo {title} {{Rotating non-Kerr black hole and energy
  extraction}},\ }\href {https://doi.org/10.1088/0004-637X/751/2/148}
  {\bibfield  {journal} {\bibinfo  {journal} {Astrophys. J.}\ }\textbf
  {\bibinfo {volume} {751}},\ \bibinfo {pages} {148} (\bibinfo {year}
  {2012})},\ \Eprint {https://arxiv.org/abs/1207.0993} {arXiv:1207.0993
  [gr-qc]} \BibitemShut {NoStop}%
\bibitem [{\citenamefont {Long}\ \emph {et~al.}(2018)\citenamefont {Long},
  \citenamefont {Chen}, \citenamefont {Wang},\ and\ \citenamefont
  {Jing}}]{Long:2017xqr}%
  \BibitemOpen
  \bibfield  {author} {\bibinfo {author} {\bibfnamefont {F.}~\bibnamefont
  {Long}}, \bibinfo {author} {\bibfnamefont {S.}~\bibnamefont {Chen}}, \bibinfo
  {author} {\bibfnamefont {S.}~\bibnamefont {Wang}},\ and\ \bibinfo {author}
  {\bibfnamefont {J.}~\bibnamefont {Jing}},\ }\bibfield  {title} {\bibinfo
  {title} {{Energy extraction from a Konoplya--Zhidenko rotating non-Kerr black
  hole}},\ }\href {https://doi.org/10.1016/j.nuclphysb.2017.10.028} {\bibfield
  {journal} {\bibinfo  {journal} {Nucl. Phys. B}\ }\textbf {\bibinfo {volume}
  {926}},\ \bibinfo {pages} {83} (\bibinfo {year} {2018})},\ \Eprint
  {https://arxiv.org/abs/1707.03175} {arXiv:1707.03175 [gr-qc]} \BibitemShut
  {NoStop}%
\bibitem [{\citenamefont {Teukolsky}(1973)}]{Teukolsky:1973ha}%
  \BibitemOpen
  \bibfield  {author} {\bibinfo {author} {\bibfnamefont {S.~A.}\ \bibnamefont
  {Teukolsky}},\ }\bibfield  {title} {\bibinfo {title} {{Perturbations of a
  rotating black hole. I\@. Fundamental equations for gravitational
  electromagnetic and neutrino field perturbations}},\ }\href
  {https://doi.org/10.1086/152444} {\bibfield  {journal} {\bibinfo  {journal}
  {Astrophys. J.}\ }\textbf {\bibinfo {volume} {185}},\ \bibinfo {pages} {635}
  (\bibinfo {year} {1973})}\BibitemShut {NoStop}%
\bibitem [{\citenamefont {Kinnersley}(1969)}]{Kinnersley:1969zza}%
  \BibitemOpen
  \bibfield  {author} {\bibinfo {author} {\bibfnamefont {W.}~\bibnamefont
  {Kinnersley}},\ }\bibfield  {title} {\bibinfo {title} {{Type D Vacuum
  Metrics}},\ }\href {https://doi.org/10.1063/1.1664958} {\bibfield  {journal}
  {\bibinfo  {journal} {J. Math. Phys.}\ }\textbf {\bibinfo {volume} {10}},\
  \bibinfo {pages} {1195} (\bibinfo {year} {1969})}\BibitemShut {NoStop}%
\bibitem [{\citenamefont {Janis}\ and\ \citenamefont
  {Newman}(1965)}]{Janis:1965tx}%
  \BibitemOpen
  \bibfield  {author} {\bibinfo {author} {\bibfnamefont {A.~I.}\ \bibnamefont
  {Janis}}\ and\ \bibinfo {author} {\bibfnamefont {E.~T.}\ \bibnamefont
  {Newman}},\ }\bibfield  {title} {\bibinfo {title} {{Structure of
  Gravitational Sources}},\ }\href {https://doi.org/10.1063/1.1704349}
  {\bibfield  {journal} {\bibinfo  {journal} {J. Math. Phys.}\ }\textbf
  {\bibinfo {volume} {6}},\ \bibinfo {pages} {902} (\bibinfo {year}
  {1965})}\BibitemShut {NoStop}%
\bibitem [{\citenamefont {Goldberg}\ \emph {et~al.}(1967)\citenamefont
  {Goldberg}, \citenamefont {MacFarlane}, \citenamefont {Newman}, \citenamefont
  {Rohrlich},\ and\ \citenamefont {Sudarshan}}]{Goldberg:1966uu}%
  \BibitemOpen
  \bibfield  {author} {\bibinfo {author} {\bibfnamefont {J.~N.}\ \bibnamefont
  {Goldberg}}, \bibinfo {author} {\bibfnamefont {A.~J.}\ \bibnamefont
  {MacFarlane}}, \bibinfo {author} {\bibfnamefont {E.~T.}\ \bibnamefont
  {Newman}}, \bibinfo {author} {\bibfnamefont {F.}~\bibnamefont {Rohrlich}},\
  and\ \bibinfo {author} {\bibfnamefont {E.~C.~G.}\ \bibnamefont {Sudarshan}},\
  }\bibfield  {title} {\bibinfo {title} {{Spin-$s$ spherical harmonics and
  $\eth$}},\ }\href {https://doi.org/10.1063/1.1705135} {\bibfield  {journal}
  {\bibinfo  {journal} {J. Math. Phys.}\ }\textbf {\bibinfo {volume} {8}},\
  \bibinfo {pages} {2155} (\bibinfo {year} {1967})}\BibitemShut {NoStop}%
\bibitem [{\citenamefont {Press}\ and\ \citenamefont
  {Teukolsky}(1973)}]{Press:1973zz}%
  \BibitemOpen
  \bibfield  {author} {\bibinfo {author} {\bibfnamefont {W.~H.}\ \bibnamefont
  {Press}}\ and\ \bibinfo {author} {\bibfnamefont {S.~A.}\ \bibnamefont
  {Teukolsky}},\ }\bibfield  {title} {\bibinfo {title} {{Perturbations of a
  Rotating Black Hole. II\@. Dynamical Stability of the Kerr Metric}},\ }\href
  {https://doi.org/10.1086/152445} {\bibfield  {journal} {\bibinfo  {journal}
  {Astrophys. J.}\ }\textbf {\bibinfo {volume} {185}},\ \bibinfo {pages} {649}
  (\bibinfo {year} {1973})}\BibitemShut {NoStop}%
\bibitem [{\citenamefont {Berti}\ \emph {et~al.}(2006)\citenamefont {Berti},
  \citenamefont {Cardoso},\ and\ \citenamefont {Casals}}]{Berti:2005gp}%
  \BibitemOpen
  \bibfield  {author} {\bibinfo {author} {\bibfnamefont {E.}~\bibnamefont
  {Berti}}, \bibinfo {author} {\bibfnamefont {V.}~\bibnamefont {Cardoso}},\
  and\ \bibinfo {author} {\bibfnamefont {M.}~\bibnamefont {Casals}},\
  }\bibfield  {title} {\bibinfo {title} {{Eigenvalues and eigenfunctions of
  spin-weighted spheroidal harmonics in four and higher dimensions}},\ }\href
  {https://doi.org/10.1103/PhysRevD.73.109902} {\bibfield  {journal} {\bibinfo
  {journal} {Phys. Rev. D}\ }\textbf {\bibinfo {volume} {73}},\ \bibinfo
  {pages} {024013} (\bibinfo {year} {2006})},\ \bibinfo {note} {[Erratum: Phys.
  Rev. D {\bf73}, 109902 (2006)]},\ \Eprint
  {https://arxiv.org/abs/gr-qc/0511111} {arXiv:gr-qc/0511111} \BibitemShut
  {NoStop}%
\bibitem [{\citenamefont {Starobinskii}(1973)}]{Starobinskii:1973}%
  \BibitemOpen
  \bibfield  {author} {\bibinfo {author} {\bibfnamefont {A.~A.}\ \bibnamefont
  {Starobinskii}},\ }\bibfield  {title} {\bibinfo {title} {{Amplification of
  waves during reflection from a rotating black hole}},\ }\href@noop {}
  {\bibfield  {journal} {\bibinfo  {journal} {J. Exp. Theor. Phys.}\ }\textbf
  {\bibinfo {volume} {37}},\ \bibinfo {pages} {28} (\bibinfo {year}
  {1973})}\BibitemShut {NoStop}%
\bibitem [{\citenamefont {Starobinskii}\ and\ \citenamefont
  {Churilov}(1974)}]{Starobinskii:1974}%
  \BibitemOpen
  \bibfield  {author} {\bibinfo {author} {\bibfnamefont {A.~A.}\ \bibnamefont
  {Starobinskii}}\ and\ \bibinfo {author} {\bibfnamefont {S.~M.}\ \bibnamefont
  {Churilov}},\ }\bibfield  {title} {\bibinfo {title} {{Amplification of
  electromagnetic and gravitational waves scattered by a rotating black
  hole}},\ }\href@noop {} {\bibfield  {journal} {\bibinfo  {journal} {J. Exp.
  Theor. Phys.}\ }\textbf {\bibinfo {volume} {38}},\ \bibinfo {pages} {1}
  (\bibinfo {year} {1974})}\BibitemShut {NoStop}%
\bibitem [{\citenamefont {Page}(1976)}]{Page:1976df}%
  \BibitemOpen
  \bibfield  {author} {\bibinfo {author} {\bibfnamefont {D.~N.}\ \bibnamefont
  {Page}},\ }\bibfield  {title} {\bibinfo {title} {{Particle Emission Rates
  from a Black Hole:\ Massless Particles from an Uncharged, Nonrotating
  Hole}},\ }\href {https://doi.org/10.1103/PhysRevD.13.198} {\bibfield
  {journal} {\bibinfo  {journal} {Phys. Rev. D}\ }\textbf {\bibinfo {volume}
  {13}},\ \bibinfo {pages} {198} (\bibinfo {year} {1976})}\BibitemShut
  {NoStop}%
\end{thebibliography}%

\end{document}